\documentclass[entropy,article,submit,pdftex,moreauthors]{mdpi} 
\renewcommand{\linenumbers}{}

\usepackage[utf8]{inputenc}
\usepackage{natbib}
\usepackage{amssymb,amsmath}
\usepackage{epsfig}
\usepackage{bm}
\usepackage{color}
\usepackage{graphicx}

\usepackage{color}
\newcommand{\vicente}[1]{{#1}}

\newcommand\beq{\begin{equation}}
\newcommand\eeq{\end{equation}}
\newcommand\beqa{\begin{eqnarray}}
\newcommand\eeqa{\end{eqnarray}}
\newcommand{\dd}{\text{d}}

\newcommand{\al}{\alpha}

\firstpage{1} 
\pubvolume{1}
\issuenum{1}
\articlenumber{0}
\pubyear{2024}
\copyrightyear{2024}
\datereceived{} 
\dateaccepted{} 
\datepublished{} 
\hreflink{https://doi.org/} 

\Title{Dynamic properties in a collisional model for confined granular fluids. A review}
\TitleCitation{Dynamic properties of confined granular fluids}

\Author{Ricardo Brito $^{1,\dagger}$\orcidA{}, 
Rodrigo Soto $^{2,\dagger}$\orcidB{},
Vicente Garz\'{o} $^{3,\dagger}$\orcidC{}}
\AuthorNames{Ricardo Brito, Rodrigo Soto and Vicente Garz\'{o}}
\AuthorCitation{Brito, R.; Soto, R.; Garz\'o, V.}

\address{%
$^{1}$ \quad Departamento de Estructura de la Materia, F\'{\i}sica T\'ermica y Electr\'onica and GISC, 
Universidad Complutense de Madrid, Spain; brito@ucm.es\\
$^{2}$ \quad Departamento de F\'{\i}sica, Facultad de Ciencias F\'{\i}sicas y Matem\'aticas, Universidad de Chile, Santiago, Chile; rsoto@uchile.cl\\
$^{3}$ \quad Departamento de F\'{\i}sica and Instituto de Computaci\'on Cient\'{\i}fica Avanzada (ICCAEx), Universidad de Extremadura, Avda. de Elvas s/n, E-06006 Badajoz, Spain; vicenteg@unex.es}

\corres{Correspondence: vicenteg@unex.es; URL: https://fisteor.cms.unex.es/investigadores/vicente-garzo-puertos/}
\firstnote{These authors contributed equally to this work.}

\abstract{
Granular systems confined in a shallow box and subjected to vertical vibration provide an attractive geometry for studying fluidized granular media. In this configuration, grains acquire kinetic energy in the vertical direction through collisions with the confining walls, and this energy is subsequently transferred to the horizontal degrees of freedom via interparticle collisions. In recent years, the so-called $\Delta$-model has been introduced as a simplified yet effective description of the dynamics of granular systems in such geometries. This review presents the results obtained from kinetic theory for the granular $\Delta$-model. To model the energy transfer mechanism, a fixed velocity increment $\Delta$ is added to the normal component of the relative velocity during collisions. In this way, the vertical motion is effectively integrated out while retaining the collisional energy injection characteristic of the confined setup. This mechanism compensates for the energy loss due to inelastic collisions and leads to stable homogeneous steady states that can be analyzed within the framework of kinetic theory. The Enskog kinetic equation is formulated for this model and first analyzed in homogeneous steady states, yielding the stationary temperature and the equation of state. The dynamics of inhomogeneous states is then investigated using the Chapman--Enskog method, from which the Navier--Stokes transport coefficients are derived. The theory is further extended to granular mixtures, in which particles may differ in mass, size, restitution coefficient, or in the value of $\Delta$. In this case, the phenomenology becomes richer; for example, energy equipartition is violated even in homogeneous steady states. The mixture dynamics is studied through the corresponding Navier--Stokes equations, and the associated transport coefficients are obtained in the low-density regime. The analysis of the hydrodynamic equations shows that, in agreement with simulations, the homogeneous state is linearly stable. Moreover, the intrinsically nonequilibrium nature of the model leads to the violation of Onsager reciprocity relations in granular mixtures. The theoretical predictions exhibit in general good agreement with both molecular dynamics simulations and direct simulation Monte Carlo results.
}


\begin{document}

\tableofcontents

\section{Introduction}
\label{sec1}

Granular materials constitute a broad class of many-body systems whose macroscopic behavior emerges from dissipative interactions of the particles that are their constituents, called {\em grains} 
\cite{Faraday1831,C90,JaegerRevModPhys,Duran,G03,AT06}. 
Unlike molecular systems, however, collisions between grains are intrinsically inelastic, leading to a continuous loss of kinetic energy~\cite{H83,BritoErnst1998}. This feature is responsible for the unusual behavior of granular systems and the main source of its rich phenomenology as opposed to their conservative counterparts 
\cite{Hermann1998,AFP13}.  
As a consequence, granular fluids are inherently nonequilibrium systems: in the absence of external energy input, they cool down monotonically and eventually come to rest in the form of sand piles~\cite{AFP13} or, in the absence of boundaries in microgravity experiments~\cite{FWEFChGB99} or in simulations with periodic boundary conditions, via a nontrivial, nonhomogeneous state 
\cite{MY94,GZ93} that generates long range correlations 
\cite{PhysRevLett.79.411}. 
Sustained dynamical states therefore require some form of driving, which compensates for collisional dissipation and maintains a continuous motion and  kinetic activity. The intrinsic nonequilibrium nature of granular matter, together with the energy input, that drives the system out of equilibrium even further, leads to a really rich behavior 
\cite{AransonBook}. 
In practice, different experimental realizations of driven granular systems correspond to different modes of energy input. Some examples of driving, such as avalanche flows on inclined plates~\cite{Pouliquen99,Ecke05}, chute flows~\cite{Savage_Lun_1988}, rotating drums~\cite{Nagel89,Ristow97}, vibrating boundaries 
\cite{MUS95}, air-fluidization beds 
\cite{Valverde03}, sheared systems 
\cite{RietzSwinney}, horizontal shaking 
\cite{Mullin2000,Ciamarra2006} or bulk forcing 
\cite{Hill97} result in different dynamical states, which are typically spatially inhomogeneous, with regions of high density, eventually in solid-like configurations.
The choice of driving is therefore not just technical: it strongly influences the stationary states, transport properties, and stability of the system \cite{AT06,K04}. 
One class of driving is obtained by forcing  via the boundaries. For instance systems where energy is supplied through collisions with vibrating or moving walls 
\cite{OU98,Ciamarra2005,Ristow97,Aumaitre03,SBKR05,MS16} or computer simulation equivalent \cite{Grossman97}. This type of driving is particularly relevant experimentally, but it introduces shock waves or boundary layers that complicate the theoretical description 
\cite{VPBTW06,VPBWT06,MB97,Kumaran98,Barrat2002}. 

\begin{figure}[h!]
\centering
\includegraphics[width=\textwidth]{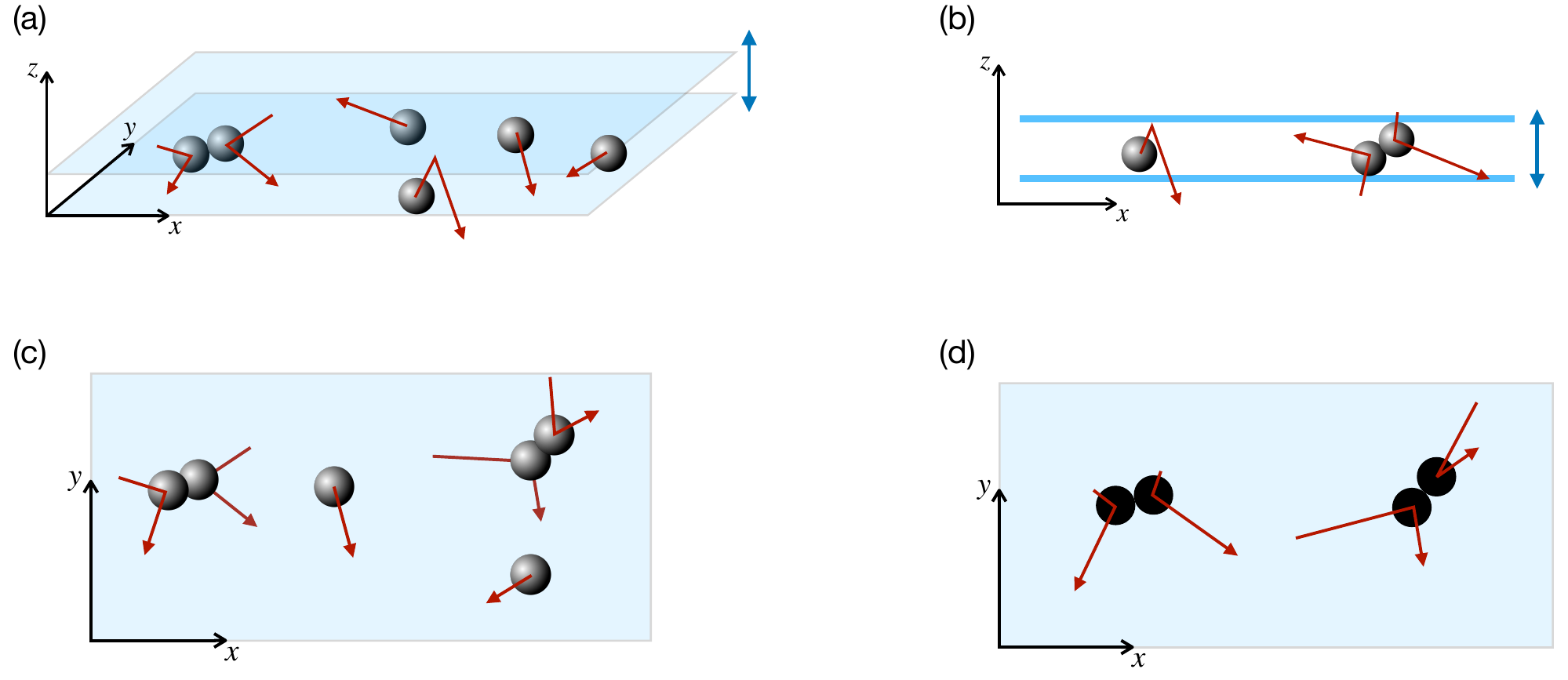}
\caption{Fig: Conceptual motivation of the $\Delta$-model. (a) Quasi two-dimensional setup, where spherical grains are placed in a vertically vibrating shallow box. Grains can collide with the vibrating walls and among themselves. (b) Lateral view of the system. Grain collisions with the top and bottom walls inject energy into the vertical degrees of freedom, which is later transferred to the horizontal ones via grain-grain oblique collisions. (c) Top view of the quasi two-dimensional system. As the height of the box is larger than the particle diameters, they can partially overlap at collisions when seeing from above.
(d) In the $\Delta$-model, the vertical motion is abstracted out keeping its effect on injecting energy into the horizontal degrees of freedom. If particles reach the collision with a small relative velocity, the net effect is to gain energy, but if their normal relative velocity is large, inelasticity overcomes the injection and the collision is dissipative. Note that in the $\Delta$-model, particles move only in $x$ and $y$, implying that there is no overlap and collisions take place when the distance is exactly equal to the particle diameter.
}
\label{fig.setup}
\end{figure}

An important class of driven granular systems corresponds to vertical vibration of quasi two-dimensional (Q2D) systems  
\cite{OU98,OU05,Prevost2004,Clerc08,Castillo12,MVPRKEU05,MS16,RCJBHS11,RCJBHS12}. In these systems, energy is injected through collisions with a vertically vibrating plate or shaker, specially in monolayers. A common way to make monolayers is to cover the experiment with a glass lid at a height slightly larger than a diameter grain. 
This configuration forces the particles to remain in the {\em quasi-}two-dimensional plane. Grain collisions with the lower and upper plate energize the $z$-component of the velocity \cite{Khain11}. While the vertical motion is directly excited by the driving, horizontal motion emerges  indirectly through grain-grain collisions, which transfer energy from vertical to horizontal degrees of freedom (see Fig.~\ref{fig.setup}). 
Usually, the vertical dynamics is fast compared to the horizontal one, and therefore it is natural to seek an effective two-dimensional description in which the net effect of confinement and vibration will be encoded in modified collision rules for the horizontal velocities. 
The advantage of these systems is twofold. On the experimental side, particles are easy to track in a monolayer, just by placing a camera on top of the experimental setting. On the theoretical description, this system can be treated as purely two-dimensional, eliminating configurations where particles stack on top of each other. 
This Q2D setup is particularly relevant for the development of theories of granular matter as it is direct to control the particle density, from very low gas-like regimes, to dense solid-like states. Also, in a wide region of parameter space, the system reaches steady states that are statistically homogeneous in the planar directions, as observed both experimentally and in computer simulations~\cite{OU98,Clerc08,MPGM23,BRS13}. They constitute, therefore, an excellent playground for studying granular hydrodynamic theories, which are normally built making gradient expansions around homogeneous states, contrary to the case of undriven granular gases, which can become uncontrollably inhomogeneous~\cite{GZ93,K99}. The Q2D experimental setup is the inspiration for the theoretical collisional model, the $\Delta$-model~\cite{BRS13}, which we analyze in this review. 

When the particles are sufficiently dilute and interact primarily through instantaneous binary collisions, granular matter can be described as a granular gas or fluid if the density is increased. 
In that spirit, and over the past decades, kinetic theory has played a central role in the theoretical understanding of granular gases. By extending the tools originally developed for molecular fluids to dissipative dynamics, kinetic theory provides a mesoscopic description that connects microscopic collision rules with macroscopic transport and collective phenomena. Starting from inelastic generalizations of the Boltzmann and Enskog equations, it has been possible to derive hydrodynamic equations, compute transport coefficients, analyze linear and nonlinear instabilities, and compare theoretical predictions with numerical simulations and experiments. Comprehensive accounts of these developments can be found in standard monographs~\cite{BP04,G19,Dorfman2021} and reviews \cite{ChGG22} of granular kinetic theory. 

However, when writing a kinetic equation for a granular fluid, one faces the problem of how to model the driving.  While the dissipative nature of collisions is well captured by a coefficient of normal restitution for the inelastic hard sphere (IHS) model, the mechanism by which energy is injected into the system is model-dependent. 
Modelization of transferal from vertical to horizontal degrees of freedom that takes place in the Q2D geometry is not an easy task. The parametrization of collisions in these confined conditions is cumbersome and hinders a straightforward form for the kinetic equation~\cite{Khain11,MGB19}. An alternative approach consists on building models that consider, in an effective way, the energy gain on the horizontal degrees of freedom in the Q2D geometry. Among these driving mechanisms are the so-called {\em thermostats}. These models are advantageous from the viewpoint of formulating kinetic equations 
\cite{NE98,Montanero00}. 
One widely used approach consists of adding external forces acting on individual particles 
\cite{WM96,PLMPV98,PLMV99,PengOhta1998}, such as stochastic (white-noise) forcing. Such models are analytically convenient when writing a kinetic equation, as energy injection acts on the particles, so they preserve homogeneity. Some of these drivings appear as additional Fokker--Planck terms in the kinetic equation and have been studied in great detail, including the derivation of steady state solutions~\cite{NE98},  velocity distributions \cite{Montanero00}, or hydrodynamic descriptions \cite{GM02,GMT13,GChV13,KG13,KG18}, validated with computer simulations. 

From a conceptual point of view, both thermostats and boundary driving introduce energy into the system through mechanisms that are external to the collisional dynamics between grains. Another option, inspired by the vertical-to-horizontal energy injection in the Q2D geometry (Fig.~\ref{fig.setup}.b-c), is to develop models in which energy injection is incorporated more directly into the collision process itself with particles moving purely in two-dimensions (Fig.~\ref{fig.setup}.d). The first of these models considers random restitution coefficients with values smaller (dissipative) or larger (energy injection) than one 
\cite{Trizac2001}. However, the system lacks an intrinsic energy scale, and the total energy of the system behaves like a random walk, and therefore no stationary state is reached. Moreover, it does not reproduce the power law decay of the velocity distribution \cite{Barrat2003}.
Such collisional models modify the binary collision rules so that collisions can either dissipate or inject energy, depending on the velocities of the colliding pair of particles. 
The advantage of these models is that they preserve the structure of the Boltzmann or Enskog equation, as they include the driving mechanism into the collision operator. 
The hope is that this modification still makes it possible to use standard techniques of kinetic theory to study such systems (driven steady states) without introducing external forces or boundary terms, and to analyze their properties within a unified kinetic theory framework. 
\vicente{An alternative  approach to modeling driven dissipative systems \cite{PNASLeiNi,PRLMaire2025} is based on a hybrid framework, in which energy is injected during collisions while dissipation, occurs during the free flight between them, instead via a normal restitution coefficient. More specifically, at each collision an amount $\Delta E > 0$ is added to the post-collisional kinetic energy. In contrast, viscous damping acts during the free-flight stage according to 
$\dot{\mathbf{v}}_i = -\gamma \mathbf{v}_i$, where $\gamma$ is the friction coefficient.
These articles derive hydrodynamic equations for the system. A key result is the emergence of hyperuniform states, which are locally disordered (fluid-like) yet exhibit long-range order akin to crystalline structures.} 

The so-called $\Delta$-model \cite{BRS13} has emerged as a suitable description for the kinetic treatment  of these confined Q2D systems. In this model, inelastic hard-sphere (or hard-disk) collisions, characterized by a normal restitution coefficient,  are supplemented by an additional velocity increment of fixed magnitude $\Delta$ along the normal collision direction. Physically, this increment represents the effective transfer of kinetic energy from vertical to horizontal motion during interparticle collisions in the confined geometry.
The $\Delta$-model can be viewed as a minimal extension of the standard IHS model. It retains a binary collision structure of the collision term while maintaining momentum conservation.  The inclusion of the $\Delta$ term that adds that amount of velocity, serves as a thermostat that balances the dissipation of the normal restitution coefficient. As a result, the energy change per collision can be either negative or positive, depending on the pre-collisional state, and leads to an stationary, non equilibrium, steady state. \vicente{As a matter of fact, some collisions can dissipate energy (dissipation dominates) while some others gain energy (due to the $\Delta$ injection mechanism). The steady state is reached when these two contributions balance on average, so that the global rate of energy change vanishes and consequently, the granular temperature attains a stationary asymptotic value.}  In this aspect, the $\Delta$-model differs from the random restitution model in that, in the latter, the energy change is uncorrelated with the pre-collisional state, resulting in the absence of a well defined steady state.

The steady state of the $\Delta$-model for a single component is stable even for long wavelength perturbations, as opposite to freely cooling granular fluids, where long enough wavelength perturbations lead to vortex formation and clustering \cite{GZ93,NEB98b,BDKS98,G03,BR13,MGH14}. The stability manifests in the equation of state of the fluid, where the dependence on density and temperature on pressure factorizes~\cite{BRS13}. It has then the inconvenience that the $\Delta$-model cannot reproduce clustering effects observed in some experiments. However, this stability allows one to control spatial gradients and to apply systematic hydrodynamic expansions in a manner closer to that of molecular fluids,  in particular Champan--Enskog--like expansions \cite{CC70}, as will be shown in the present review. An extension of the $\Delta$-model considers that each particle carries an internal variable which models the energy gained in the vertical direction since the last collision and the  value of $\Delta$ depends on this variable. This results in an equation of state that presents a van der Waals loop, leading to a clustering instability~\cite{RSG18}. 

Prior to a formal kinetic study of the $\Delta$-model, its basic physical mechanisms and macroscopic equations were derived in Ref.~\cite{BRS13}. There it was demonstrated that the system reaches a  non-equilibrium steady  state.  The study of fluctuations around that state, via Landau--Placzeck theory was carried out for the density and velocity fluctuations, putting emphasis on the relevance of the energy, strictly nonconserved, but that can be considered a quasi-conserved quantity. In a second study\cite{SRB14}, the shear viscosity was derived by a simple linear response theory. Parallel to these developments, Brey and coworkers 
analyzed several aspects of the model, like the velocity distribution function \cite{BMGB14}, the hydrodynamic behavior \cite{BMGB14,BBMG15}, with special emphasis on the structure and stability of homogeneous steady states \cite{BBGM16} and the existence of a normal or hydrodynamic solution \cite{BMGB14}. \vicente{The evolution equations for both the in-plane temperature and the $z$-component of the temperature were derived in Refs.~\cite{MGB19a,MGB19,MPGM23}, yielding explicit expressions that depend on the vibration frequency and the separation between the plates.  Remarkably, the stationary temperature of the vibrated system qualitative resembles that of the $\Delta$ model (see, e.g. Fig.~4 of Ref.~\cite{MGB19}). References~\cite{BGMB14bis,BBGM17} compare the predictions of the $\Delta$ model  with computer simulations of the vertically driven system, and excellent agreement between simulations and the results of $\Delta$ model is found. In contrast, alternative models such as the stochastic thermostat model \cite{WM96} 
show a significantly worse agreement. These results support the conclusion that the $\Delta$ model provides an accurate description of the vibrated monolayer outside the clustering regime. As a side remark, Ref.~\cite{MGB19a} also reports  the appearance of Mpemba effect in thin vibrated granular gases, in agreement with observation in other dissipative systems \cite{Mpemba}.}

The $\Delta$-model was then extended to mixtures of granular particles, where two or more species coexist \cite{GBS18}. The species may be distinguished by material properties, such as mass or diameter, or by dynamical ones, such as restitution coefficient or different values of the $\Delta$ parameter. The study of mixtures of granular materials was first addressed by Jenkins and Mancini in Ref.~\cite{JM87} by assuming the equipartition of energy (i.e., $T_i=T$, where $T_i$ is the partial temperature of species $i$ and $T$ is the global granular temperature). However, later studies \cite{GD99b} clearly show that granular mixtures present a phenomenology much richer than that of (equilibrium) molecular mixtures since under several types of forcing, energy equipartition is broken and each species reaches a different granular temperature~\cite{Feitosa2002,CH02,WP02}. 
Such lack of equipartition is quite general and is observed even in a single component between the translational and rotational degrees of freedom 
\cite{Huthmann1997,McNamara1998,Cafiero2002}. As expected, the $\Delta$-model also displays this remarkable phenomenon~\cite{BSG20}.
Granular mixtures also exhibit various segregation phenomena, in which particles with similar properties may cluster \cite{Aumaitre03} or preferentially migrate to different regions of the container~\cite{SBKR05}, giving rise to effects such as the Brazil nut and reverse Brazil nut effects~\cite{Rosato1987,Huerta2004,Breu2003,Shinbrot2004,G08}. The $\Delta$-model for mixtures shows analogous Brazil and reverse Brazil nut behavior~\cite{GBS24,GGBS24}. 

More recently, the $\Delta$-model has attracted renewed attention through a series of studies due to Foffi and coworkers. In a recent publication~\cite{PMFBRSF24} they showed that a mixture of vibrated grains can form quasicrystals, and the $\Delta$-model is a {\em bona fide} model to describe them. Moreover, the model presents long range order~\cite{joyce2016attractor,MP24}, as evidenced by hyperuniformity~\cite{Maire_2025}. Finally, the model has been used to study certain absorbing phases in granular systems \cite{MPSTSF24,Maire_2025} \vicente{(introducing a friction term as  in Refs.\  \cite{PNASLeiNi,PRLMaire2025})},  the coexistence between a fluid and a crystalline phase in granular fluids \cite{MPSF25}, \vicente{and the dynamics of non-equilibrium interfaces~\cite{PRLMaire2025}}. 
\vicente{Variations of the $\Delta$-model include models where instead of adding a fixed velocity, a fixed amount of energy is given at the collision~\cite{PNASLeiNi}. The collision rules change but the main phenomenology of the $\Delta$-model is preserved. An important variation is when $\Delta$ is made to depend on the time since last collision to induce phase separation~\cite{RSG18,MPSTSF24,PRLMaire2025}.
}
\vicente{The $\Delta$-model can also be used as an appropriate description for a class of active matter where activity is not in the form of self-propulsion, but in the capacity to inject energy into the system. For example, when active spinners collide, the translational degrees of freedom effectively gain  energy at the collision~\cite{PNASLeiNi,liu2025hyperuniform}. This energy injection has been casted into a variation of the $\Delta$-model, where the additional velocity is in the tangential rather than in the normal direction, generating a chiral fluid with odd rheological properties~\cite{maire2026kinetic}.}
\vicente{These results demonstrate} the versatility of the $\Delta$ model in linking microscopic driving mechanisms with the emergence of complex macroscopic  behavior.

The present review is organized as follows. In Section \ref{sec2} we introduce the $\Delta$-collisional model and formulate the corresponding Enskog kinetic equation, from where the balance equations are obtained. Section \ref{sec3}  is devoted to the analysis of homogeneous states, including the properties of the rate of energy and the existence of steady solutions. In Section~\ref{sec4} we apply the Chapman--Enskog method to derive the Navier--Stokes hydrodynamic equations and obtain explicit expressions for the transport coefficients.  
The extension of the kinetic equation to granular mixtures is presented in Section \ref{sec6}, where the analysis of time dependent homogeneous states is performed. The derivation of the Navier--Stokes equations for mixtures and the calculation of the transport coefficients is performed in Section \ref{sec8}. Issues such as the breakdown of Onsager relations on granular mixtures and the stability of homogeneous states are addressed in Section \ref{sec9}. Finally, we summarize the main results and discuss open problems and possible directions for future research in Section \ref{sec10}.

\section{Enskog kinetic equation for collisional model of confined granular fluids}
\label{sec2}

\subsection{Collisional model}

We consider a granular fluid modeled as a gas of inelastic hard spheres of mass $m$ and diameter $\sigma$. For the sake of simplicity, henceforth we will assume that the spheres are completely smooth and so, the inelasticity of binary collisions is only characterized by a constant positive coefficient of normal restitution $\al\leq 1$. The case $\al=1$ corresponds to elastic collisions. In the case of smooth particles, the inelastic character of collisions only affects to the translational degrees of freedom of grains. As mentioned in Section \ref{sec1}, we are interested here in analyzing the dynamic properties in \textit{confined} granular fluids. However, due to the technical difficulties associated with the restrictions imposed by the confinement in the Boltzmann or Enskog collision operators~\cite{Khain11,MGB19}, is it quite usual in the granular literature to adopt a coarse-grained approach in which the effect of confinement on grain dynamics is accounted for in an effective way. In this context, we consider in this paper a collisional model (the $\Delta$-collisional model) proposed years ago by Brito \textit{et al.} \cite{BRS13}. In this model, the factor $\Delta>0$ is introduced in the scattering rules to mimic the transfer of kinetic energy from the vertical degrees of freedom of grains (which has been gained by the collisions of particles with the vibrating walls) to the horizontal ones. The relationship between the pre-collisional $(\mathbf{v}_1, \mathbf{v}_2)$ and post-collisional $(\mathbf{v}_1',\mathbf{v}_2')$ velocities in the $\Delta$-model is~\cite{BRS13}
\beq
\label{2.1}
\mathbf{v}_1'=\mathbf{v}_1-\frac{1}{2}\left(1+\alpha\right)(\widehat{{\boldsymbol {\sigma}}}\cdot \mathbf{g}_{12})\widehat{{\boldsymbol {\sigma }}}-\Delta \widehat{{\boldsymbol {\sigma }}},\quad 
{\bf v}_{2}'=\mathbf{v}_{2}+\frac{1}{2}\left(1+\alpha\right)
(\widehat{{\boldsymbol {\sigma}}}\cdot \mathbf{g}_{12})
\widehat{\boldsymbol {\sigma}}+\Delta \widehat{{\boldsymbol {\sigma }}}.
\eeq
In Equation~\eqref{2.1}, $\mathbf{g}_{12}=\mathbf{v}_1-\mathbf{v}_2$ is the relative velocity of the two colliding spheres, $\widehat{{\boldsymbol {\sigma}}}$ is a unit vector pointing from the center of particle 1 to the center of particle 2, and particles are approaching if $\widehat{{\boldsymbol {\sigma}}}\cdot \mathbf{g}_{12}>0$. In addition, the parameter $\Delta$ is an extra velocity added to the relative motion. This extra velocity points outward in the normal direction $\widehat{\boldsymbol {\sigma}}$, as required by the conservation of angular momentum~\cite{L04bis}. The relative velocity after collision is
\beq
\label{2.2}
\mathbf{g}_{12}'=\mathbf{v}_1'-\mathbf{v}_2'=\mathbf{g}_{12}-(1+\al)(\widehat{{\boldsymbol {\sigma}}}\cdot \mathbf{g}_{12})
\widehat{\boldsymbol {\sigma}}-2\Delta \widehat{{\boldsymbol {\sigma }}},
\eeq
so that it is quite simple to get the relation 
\beq
\label{2.3}
(\widehat{{\boldsymbol {\sigma}}}\cdot \mathbf{g}_{12}')=-\al (\widehat{{\boldsymbol {\sigma}}}\cdot \mathbf{g}_{12})-2\Delta.
\eeq

According to the collision rules \eqref{2.1}, the total momentum is conserved in a binary collision ($\mathbf{v}_1+\mathbf{v}_2=\mathbf{v}_1'+\mathbf{v}_2'$) but the total kinetic energy is not conserved as expected. The change in kinetic energy upon collision is
\beq
\label{2.4}
\Delta E\equiv \frac{m}{2}\left(v_1^{'2}+v_2^{'2}-v_1^2-v_2^2\right)=m\left[\Delta^2+\al \Delta (\widehat{{\boldsymbol {\sigma}}}\cdot \mathbf{g}_{12})-\frac{1-\al^2}{4}(\widehat{{\boldsymbol {\sigma}}}\cdot \mathbf{g}_{12})^2\right].
\eeq
It is quite apparent that (i) the right-hand side of Equation\ \eqref{2.4} vanishes for elastic collisions ($\al=1$) and $\Delta=0$ and that (ii) $\Delta E>0$ (energy can be gained in collisions) or $\Delta E<0$ (energy can be lost in collisions) depending on whether $\widehat{{\boldsymbol {\sigma}}}\cdot \mathbf{g}_{12}$ is smaller than or larger than $2\Delta /(1-\al)$. \vicente{Moreover, as we will show later, the average value of change in kinetic energy vanishes ($\left<\Delta E \right>=0$) in the steady state. Thus,  the injection of energy due to the parameter $\Delta$ and collision dissipation cancels out on average in the asymptotic steady state.}

It is also convenient to consider the \textit{inverse} or restituting collision where $\left(\mathbf{v}_1'',\mathbf{v}_2''\right)$ are the pre-collisional velocities while $\left(\mathbf{v}_1,\mathbf{v}_2\right)$ are the post-collisional velocities with the same collision vector $\widehat{{\boldsymbol {\sigma }}}$:
\beq
\label{2.5}
\mathbf{v}_1''=\mathbf{v}_1-\frac{1}{2}\left(1+\alpha^{-1}\right)(\widehat{{\boldsymbol {\sigma}}}\cdot \mathbf{g}_{12})\widehat{{\boldsymbol {\sigma}}}-\alpha^{-1}\Delta \widehat{{\boldsymbol {\sigma}}},\quad
\mathbf{v}_2''=\mathbf{v}_2+\frac{1}{2}\left(1+\alpha^{-1}\right)(\widehat{{\boldsymbol {\sigma}}}\cdot \mathbf{g}_{12})\widehat{{\boldsymbol {\sigma}}}+\alpha^{-1}\Delta \widehat{{\boldsymbol {\sigma }}}. 
\eeq
According to Equation \eqref{2.5}, the relationship between the relative velocities $\mathbf{g}_{12}''=\mathbf{v}_1''-\mathbf{v}_2''$ and $\mathbf{g}_{12}=\mathbf{v}_1-\mathbf{v}_2$ is
\beq
\label{2.6}
\mathbf{g}_{12}''=\mathbf{g}_{12}-(1+\al^{-1})(\widehat{{\boldsymbol {\sigma}}}\cdot \mathbf{g}_{12})
\widehat{\boldsymbol {\sigma}}-2\al^{-1}\Delta \widehat{{\boldsymbol {\sigma }}}.
\eeq
From Equation \eqref{2.6}, one gets 
\beq
\label{2.7}
(\widehat{{\boldsymbol {\sigma}}}\cdot \mathbf{g}_{12}'')=-\al^{-1} (\widehat{{\boldsymbol {\sigma}}}\cdot \mathbf{g}_{12})-2\al^{-1}\Delta.
\eeq

Additionally, the volume transformation in velocity space for the \textit{direct} collision $\left(\mathbf{v}_1,\mathbf{v}_2\right)\to \left(\mathbf{v}_1',\mathbf{v}_2'\right)$ is
\beq\label{2.8}
d\mathbf{v}_1'd \mathbf{v}_2'=\al d \mathbf{v}_1 d \mathbf{v}_2,
\eeq 
while for the \textit{inverse} collision $\left(\mathbf{v}_1'',\mathbf{v}_2''\right)\to \left(\mathbf{v}_1,\mathbf{v}_2\right)$ is
\beq
\label{2.9}
d\mathbf{v}_1''d \mathbf{v}_2''=\al^{-1} d \mathbf{v}_1 d \mathbf{v}_2.
\eeq

\subsection{Enskog kinetic equation}

It is well known that granular materials under rapid flow conditions admit a
hydrody\-namic-like description. The corresponding granular hydrodynamic equations can be obtained from a more fundamental point of view by using  the tools of the classical kinetic theory of gases \cite{CC70, FK72,Dorfman2021} conveniently adapted to dissipative dynamics \cite{BP04,G19}. Kinetic theory provides a \textit{mesoscopic} description of matter, midway between a formal treatment based on Newton's equations and a more phenomenological approach based on continuum mechanics. It has been widely employed by the engineering and physics community in the past decades to attempt to understand the behavior of granular matter. At a kinetic level, it is assumed that all the relevant information on the state of the granular fluid system is provided by the knowledge of the one-particle velocity distribution function $f(\mathbf{r}, \mathbf{v}, t)$. This quantity is defined in such a way that $f(\mathbf{r}, \mathbf{v}, t) d\mathbf{r} d\mathbf{v}$ gives the \textit{average} number of particles which at time $t$ are located in $d\mathbf{r}$ around the point $\mathbf{r}$ and with velocities in the range $d\mathbf{v}$ around $\mathbf{v}$.

For moderate densities, the Enskog kinetic equation is the natural extension of the usual Boltzmann equation for dilute gases. The former equation accounts for effects of finite density in the dynamic properties of the gas. In the $\Delta$-model and in the presence of the gravity acceleration $\mathbf{g}$, the inelastic version of the Enskog equation is ~\cite{BGMB13}
\beq
\label{2.10}
\frac{\partial f}{\partial t}+\mathbf{v}\cdot \nabla f+\mathbf{g}\cdot \frac{\partial f}{\partial \mathbf{v}}=J_\text{E}[\mathbf{r},\mathbf{v}|f,f],
\eeq
where the Enskog collision operator $J_\text{E}$ of the model reads
\beqa
\label{2.11}
& & J_\text{E}[\mathbf{r},\mathbf{v}_1|f,f]\equiv \sigma^{d-1}\int d{\bf v}_{2}\int d \widehat{\boldsymbol{\sigma}}
\Theta (-\widehat{{\boldsymbol {\sigma }}}\cdot {\bf g}_{12}-2\Delta)
(-\widehat{\boldsymbol {\sigma }}\cdot {\bf g}_{12}-2\Delta)
 \nonumber\\
& & \times \al^{-2} f_2(\mathbf{r}, \mathbf{r}+{\boldsymbol {\sigma }},\mathbf{v}_1'', \mathbf{v}_2'';t)-\sigma^{d-1}\int\ d{\bf v}_{2}\int d\widehat{\boldsymbol{\sigma}}
\Theta (\widehat{{\boldsymbol {\sigma}}}\cdot {\bf g}_{12})
(\widehat{\boldsymbol {\sigma }}\cdot {\bf g}_{12})\nonumber\\
& & \times f_2(\mathbf{r}, \mathbf{r}+{\boldsymbol {\sigma}},\mathbf{v}_1, \mathbf{v}_2;t).
\eeqa
In Equation \eqref{2.11},
\beq
\label{2.12}
f_2(\mathbf{r}_1,\mathbf{r}_2,\mathbf{v}_1,\mathbf{v}_2; t)\equiv 
\chi(\mathbf{r}_1,\mathbf{r}_2) f(\mathbf{r}_1,\mathbf{v}_1;t)f(\mathbf{r}_2,\mathbf{v}_2;t),
\eeq
$\chi(\mathbf{r}_1,\mathbf{r}_2)$ denotes the pair distribution function, 
$\Theta(x)$ is the Heaviside step function and  $d$ is the dimensionality of the system ($d=2$ for hard disks and $d=3$ for hard spheres). Note that although the $\Delta$-model attempts to describe confined quasi-two-dimensional systems ($d=2$), the kinetic theory exposed in this review is performed for an arbitrary number of dimensions $d$.

Similar to the Boltzmann equation, the Enskog equation assumes the molecular chaos hypothesis, which means it neglects velocity correlations among particles about to collide. One consequence of this hypothesis is that the two-body distribution function $f_2$ factorizes into the product of one-particle velocity distribution functions. However, unlike the Boltzmann equation, the Enskog equation accounts for (i) the spatial correlations between colliding pairs via the pair distribution function at contact $\chi(\mathbf{r},\mathbf{r}+\boldsymbol{\sigma})$, and (ii) the variation of distribution functions over a distance equal to the diameter of grains (excluded volume effects). These two factors yield corrections to the Boltzmann results. In particular, there are non-vanishing collisional transfer contributions to the fluxes due to the spatial difference of the colliding spheres. 

Given that here our main objective is to determine the dynamic properties of the granular fluid, we are interested in evaluating the collisional moments of the Enskog collision operator. In other words, we want to get an expression for $I(\psi)$ where $\psi$ is an arbitrary function of velocity, and $I(\psi)$ is defined as  
\beq
\label{2.12.0}
I(\psi)= \int\; d \mathbf{v}_1\; \psi(\mathbf{v}_1) J_\text{E}[\mathbf{r},\mathbf{v}_1|f,f].
\eeq
By following similar mathematical steps as those made for the conventional IHS model \cite{BP04,G19}, $I(\psi)$ can be rewritten in a more convenient way as~\cite{BGMB13,SRB14} 
\beq
\label{2.13}
I_\psi=\sigma^{d-1}\int \dd \, \mathbf{v}_1\int\ \dd{\bf v}_{2}\int \dd\widehat{\boldsymbol{\sigma}}\,
\Theta (\widehat{{\boldsymbol {\sigma }}}\cdot {\bf g}_{12})(\widehat{\boldsymbol {\sigma }}\cdot {\bf g}_{12})
f_2(\mathbf{r},\mathbf{r}+\boldsymbol{\sigma},\mathbf{v}_1,\mathbf{v}_2; t)
\left[\psi(\mathbf{v}_1')-\psi(\mathbf{v}_1)\right],
\eeq
where $\mathbf{v}_1'$ is defined by Equation \eqref{2.1}. Equation \eqref{2.13} gives the same result as for the IHS model~\cite{BP04}.

\subsection{Hydrodynamic balance equations}

The relevant hydrodynamic fields of the granular gas can be defined as the first few velocity moments of the velocity distribution function $f({\bf r},{\bf v},t)$. The number density of particles $n({\bf r},t)$, the mean flow velocity $\mathbf{U}(\mathbf{r},t)$, and the granular temperature $T(\mathbf{r},t)$ are given, respectively, by  
\beq
\label{2.14}
n({\bf r},t) = \int d \mathbf{v} f({\bf r},{\bf v},t),
\eeq
\beq
\label{2.15}
\mathbf{U}({\bf r},t) = \frac{1}{n({\bf r},t) }\int d\mathbf{v} {\bf v} f({\bf r},{\bf v},t),
\eeq
\beq
\label{2.16}
T({\bf r},t)= \frac{1}{d n({\bf r},t)}
\int d\mathbf{v} m V^2 f({\bf r},{\bf v},t),
\eeq
where $\mathbf{V}=\mathbf{v}-\mathbf{U}$ is the peculiar velocity.

The corresponding balance equations for the densities of mass, momentum and energy can be derived by using the relation \eqref{2.13}. Its derivation follows similar mathematical steps as those made for the IHS model and adopt the standard form for rapid granular flows~\cite{GD99a,L05}.   
They are given by
\beq
\label{2.17}
D_t n+n\nabla \cdot \mathbf{U}=0,
\eeq
\beq
\label{2.18}
\rho D_t \mathbf{U}+\nabla \cdot \mathsf{P}=\rho \mathbf{g},
\eeq
\beq
\label{2.19}
D_t T+\frac{2}{dn}\left(\nabla \cdot \mathbf{q}+\mathsf{P}:\nabla \mathbf{U} \right)=-\zeta T.
\eeq
In Equations \eqref{2.17}--\eqref{2.19}, $D_t\equiv \partial_t+\mathbf{U}\cdot \nabla$ is the material derivative, $\rho=mn$ is the mass density, and $\nabla_i\equiv \partial/\partial r_i$. As with molecular (elastic) fluids \cite{CC70,FK72}, the pressure tensor
$\mathsf{P}({\bf r},t)$ and
the heat flux $\mathbf{q}({\bf r},t)$ have both {\em kinetic} and {\em collisional transfer} contributions. Thus, ${\sf P}={\sf P}_k+{\sf P}_c$ and ${\bf q}={\bf q}_k+{\bf q}_c$. The kinetic contributions
are given as usual by
\beq
\label{2.20}
{\sf P}_k({\bf r}, t)=\int \; d{\bf v} m{\bf V}{\bf V}f({\bf r},{\bf v},t),
\eeq
\beq
\label{2.21}
{\bf q}_k({\bf r}, t)=\int \; d{\bf v} \frac{m}{2}V^2{\bf V}f({\bf r},{\bf v},t).
\eeq
The collisional transfer contributions are \cite{GBS18}
\beqa
\label{2.22}
\mathsf{P}_{c}&=&\frac{1+\alpha}{4}m \sigma^{d}
\int d\mathbf{v}_{1}\int d\mathbf{v}_{2}\int
d\widehat{\boldsymbol {\sigma}}\Theta (\widehat{\boldsymbol{\sigma}}\cdot
\mathbf{g}_{12})(\widehat{\boldsymbol {\sigma}}\cdot \mathbf{g}_{12})
\widehat{\boldsymbol {\sigma}}\widehat{\boldsymbol {\sigma }}\left[(\widehat{\boldsymbol {\sigma}}\cdot \mathbf{g}_{12})+\frac{2\Delta}{1+\al}\right]\nonumber\\
&  & \times
\int_{0}^{1} \dd \lambda f_2\Big(\mathbf{r}-\lambda{\boldsymbol{\sigma}},\mathbf{r}+(1-\lambda)
{\boldsymbol {\sigma}},\mathbf{v}_{1},\mathbf{v}_{2},t\Big),
\nonumber\\
\eeqa
\beqa
\label{2.23}
{\bf q}_{c}&=&\frac{1+\alpha}{4}m \sigma^{d}
\int d\mathbf{v}_{1}\int d\mathbf{v}_{2}\int
\dd\widehat{\boldsymbol {\sigma}}\Theta (\widehat{\boldsymbol{\sigma}}\cdot
\mathbf{g}_{12})(\widehat{\boldsymbol {\sigma}}\cdot \mathbf{g}_{12})^{2}(\widehat{\boldsymbol {\sigma}}\cdot {\bf G})
\widehat{\boldsymbol {\sigma}}
\nonumber\\
& &\times
\int_{0}^{1} d\lambda f_2\left[\mathbf{r}-
\lambda{\boldsymbol{\sigma}},\mathbf{r}+(1-\lambda)
{\boldsymbol {\sigma}},\mathbf{v}_{1},\mathbf{v}_{2},t\right]-\Delta \frac{m \sigma^d}{4}
\nonumber\\
& &\times
\int d\mathbf{v}_{1}\int d\mathbf{v}_{2}\int
d\widehat{\boldsymbol {\sigma}}\Theta (\widehat{\boldsymbol{\sigma}}\cdot
\mathbf{g}_{12})(\widehat{\boldsymbol {\sigma}}\cdot \mathbf{g}_{12})\widehat{\boldsymbol {\sigma}}\left[\Delta +\al (\widehat{\boldsymbol {\sigma}}\cdot \mathbf{g}_{12})-2 (\widehat{\boldsymbol {\sigma}}\cdot \mathbf{G})\right]\nonumber\\
& & \times
\int_{0}^{1} d \lambda f_2\Big(\mathbf{r}-
\lambda {\boldsymbol{\sigma}},\mathbf{r}+(1-\lambda)
{\boldsymbol {\sigma}},\mathbf{v}_{1},\mathbf{v}_{2},t\Big).
\eeqa
Here, ${\bf G}=\frac{1}{2}({\bf V}_1+{\bf V}_2)$ is the velocity of the center of mass. Finally, \vicente{the rate of energy $\zeta$} is given by
\beqa
\label{2.24}
\zeta&=&-\frac{m}{d n T}\sigma^{d-1}
\int d\mathbf{v}
_{1}\int d\mathbf{v}_{2}\int d\widehat{\boldsymbol {\sigma}}
\Theta (\widehat{\boldsymbol {\sigma}}\cdot
\mathbf{g}_{12})(\widehat{ \boldsymbol {\sigma}}\cdot
\mathbf{g}_{12})\nonumber\\
& &\times \Big[\Delta^2+\al \Delta (\widehat{{\boldsymbol {\sigma}}}\cdot \mathbf{g}_{12})-\frac{1-\al^2}{4}(\widehat{{\boldsymbol {\sigma}}}\cdot \mathbf{g}_{12})^2\Big] f_2(\mathbf{r}, \mathbf{r}+\boldsymbol {\sigma},\mathbf{v}_{1},\mathbf{v}_{2},t).
\eeqa
The \vicente{rate of energy} is due to \vicente{
competing effects of the energy injected by $\Delta$ and the energy lost by dissipative collisions.} Thus, in contrast to the conventional IHS model where $\zeta$ it is always positive, in the $\Delta$-model $\zeta$ can take negative values for small temperatures [see Equation~\eqref{3.16} below]. This property allows the system to reach stable steady states. When $\Delta=0$, Equations \eqref{2.22}--\eqref{2.24} reduce to those obtained in the IHS model \cite{G19}.

\vicente{It must be noted that in this paper will assume the Einstein summation convention over repeated Greek indices. Additionally, when studying multicomponent granular systems, Latin indices will be used to label the particle species (running from 1 to $s$) and Greek indices will be used to label the spatial dimensions ($d=2$ for disks and $d=3$ for spheres). Also, Greek indices will be used to label the hydrodynamic modes when studying the linear stability of the homogeneous states.}

As is well known, the macroscopic balance equations \eqref{2.17}--\eqref{2.19} provide the basis for developing a hydrodynamic description of confined, dense granular fluids. However, as with elastic collisions \cite{CC70,FK72}, these equations are not a closed set of equations for the hydrodynamic fields $n$, $\mathbf{U}$ and $T$. To become a closed set, one has to express the momentum $\mathsf{P}$ and heat $\mathbf{q}$ fluxes as well as the rate of energy $\zeta$ in terms of the hydrodynamic fields and their spatial gradients. These types of equations are referred to as the constitutive equations for the fluxes and the rate of energy. To first order in spatial gradients, these equations are the Navier--Stokes--Fourier equations,  and the corresponding expressions of the transport coefficients are obtained by solving the Enskog kinetic equation \eqref{2.10} by means of the Chapman--Enskog method~\cite{CC70} conveniently adapted to account for inelastic collisions.

\section{Homogeneous states}
\label{sec3}
\subsection{General results}

Before considering inhomogeneous states, it is convenient to analyze first homogeneous situations ($\nabla\to 0$). In this state and in the absence of gravity field ($\mathbf{g}=\mathbf{0}$), the Enskog equation \eqref{2.1} simply reduces to 
\beq
\label{3.1}
\frac{\partial f}{\partial t}=J_\text{E}[\mathbf{v}|f,f]
\eeq
where here 
\beqa
\label{3.2}
& & J_\text{E}[\mathbf{v}_1|f,f]\equiv \sigma^{d-1}\chi \int d{\bf v}_{2}\int d \widehat{\boldsymbol{\sigma}}
\Theta (-\widehat{{\boldsymbol {\sigma }}}\cdot {\bf g}_{12}-2\Delta)
(-\widehat{\boldsymbol {\sigma }}\cdot {\bf g}_{12}-2\Delta)
\al^{-2} f(\mathbf{v}_1'',t)
 \nonumber\\
& & \times f(\mathbf{v}_2'',t)-\sigma^{d-1}\chi
\int\ d{\bf v}_{2}\int d\widehat{\boldsymbol{\sigma}}
\Theta (\widehat{{\boldsymbol {\sigma}}}\cdot {\bf g}_{12})
(\widehat{\boldsymbol {\sigma }}\cdot {\bf g}_{12})f(\mathbf{v}_1,t)f(\mathbf{v}_2,t)
\eeqa
is the Enskog collision operator for homogeneous states. 
According to  Equation \eqref{3.2}, since the pair correlation $\chi$ is constant, the Enskog collision operator \eqref{3.2} can be recognized as the Boltzmann collision operator for the $\Delta$-model multiplied by $\chi$. For homogeneous isolated systems, the mass and momentum balance equations \eqref{2.17} and \eqref{2.18} are trivially satisfied and the energy balance equation \eqref{2.19} becomes 
\beq
\label{3.3}
\frac{\partial T}{\partial t}=-T\zeta.
\eeq
The rate of energy $\zeta$ for homogeneous states is given by 
\beq
\label{3.4}
\zeta=-\frac{m}{d n T}\sigma^{d-1}\chi 
\int d\mathbf{v}_{1}\int d\mathbf{v}_{2}
\Big[B_1 g_{12}\Delta^2+B_2 g_{12}^2 \al \Delta -B_3 g_{12}^3 \frac{1-\al^2}{4}\Big]f(\mathbf{v}_1,t)f(\mathbf{v}_2,t),
\eeq
where for the angular integrations use has been made of the relation~\cite{NE98} 
\beq
\label{3.5}
B_k\equiv \int\; \dd\widehat{\boldsymbol{\sigma}}\,
\Theta (\widehat{{\boldsymbol {\sigma }}}\cdot \mathbf{g})(\widehat{\boldsymbol {\sigma }}\cdot \widehat{\mathbf{g}})^k=\pi^{(d-1)/2} \frac{\Gamma\left(\frac{k+1}{2}\right)}{\Gamma\left(\frac{k+d}{2}\right)}
\eeq
for positive integers $k$. In the IHS model ($\Delta=0$), $\zeta(t)\propto \sqrt{T(t)}$ and the integration of Equation \eqref{3.3} leads to the well-known Haff's cooling law \cite{H83}: $T(t)=T(0)/(1+\frac{1}{2}\zeta(0)t)^2$, $T(0)$ being the initial temperature and $\zeta(0)$ is the energy rate at $t=0$. However, when $\Delta\neq 0$, the time-dependence of $\zeta$ is more complex and  so, the time-dependence of the granular temperature cannot analytically be obtained.      

As in the homogeneous cooling state (HCS) for the IHS model, although the solution to the Enskog equation \eqref{3.1} is not known to date, dimensional analysis and symmetry considerations suggest the existence of an isotropic in velocity space scaling solution where $f(\mathbf{v},t)$ depends on time through the granular temperature $T(t)$. This scaling solution is \cite{BGMB13,BMGB14}  
\beq
\label{3.6}
f(\mathbf{v},t)=n v_\text{th}(t)^{-d}
\varphi(\mathbf{c},\Delta^*),
\eeq
where $v_\text{th}(t)=\sqrt{2T(t)/m}$ is the thermal velocity and $\varphi$ is a reduced distribution whose dependence on $T$ is encoded through the dimensionless velocity $\mathbf{c}\equiv \mathbf{v}/v_\text{th}$ and the dimensionless parameter $\Delta^* \equiv \Delta/v_\text{th}\propto T(t)^{-1/2}$. Thus, in contrast to the HCS, the unknown scaled distribution $\varphi$ depends on the granular temperature $T$ not only through the scaled velocity $\mathbf{c}$ but also through $\Delta^*(t)$. This is an additional intricacy of the $\Delta$-model in comparison with the IHS model.    

According to the solution \eqref{3.6}, since the time-dependence of the distribution $f$ is through $T$, then 
\beq
\label{3.7}
\frac{\partial f}{\partial t}=\frac{\partial f}{\partial T}\frac{\partial T}{\partial t}=-\zeta T \frac{\partial f}{\partial T}.
\eeq
Additionally, $f$ depends explicitly on $T$ through the thermal velocity and implicitly through $\mathbf{c}$ and $\Delta^*$. As a consequence, 
\beq
\label{3.8}
T\frac{\partial f}{\partial T}=-\frac{1}{2}\frac{\partial}{\partial \mathbf{v}}\cdot \left(\mathbf{v}f\right)
-\frac{1}{2}\Delta^*\frac{\partial f}{\partial \Delta^*},
\eeq
and the Enskog equation \eqref{3.1} reads 
\beq
\label{3.9} 
\frac{1}{2}\zeta \frac{\partial }{\partial \mathbf{v}}\cdot \left(\mathbf{v}f\right)
+\frac{1}{2}\zeta\Delta^*\frac{\partial f}{\partial \Delta^*}=J_\text{E}[\mathbf{v}|f,f].
\eeq

As said before, an exact solution to Equation \eqref{3.9} has not been found so far. However, a very good approximation can be obtained from an expansion
in Sonine polynomials. In particular, the time-dependence of the kurtosis
\beq
\label{3.10}
a_2=\frac{4}{d(d+2)}\int\; d \mathbf{c}\; c^4 \varphi(c)-1
\eeq
of the scaled distribution $\varphi$ has been widely studied in Refs.\ \cite{BGMB13,BMGB14}. The analytical results derived in those works (which are based on the scaling solution \eqref{3.6}) exhibit good agreement with the numerical results obtained from the direct simulation Monte Carlo (DSMC) method \cite{B94}.

Since the distribution function $f$ is isotropic in velocity space, according to Equations \eqref{2.20}--\eqref{2.23}, the pressure tensor is diagonal and 
the heat flux vanishes:
\beq
\label{3.11}
P_{ij}=p\delta_{ij}, \quad \mathbf{q}=\mathbf{0}.
\eeq
The hydrostatic pressure $p$ can be written as $p=n T p^*$ where
\beq
\label{3.12}
p^*=1+2^{d-2}\chi \phi (1+\al)+\frac{2^{d-1}\Gamma\left(\frac{d}{2}\right)}{\sqrt{\pi}
\Gamma\left(\frac{d+1}{2}\right)}\chi \phi \Delta^*
\int d\mathbf{c}_1\int d\mathbf{c}_2  g_{12}^*
\varphi(\mathbf{c}_1,t)\varphi(\mathbf{c}_2,t),
\eeq
where $\mathbf{g}_{12}^*\equiv \mathbf{g}_{12}/v_\text{th}$ and 
\beq
\label{3.13}
\phi=\frac{\pi^{d/2}}{2^{d-1}d \Gamma(d/2)} n\sigma^d
\eeq
is the solid volume fraction. Note that, besides the standard ideal gas and excluded volume contributions to the pressure, there is a new term  proportional to $\Delta$ in Equation \eqref{3.12}. This term is due to the additional momentum transfer at collisions. Moreover, the expression \eqref{3.4} of $\zeta$ can be rewritten as 
\beq
\label{3.14}
\zeta=-\frac{2}{d}n\sigma^{d-1}v_\text{th} \chi \int d\mathbf{c}_1
\int d\mathbf{c}_2 \varphi(\mathbf{c}_1)\varphi(\mathbf{c}_2)\left(B_1 g^{*}\Delta^{*2}+B_2 \al g^{*2}\Delta^*-\frac{1-\al^2}{4}B_3 g^{*3}\right).
\eeq

\subsection{Homogeneous steady states}

As has been clearly demonstrated, numerical computations are generally required to solve the $\Delta$-model in a homogeneous time-dependent state, essentially due to the intricate dependence of the scaling distribution $\varphi$ on $\Delta^*(t)$. A detailed study of the $\Delta^*$-dependence of $\varphi$ and $a_2$ for different initial conditions has been carried out in Refs.\ \cite{BGMB13,BMGB14}.         

To obtain analytical results, \vicente{one typically considers the long-time limit where $\Delta^*(t)$ achieves a constant value $\Delta^*$ independent of time}.  
In this limiting case (\vicente{homogeneous steady state, HSS}), an explicit expression of the kurtosis in the vicinity of the steady state can be derived. 
For a two-dimensional granular gas, the dependence of $a_2$ on $\al$ was studied in Ref.\ \cite{GBS18} showing that the magnitude of $a_2$ never exceeds 0.103. Thus, in the steady state, contributions to $\zeta$ and $p^*$ coming from terms proportional to $a_2$ are generally negligible compared to the remaining contributions. As a consequence, for practical purposes, the integrals \eqref{3.12} and \eqref{3.14} involving the distribution $\varphi$ can be computed by replacing it by its Maxwellian form $\varphi_\text{M}$: 
\beq
\label{3.15}
\varphi(\mathbf{c},\Delta^*)\to \varphi_\text{M}(\mathbf{c})=\pi^{-d/2}
e^{-c^2}.
\eeq
Within the Maxwellian approximation, the rate of energy $\zeta$ is given by 
\beq
\label{3.16}
\zeta_\text{M}=\frac{\sqrt{2}\pi^{\frac{d-1}{2}}}{d\Gamma\left(\frac{d}{2}\right)}n\sigma^{d-1}v_\text{th} \chi \left(1-\al^2-2\Delta^{*2}-\sqrt{2\pi}\al \Delta^*\right),
\eeq
while the (reduced) hydrostatic pressure $p^*$ is 
\beq
\label{3.17}
p^*_\text{M}=1+2^{d-2}\chi \phi (1+\al)+\frac{2^d}{\sqrt{2\pi}}\chi \phi \Delta^*.
\eeq
In the steady state, $\partial_t T=0$, and so Equation \eqref{3.3} implies that $\zeta=0$. According to Equation \eqref{3.16}, the condition $\zeta_\text{M}=0$ yields a quadratic equation in $\Delta^*$ whose physical solution (i.e., $\Delta^*=0$ if $\al=1$) provides the $\al$-dependence of $\Delta^*$ in the Maxwellian approximation. This solution is
\beq
\label{3.18}
\Delta^*_\text{M}(\al)=\frac{1}{2}\sqrt{\frac{\pi}{2}}\al\left[\sqrt{1+
\frac{4(1-\al^2)}{\pi \al^2}}-1\right].
\eeq
Since $\Delta^*=\Delta/\sqrt{2T/m}$, at given values of $\al$ and $\Delta$, Equation \eqref{3.18} gives the value of the stationary temperature. As expected, according to Equation \eqref{3.18}, for elastic collisions ($\al=1$) the steady state is only achieved for $\Delta_\text{M}^*=0$. The relationship \eqref{3.18} has been tested against molecular dynamics (MD) simulations showing excellent agreement with deviations smaller than 2 \%, except for small values of the coefficient of restitution and/or high densities  \cite{BRS13}. At a value fixed of $\Delta$ note that Equation \eqref{3.18} predicts that the granular temperature diverges when $\al \to 1$. This result has been verified in MD simulations of the $\Delta$-model (see Figure 2 of Ref.\ \cite{BRS13}). It must be remarked that this sort of divergence has been also observed in MD simulations carried out in three-dimensional systems with vibrating walls (see Figure 4 of Ref.\ \cite{MGB19}), with the stationary temperature scaling as the wall velocity squared with a prefactor that depends on the height of the box. In addition, there is a qualitative agreement between the stationary temperature $T$ obtained from MD simulations and its theoretical prediction derived from the $\Delta$-model \cite{BRS13}.

\vicente{Since the dependence of $\zeta_\text{M}$ on the volume fraction $\phi$ is only through $\chi(\phi)$ (see Equation \eqref{3.16}), $\Delta^*_\text{M}(\al)$ is independent of $\phi$. Beyond the Maxwellian approximation to $\zeta$, one expects that the energy rate can be also written as $\zeta=\chi(\phi)\overline{\zeta}(\Delta^*,\al)$. Hence, the steady condition ($\zeta=0$) provides an expression of $\Delta^*$ independent of density.}

\begin{figure}[h!]
\centering
\includegraphics[width=0.45\textwidth]{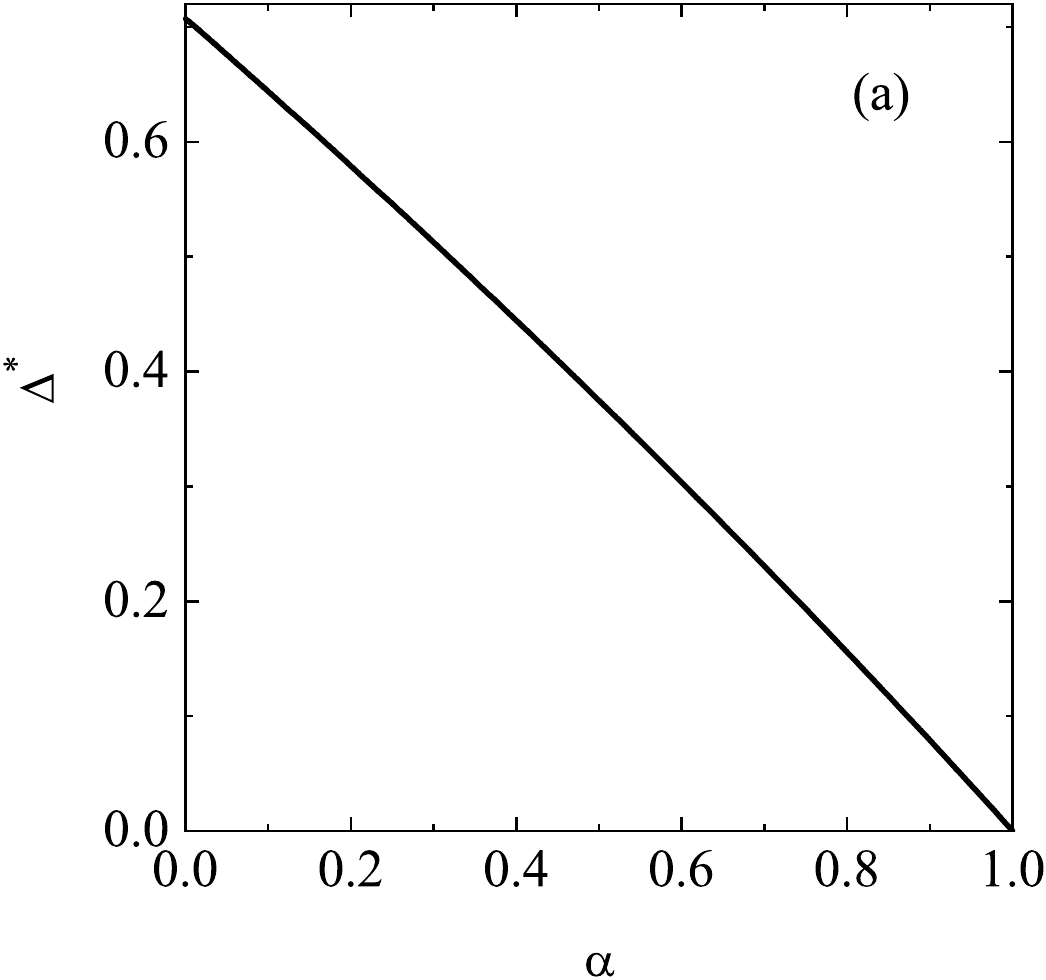}
\includegraphics[width=0.45\textwidth]{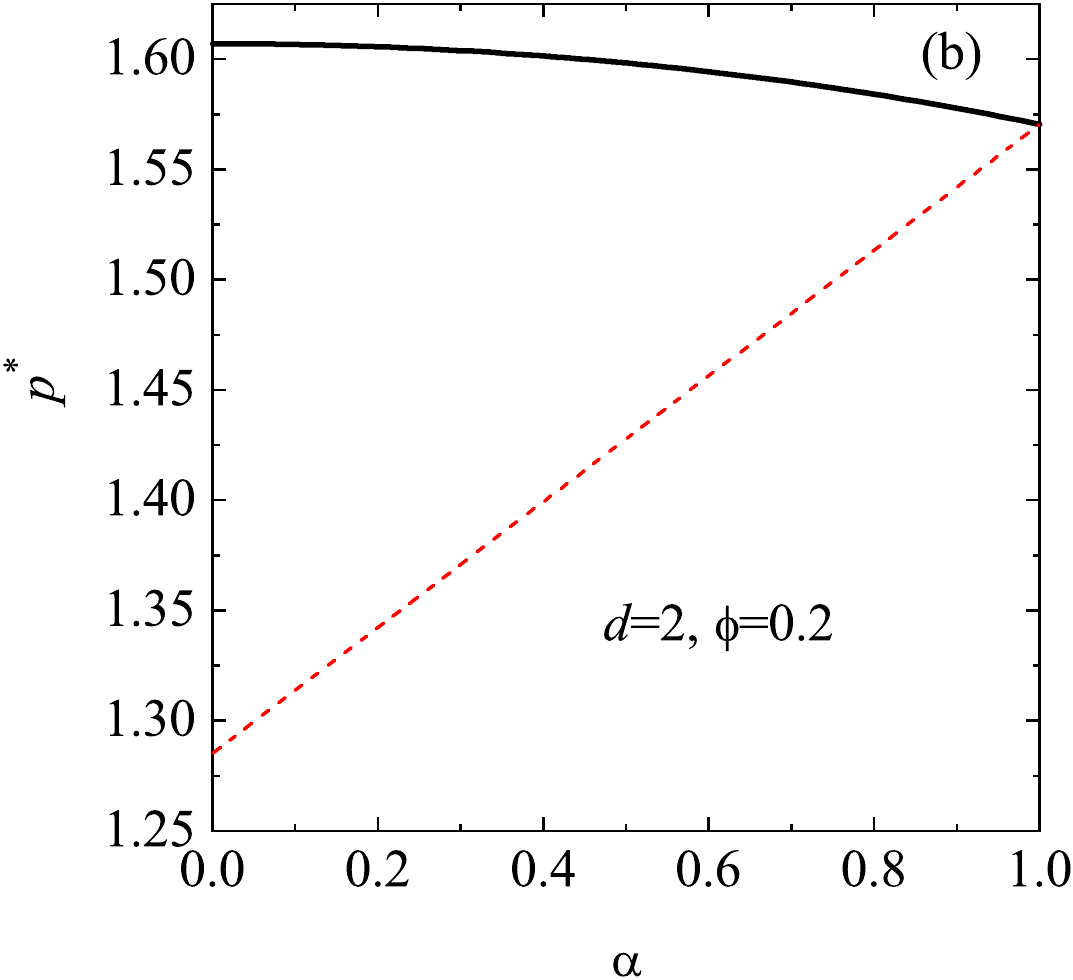}
\caption{Panel (a): Plot of $\Delta_\text{M}^*$ versus the coefficient of restitution $\al$ in the steady state. Panel (b): Plot of the (reduced) pressure $p_\text{M}^*$ versus the coefficient of restitution $\al$ for a two-dimensional ($d=2$) system with a solid volume fraction $\phi=0.2$. The solid line corresponds to the result obtained in the $\Delta$-model while the dashed line refers to the result obtained in the IHS model ($\Delta^*=0$).
}
\label{fig1}
\end{figure}

Panel (a) of Figure \ref{fig1} shows the $\al$-dependence of the (dimensionless) extra velocity $\Delta_\text{M}^*$. As expected, $\Delta_\text{M}^*$ increases with decreasing $\alpha$. To complement the panel (a) of Figure \ref{fig1}, the dependence of the (reduced) pressure $p_\text{M}^*$ on the coefficient of restitution $\al$ is illustrated by the panel (b) of Figure \ref{fig1} for $d=2$ and $\phi=0.2$. A good approximation to the pair correlation $\chi$ for a two-dimensional gas is \cite{JM87}
\beq
\label{3.19}
\chi(\phi)=\frac{1-\frac{7}{16}\phi}{(1-\phi)^2}.
\eeq
We have also included the prediction of the (reduced) pressure $p^*$ given by the IHS model \cite{GD99a,L05}. We observe that the effect of inelasticity on the pressure is much more significant in the freely cooling gas of IHS than in the $\Delta$-model.

\section{Chapman--Enskog method applied to the $\Delta$-model}
\label{sec4}

Once the homogeneous time-dependent state is well-characterized, the next step is to obtain the Navier--Stokes hydrodynamic equations of the confined granular gas with explicit forms for the transport coefficients. To achieve this goal we solve the Enskog equation \eqref{2.10} to first order in spatial gradients by means of a generalization of the conventional Chapman--Enskog method \cite{CC70} to dissipative dynamics.    

As widely discussed in many textbooks (see for instance, Refs.\ \cite{CC70,FK72,GS03,soto2016kinetic}), there are two separate stages in the relaxation of a \textit{molecular} (elastic) fluid toward equilibrium. For times of the order of the mean free time, a first stage (\textit{kinetic} regime) is identified where the effect of collisions is to relax quickly the fluid toward a local equilibrium state. This stage depends on the initial preparation of the system. A second stage is then identified in which the gas slowly evolves toward the total equilibrium. 
In this stage (referred to as the \textit{hydrodynamic} regime), the gas has forgotten the microscopic details of its initial condition and its state is governed solely by the hydrodynamic fields. 
This special solution is referred to as a \textit{normal} or hydrodynamic solution. These two stages are also expected in \textit{granular} gases, except that in the kinetic regime  relaxation occurs toward a time-dependent non-equilibrium distribution rather than a local equilibrium distribution. It is worth noting that although the granular temperature $T$ is not a conserved field due to the inelastic collisions, it is still assumed to be a slow hydrodynamic field (i.e., its time evolution is much slower than the remaining kinetic excitations). This assumption has been clearly confirmed by the good agreement found between theoretical predictions based on this hypothesis and computer simulations in different nonequilibrium problems \cite{BRC99,BRCG00,BRM01,DHGD02,LBD02,MG02,MG03,GM03,BRM05,LLC07,MDCPH11,BR13,ChS13,MGH14,ChG23}.

According to the above scenario, in the hydrodynamic regime it is expected that the distribution function $f(\mathbf{r}, \mathbf{v}, t)$ qualifies as a normal solution and hence, it depends on space and time through a functional dependence on the hydrodynamic fields $n$, $\mathbf{U}$, and $T$: 
\beq
\label{4.1}
f(\mathbf{r}, \mathbf{v}, t)=f[\mathbf{v}|n(\mathbf{r},t), \mathbf{U}(\mathbf{r},t), T(\mathbf{r},t)].
\eeq
As discussed previously, although the temperature is not strictly a slow field, it has been shown in Ref.~\cite{BBMG15} in the context of the $\Delta$-model for dilute granular gases that after a short transient the distribution function does adopt a normal solution. A similar behavior is expected for dense granular fluids. As usual, the functional dependence \eqref{4.1} can be made local in space by means of an expansion in spatial gradients of the hydrodynamic fields. To
generate it, $f$ is written as a series expansion in a formal parameter $\epsilon$ measuring the nonuniformity of the system:
\beq
\label{4.2}
f=f^{(0)}+\epsilon f^{(1)}+\epsilon^2 f^{(2)}+\cdots,
\eeq
where each factor of $\epsilon$ means an implicit gradient of a
hydrodynamic field. The fact that in the $\Delta$-model the homogeneous steady state is stable for any inelasticity (see Refs.\ \cite{BRS13,BBGM16,GBS21a}) makes it possible to control the strength of spatial gradients through initial or boundary conditions, as occurs with molecular fluids. Thus, although the results obtained in the Navier--Stokes domain apply to sufficiently small gradients (low Knudsen number), they are not restricted \emph{a priori} to small degree of dissipation. 

Furthermore, in the presence of the gravity field $\mathbf{g}$, it is also necessary to characterize the magnitude of gravity relative to spatial gradients. As for elastic collisions \cite{CC70,FK72}, the magnitude of $\mathbf{g}$ is assumed to be at least to first-order in the perturbation expansion.

According to the expansion \eqref{4.2} for the distribution
function, the Enskog collision operator and time derivative
must also be expanded in powers of $\epsilon$:
\beq
\label{4.3}
J_\text{E}=J_\text{E}^{(0)}+\epsilon J_\text{E}^{(1)}+\cdots, \quad
\partial_t=\partial_t^{(0)}+\epsilon\partial_t^{(1)}+\cdots.
\eeq
The coefficients in the time derivative expansion are identified
by a representation of the fluxes and the rate of energy in the
macroscopic balance equations as a similar series through their
definitions as functionals of $f$. The expansion \eqref{4.2} yields
similar expansions for the momentum and heat fluxes, and the rate of energy when
substituted into their definitions \eqref{2.20}--\eqref{2.24}, respectively:
\beq
\label{4.4}
P_{\lambda \beta}=P_{\lambda \beta}^{(0)}+\epsilon P_{\lambda \beta}^{(1)}+\cdots, \quad
\mathbf{q}=\mathbf{q}^{(0)}+\epsilon \mathbf{q}^{(1)}+\cdots,\\
\eeq
\beq
\label{4.5}
\zeta=\zeta^{(0)}+\epsilon \zeta^{(1)}+\cdots.
\eeq

In the zeroth-order approximation, $\partial_t^{(0)}n=\partial_t^{(0)}U_\lambda=0$ and $\partial_t^{(0)}T=-T\zeta^{(0)}$. Here, $\zeta^{(0)}$ is the zeroth-order contribution to the rate of energy. An approximate form of this quantity is given by Equation \eqref{3.16} in the HSS. Since the distribution $f^{(0)}(\mathbf{r},\mathbf{v},t)$ formally verifies the same equation  \eqref{3.9} for a strictly homogeneous state, $f^{(0)}$ is nothing more than the \textit{local} version of the scaling solution \eqref{3.6}, namely, it is given by Equation \eqref{3.6} except by the replacements $n\to n(\mathbf{r},t)$, $\mathbf{v}\to \mathbf{v}-\mathbf{U}(\mathbf{r},t)$, and $T\to T(\mathbf{r},t)$. As a consequence, in the steady state, the local versions of the (approximate) expressions \eqref{3.16} and \eqref{3.17} provide the forms of $\zeta^{(0)}$ and $p^*$, respectively.

\subsection{First-order approximation}

The determination of the first-order distribution $f^{(1)}$ follows similar steps as those made in the IHS model (see for instance, chapter 3 of the textbook \cite{G19}), except that in the $\Delta$-model there are new terms coming from the additional temperature-dependence of  $f^{(0)}$ through $\Delta^*$. The first-order velocity distribution function $f^{(1)}(\mathbf{r}, \mathbf{v}, t)$ is given by
\beq
\label{4.6}
f^{(1)}=\boldsymbol{\mathcal{A}}\cdot  \nabla \ln
T+\boldsymbol{\mathcal{B}} \cdot \nabla \ln n
+\mathcal{C}_{\lambda \beta}\frac{1}{2}\left(\nabla_{\lambda}U_{\beta}+\nabla_{\beta
}U_{\lambda}-\frac{2}{d}\delta_{\lambda\beta}\nabla \cdot
\mathbf{U} \right)+\mathcal{D}\nabla \cdot\mathbf{U}.
\eeq
The quantities $\boldsymbol{\mathcal{A}}(\mathbf{V})$, $\boldsymbol{\mathcal{B}}(\mathbf{V})$, $\mathcal{C}_{\lambda\beta}(\mathbf{V})$ and $\mathcal{D}(\mathbf{V})$ are the solutions of the following linear integral equations \cite{GBS18}:
\beq
\label{4.7}
-\zeta^{(0)}T\frac{\partial \boldsymbol{\mathcal{A}}}{\partial T}-\boldsymbol{\mathcal{A}}T\frac{\partial \zeta^{(0)}}
{\partial T} +
\mathcal{L}\boldsymbol{\mathcal{A}}=\mathbf{A},
\eeq
\beq
\label{4.8}
-\zeta^{(0)}T\frac{\partial \boldsymbol{\mathcal{B}}}{\partial T}+
\mathcal{L}\boldsymbol{\mathcal{B}}=\mathbf{B}+\zeta^{(0)}\left(1+\phi\frac{\partial}{\partial \phi}\ln \chi\right)\boldsymbol{\mathcal{A}}, 
\eeq
\beq
\label{4.9}
-\zeta^{(0)}T\frac{\partial \mathcal{C}_{\lambda\beta}}{\partial T}
+\mathcal{L}\mathcal{C}_{\lambda\beta}=C_{\lambda\beta}, 
\eeq
\beq
\label{4.10}
-\zeta^{(0)}T\frac{\partial \mathcal{D}}{\partial T}+
\mathcal{L}\mathcal{D}=D. 
\eeq
In Equations \eqref{4.7}--\eqref{4.10}, we have introduced the linear operator $\mathcal{L}$ given by
\begin{equation}
\mathcal{L}X=-\left(J_\text{E}^{(0)}[f^{(0)},X]+J_\text{E}^{(0)}[X,f^{(0)}]\right),  \label{4.11}
\end{equation}
where the operator $J_\text{E}^{(0)}$ is defined in Equation \eqref{3.2} with the replacements $\chi \to \chi(\mathbf{r}, t)$ and $f(\mathbf{v};t)\to f^{(0)}(\mathbf{r}, \mathbf{v};t)$.
The inhomogeneous terms (which depend on $f^{(0)}$) in Equations \eqref{4.7}--\eqref{4.10} are 
\beq
\label{4.12}
{\bf A}\left( \mathbf{V}\right)=-\mathbf{V}T\frac{\partial f^{(0)}}{\partial T}
-\frac{p}{\rho}\left(1+T\frac{\partial}{ \partial T}\ln p^*\right)\frac{\partial f^{(0)}}{\partial \mathbf{V}}
-\boldsymbol{\mathcal{K}}\left[T\frac{\partial f^{(0)}}{\partial T}\right],
\eeq
\beq
\label{4.13}
{\bf B}\left(\mathbf{V}\right)= -{\bf V}f^{(0)}-\frac{p}{\rho}
\left(1+\phi\frac{\partial}{\partial \phi}\ln p^*\right)
\frac{\partial f^{(0)}}{\partial \mathbf{V}}-\left(1+\frac{1}{2}\phi\frac{\partial}{\partial \phi}\ln \chi\right)
\boldsymbol{\mathcal{K}}\left[f^{(0)}\right],
\eeq
\beq
\label{4.14}
C_{\lambda\beta}\left(\mathbf{V}\right)=V_\lambda\frac{\partial f^{(0)}}{\partial V_\beta}+\mathcal{K}_\lambda\left[\frac{\partial f^{(0)}}{\partial V_\beta}
\right],
\eeq
\beq
\label{4.15}
D\left(\mathbf{V}\right)=\frac{1}{d}\frac{\partial}{\partial \mathbf{V}}\cdot \left( \mathbf{V} f^{(0)}\right)
+\left(\zeta_{U}+\frac{2}{d} p^*\right)T\frac{\partial f^{(0)}}{\partial T}+\frac{1}{d}\mathcal{K}_{\lambda}\left[\frac{\partial f^{(0)}}{\partial V_\lambda}\right]. \eeq
The operator $\boldsymbol{\mathcal{K}}$ is given by \cite{GBS18}
\beqa
\label{4.16}
\boldsymbol{\mathcal{K}}[X] &=&-\sigma^{d}\chi\int d \mathbf{v}_{2}\int 
d\widehat{\boldsymbol {\sigma}}\Theta (-\widehat{\boldsymbol {\sigma}} \cdot
\mathbf{g}_{12}-2\Delta)
(-\widehat{\boldsymbol {\sigma }}\cdot
\mathbf{g}_{12}-2\Delta)
\widehat{\boldsymbol{\sigma}} \alpha^{-2}f^{(0)}(\mathbf{v}_{1}'')X(\mathbf{v}_{2}'')\nonumber\\
& & 
+\sigma^{d}\chi\int d \mathbf{v}_{2}\int d\widehat{\boldsymbol {\sigma
}}\Theta (\widehat{\boldsymbol {\sigma}} \cdot
\mathbf{g}_{12})(\widehat{\boldsymbol {\sigma }}\cdot
\mathbf{g}_{12})
\widehat{\boldsymbol{\sigma}} f^{(0)}(\mathbf{v}_{1})X(\mathbf{v}_{2}).  
\eeqa
In Equation \eqref{4.15}, $\zeta_U$ is defined through the expression
\begin{equation}
\label{4.17}
\zeta^{(1)}=\zeta_U\nabla \cdot {\bf U}.
\end{equation}

In the low-density limit ($\phi=0$), $p^*=1$, $\boldsymbol{\mathcal{K}}[X] \to 0$, and the integral equations \eqref{4.7}--\eqref{4.15} reduce to those obtained in Ref.\ \cite{BBMG15} for dilute granular gases. With respect to the rate of energy, in the limit $\phi\to 0$, the quantity $D$ becomes 
\beq
\label{4.18}
D=\zeta_U T\frac{\partial f^{(0)}}{\partial T}-\frac{1}{d}\Delta^*\frac{\partial f^{(0)}}{\partial \Delta^*},
\eeq
and hence, $\zeta_U\neq 0$ even for dilute granular gases. This contrasts with the results obtained in the IHS model \cite{BDKS98}. However, for dense gases, $\zeta_U \neq 0$ for the IHS model~\cite{GD99a,L05}.

\subsection{Navier--Stokes transport coefficients.}
\label{sec5}

Based on symmetry considerations, the first-order contributions to the pressure tensor $P_{ij}^{(1)}$ and the heat flux $\mathbf{q}^{(1)}$ are given, respectively, by
\begin{equation}
\label{4.19}
P_{\lambda\beta}^{(1)}=-\eta\left( \nabla_{\lambda}U_{\beta}+\nabla_{\beta}U_{\lambda}-\frac{2}{d}\delta _{\lambda\beta}\nabla \cdot
\mathbf{U} \right) -\eta_\text{b}  \nabla \cdot \mathbf{U}\; \delta_{\lambda\beta},
\end{equation}
\begin{equation}
\label{4.20}
{\bf q}^{(1)}=-\kappa \nabla T-\mu \nabla n.
\end{equation}
In Equations \eqref{4.19}--\eqref{4.20}, $\eta$ is the shear viscosity, $\eta_\text{b}$ is the bulk viscosity, $\kappa$ is the thermal conductivity, and $\mu$ is the diffusive heat conductivity coefficient. The coefficient $\mu$ is an additional transport coefficient not present in the elastic case. The contribution to the heat flux coming from the density gradient is also present in relativistic gases \cite{GLW80,CK02} as well as in ordinary (elastic) gases subjected to a drag force proportional to the particle velocity \cite{PG14}.

While the coefficients $\eta$, $\kappa$, and $\mu$ have kinetic and collisional contributions, the bulk viscosity has only collisional contributions, and hence it vanishes in the low-density limit ($\phi\to 0$). The kinetic contributions to the transport coefficients  $\eta$, $\kappa$, and $\mu$ can be expressed in terms of the solutions of the set of linear integral equations \eqref{4.7}--\eqref{4.9}.

Given that the calculations to determine the Navier--Stokes transport coefficients are very long, here only some partial steps are offered in the calculation of the shear and bulk viscosities. Technical details to evaluate the remaining transport coefficients and the rate of energy can be found in Refs. \cite{GBS18,GBS20,GBS26}.

\subsection{Shear and bulk viscosities}

As mentioned before, the shear viscosity $\eta$ has 
kinetic and collisional contributions, i.e., $\eta=\eta_\text{k}+\eta_\text{c}$. However, the bulk viscosity $\eta_\text{b}=\eta_\text{b,c}$ since its kinetic contribution $\eta_\text{b,k}$ vanishes. The collisional contributions $\eta_\text{c}$ and $\eta_\text{b,c}$ can be obtained by expanding the expression \eqref{2.22} for the collisional pressure tensor to first order in spatial gradients. After some algebra, one gets the expressions \cite{GBS18}  
\beq
\label{4.21}
\eta_\text{b}=\frac{\pi^{d/2}}{2d^2\Gamma\left(\frac{d}{2}\right)}n^2\sigma^{d+1}m \chi v_\text{th}\Bigg[\frac{(d+1)}{2\sqrt{\pi}}\frac{\Gamma\left(\frac{d}{2}\right)}{\Gamma\left(\frac{d+3}{2}\right)}
(1+\al)I_{\eta_\text{b}}+\Delta^*\Bigg],
\eeq
\beq
\label{4.22}
\eta_{c}=\frac{\pi^{d/2}}{d\Gamma\left(\frac{d}{2}\right)}n\sigma^d \chi
\Bigg[\frac{1+\al}{d+2}+\frac{d}{\sqrt{\pi}(d+1)}
\frac{\Gamma\left(\frac{d}{2}\right)}{\Gamma\left(\frac{d+1}{2}\right)}
I_{\eta_c}\Delta^*\Bigg]\eta_\text{k}+
\frac{d}{d+2}\eta_\text{b},
\eeq
where we have introduced the dimensionless integrals 
\beq
\label{4.23}
I_{\eta_\text{b}}=\int d \mathbf{c}_1
\int d\mathbf{c}_2\; g_{12}^*\; \varphi(\mathbf{c}_1)\varphi(\mathbf{c}_2),
\eeq
\beq
\label{4.24}
I_{\eta_c}=\int d\mathbf{c}_1
\int d\mathbf{c}_2\; g_{12}^{*-1}g_{12,x}^{*2}g_{12,y}^{*2}\varphi_\text{M}(\mathbf{c}_1)\varphi_\text{M}(\mathbf{c}_2).
\eeq
It must be remarked that upon obtaining Equations \eqref{4.21}--\eqref{4.24} we have neglected the contributions proportional to $\zeta_U$ (it is expected that this quantity is in general very small). 
\vicente{Furthermore, given that the unknown $\mathcal{C}_{\lambda\beta}$ is also involved in the determination of $\eta_c$, we have replaced $\mathcal{C}_{\lambda\beta}$ by its corresponding leading Sonine approximation. According to Equation \eqref{4.9}, $\mathcal{C}_{\lambda\beta}\propto C_{\lambda\beta}\propto m V_\lambda V_\beta$. Thus, the leading Sonine approximation to $\mathcal{C}_{\lambda\beta}$ is given by} 
\beq
\label{4.25}
\mathcal{C}_{\lambda\beta}(\mathbf{V})\to -\frac{\eta_\text{k}}{n T^2}R_{\lambda\beta}(\mathbf{V})f_\text{M}(\mathbf{V}),
\eeq
where 
\beq
\label{4.25.1}
f_\text{M}(\mathbf{V})=n \left(\frac{m}{2\pi T}\right)^{d/2}
e^{-\frac{mV^2}{2T}}
\eeq
is the Maxwellian distribution and $R_{\lambda\beta}(\mathbf{V})$ is the traceless tensor
\beq
\label{4.26}
R_{\lambda\beta}(\mathbf{V})=m\left(V_\lambda V_\beta-\frac{1}{d}\delta_{\lambda\beta}V^2\right).
\eeq
\vicente{We note that the Sonine expansion of $\mathcal{C}_{\lambda\beta}(\mathbf{V})$ is different from the one usually employed for the zeroth-order distribution function $f^{(0)}(\mathbf{V})$ because the latter is isotropic in velocity space.}

It only remains to evaluate the kinetic shear viscosity $\eta_\text{k}$. To get it, as usual, we multiply both sides of Equation \eqref{4.9} by $R_{ij}(\mathbf{V})$ and integrate over velocity. After some algebra, one achieves the result  
\beq
\label{4.27}
\left(-\zeta^{(0)}T\partial_T+\nu_\eta\right)\eta_\text{k}=-\frac{\int d\mathbf{V}\; R_{\lambda\beta}(\mathbf{V}) C_{\lambda\beta}(\mathbf{V})}{(d-1)(d+2)},
\eeq
where
\begin{equation}
\label{4.28}
\nu_\eta=\frac{\int d{\bf v} R_{\lambda\beta}({\bf V}){\cal L}{\cal C}_{\lambda\beta}({\bf V})}
{\int d{\bf v}R_{\lambda\beta}({\bf V}){\cal C}_{\lambda\beta}({\bf V})}.
\end{equation}
In the hydrodynamic regime, the kinetic coefficient $\eta_\text{k}$ can be written as
\beq
\label{4.29}
\eta_\text{k}(T)=\eta_0(T)\eta_\text{k}^*(\alpha,\phi,\Delta^*),
\eeq
where
\begin{equation}
\label{4.30}
\eta_0(T)=\frac{d+2}{8}\Gamma\left(\frac{d}{2}\right)
\pi^{-\frac{d-1}{2}}\sigma^{1-d}\sqrt{mT}
\end{equation}
is the low density value of the shear viscosity in the elastic limit. According to Equation \eqref{4.29}, one has
the identity
\beq
\label{4.31}
T\partial_T \eta_\text{k}=\left(T\partial_T \eta_0\right)\eta_\text{k}^*-\frac{1}{2}\eta_\text{k} \Delta^*
\frac{\partial \ln \eta_\text{k}^*}{\partial \Delta^*}= 
\frac{1}{2}\eta_\text{k}\left(1- \Delta^*
\frac{\partial \ln \eta_\text{k}^*}{\partial \Delta^*}\right).
\eeq
Thus, Equation \eqref{4.27} reads
\beq
\label{4.32}
\frac{1}{2}\zeta^{(0)}\eta_\text{k} \Delta^* \frac{\partial \ln \eta_\text{k}^*}{\partial \Delta^*}+\left(\nu_\eta-\frac{1}{2}\zeta^{(0)}\right)\eta_\text{k}=n T
 -\frac{1}{(d-1)(d+2)}\int\; \dd \mathbf{v}
R_{\lambda\beta}(\mathbf{V}) {\cal K}_\lambda\left[\frac{\partial f^{(0)}}{\partial V_\beta}
\right],
\eeq
where use has been made of the explicit form \eqref{4.14} of $C_{\lambda\beta}$. As occurs for dilute granular gases \cite{BBMG15}, in contrast to the conventional IHS model, $\eta_\text{k}$ is given as the solution of an intricate first-order differential equation. The integral appearing in the right-hand side of Equation \eqref{4.32} can be computed as \cite{GBS18}
\beqa
\label{4.33}
\int\; d \mathbf{v}
R_{\lambda\beta}(\mathbf{V}) {\cal K}_\lambda \left[\frac{\partial f^{(0)}}{\partial V_\beta}\right]
&=&
2^{d-2}(d-1)\chi \phi (1+\al)(1-3\al)n T\nonumber\\
& & 
+2^d (d-1)\chi \phi \Delta^* n T\left[\frac{\Gamma\left(\frac{d}{2}\right)}{\sqrt{\pi}
\Gamma\left(\frac{d+1}{2}\right)} {I}_{\eta_\text{k}} - \Delta^{*}\right],
\eeqa
where \footnote{The expression \eqref{4.34} displayed here corrects a typo found in the previous result obtained in Ref.\ \cite{GBS24a}. Moreover, Table \ref{table1} provides the correct forms for the complete set of Navier--Stokes transport coefficients.}
\beq
\label{4.34}
{I}_{\eta_\text{k}}=2\int d{\bf c}_1\int d{\bf c}_2
\varphi({\bf c}_1)\varphi({\bf c}_2)
\left[g_{12}^{*-1} (\mathbf{g}_{12}^*\cdot \mathbf{c}_1)-(1+\al)g_{12}^*\right].
\eeq

\begin{figure}[h!]
\centering
\includegraphics[width=0.45\textwidth]{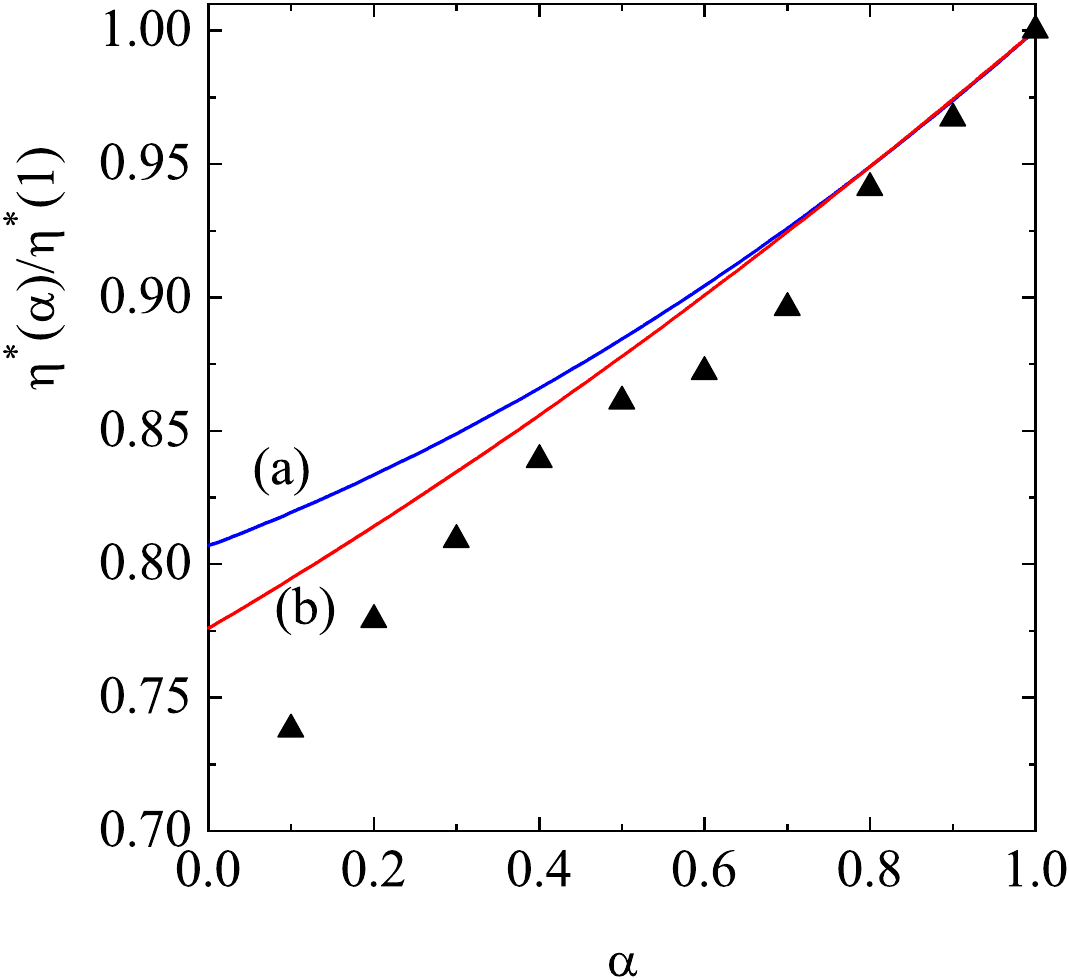}
\caption{Plot of the (scaled) shear viscosity coefficient $\eta^*(\al)/\eta^*(1)$ versus the coefficient of restitution $\al$ for a two-dimensional granular gas ($d=2$) and two different values of the solid volume fraction $\phi$: $\phi=0.1$ (a) and $\phi=0.314$ (b). The solid lines correspond to the kinetic theory results while symbols refer to MD simulations performed in Ref.\ \cite{SRB14} for $\phi=0.314$.  
}
\label{fig2}
\end{figure}
\begin{figure}[h!]
\centering
\includegraphics[width=0.45\textwidth]{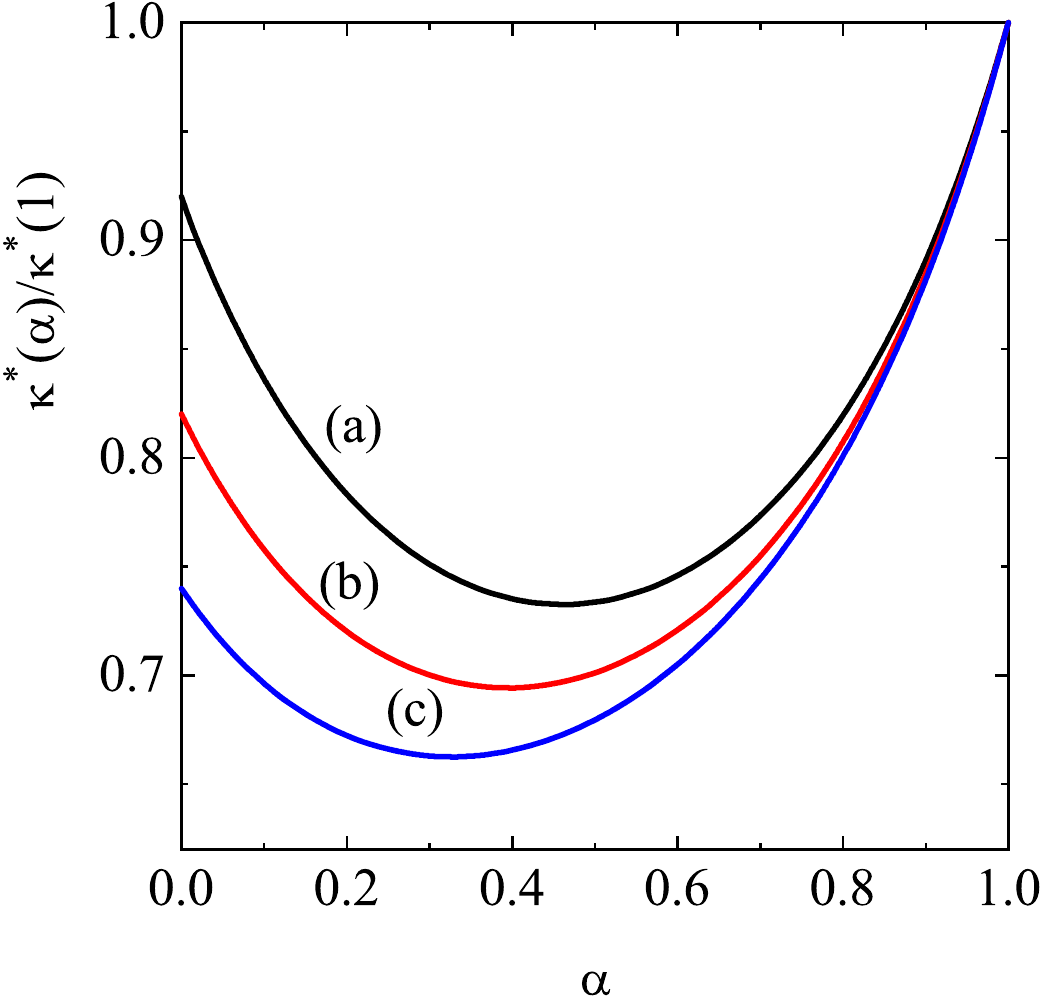}
\includegraphics[width=0.46\textwidth]{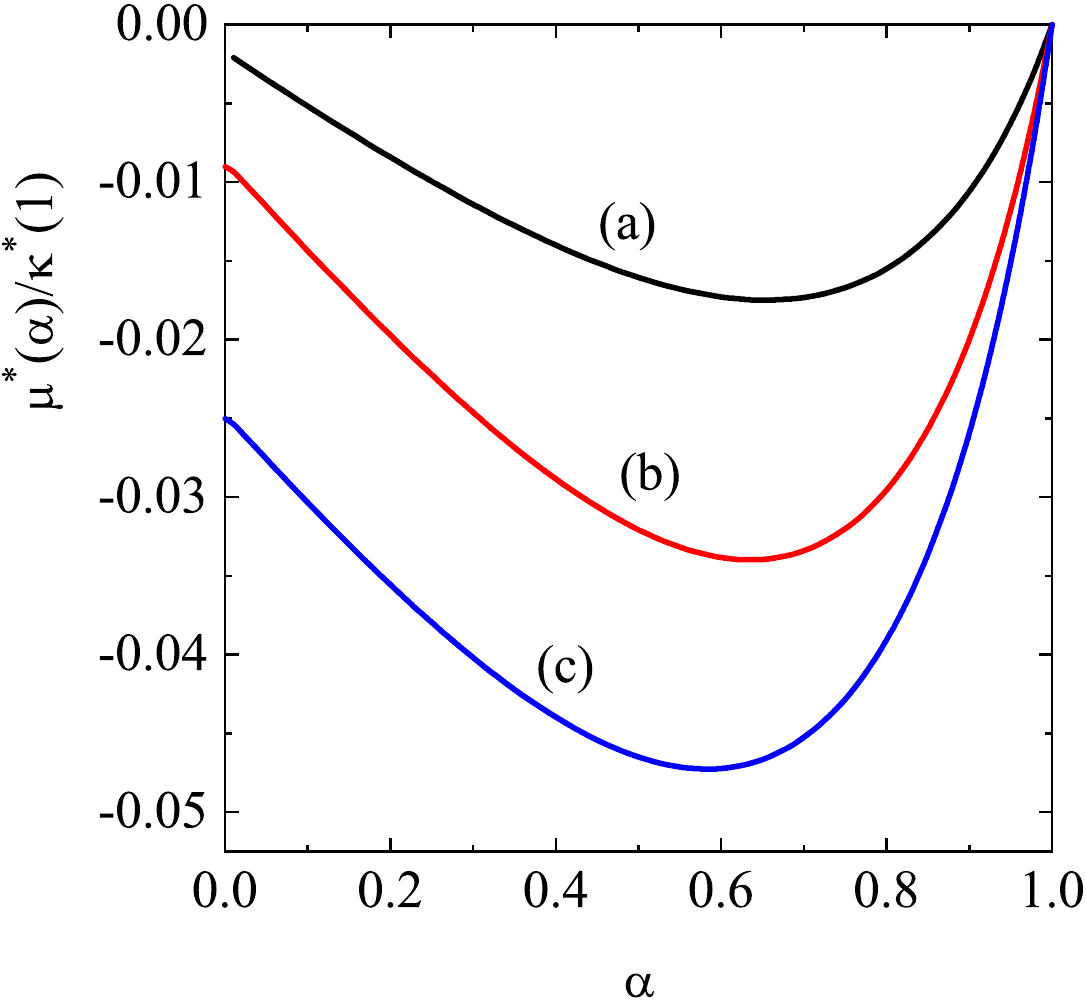}
\caption{Plot of the (scaled) thermal conductivity $\kappa^*(\al)/\kappa^*(1)$ and diffusive heat conductivity $\mu^*(\al)/\kappa^*(1)$ coefficients  
versus the coefficient of restitution $\al$ for a two-dimensional granular gas ($d=2$) and three different values of the solid volume fraction $\phi$: $\phi=0.1$ (a), $\phi=0.2$ (b), and $\phi=0.3$ (c).
}
\label{fig3}
\end{figure}

\begin{table*}[t]
\caption{Explicit expressions of the scaled transport coefficients for a two-dimensional monocomponent granular gas ($d=2$) at the stationary temperature.}
\label{table1}
\begin{tabularx}{\textwidth}{C}
\hline
\hline
\vspace{0.5cm}\\
$\eta^*=\left[1+\frac{1}{2}\phi \chi \left(1+\alpha+\sqrt{\frac{2}{\pi}}\Delta_\text{M}^*\right)\right]\eta_\text{k}^*
+\frac{1}{2}\eta_\text{b}^*$,\\
\vspace{0.2cm}\\
$\eta_\text{k}^*=\nu_\eta^{^*-1}\left\{1-\frac{1}{4}\phi \chi\left[(1+\al)
(1-3 \alpha) -4\sqrt{\frac{2}{\pi}}(1+2\al)\Delta_\text{M}^*-4\Delta_\text{M}^{*2}\right]\right\}$,\\
\vspace{0.2cm}\\
$\eta_\text{b}^*=\frac{8}{\pi}\phi^2 \chi \left(1+\alpha+\sqrt{\frac{\pi}{2}}\Delta_\text{M}^*\right)$,\\
\vspace{0.2cm}\\
$\kappa^*=\left[1+\frac{3}{4}\phi \chi \left( 1+\alpha+\sqrt{\frac{2}{\pi}}
\Delta_\text{M}^*\right)\right]\kappa_\text{k}^*
+\frac{2}{\pi}\phi^2 \chi \left(1+\alpha+\sqrt{\frac{\pi}{2}}
\Delta_\text{M}^*\right)$,\\
\vspace{0.2cm}\\
$\begin{aligned}
\kappa_\text{k}^*=\frac{1}{2\nu_\kappa^*+\Delta_\text{M}^*\Big(\frac{\partial \zeta_0^*}{\partial \Delta^*}\Big)}
\Bigg\{1+\frac{3}{8}\phi \chi(1+\alpha)^2(2\alpha-1)-\frac{\Delta_\text{M}^*}{\sqrt{2\pi}}\phi \chi \nonumber\\
\times\left[\frac{3}{4}+3(1+\al)\left(1-\frac{1}{2}\sqrt{2\pi}\Delta_\text{M}^*\right)-\frac{9}{2}(1+\al)^2-\Delta_\text{M}^{*2}\right]\Bigg\},\nonumber
\end{aligned}$\\
\vspace{0.2cm}\\
$\mu^*=\left[1+\frac{3}{4}\phi \chi \left( 1+\alpha+\sqrt{\frac{2}{\pi}}
\Delta_\text{M}^*\right)\right]\mu_\text{k}^*$,\\
\vspace{0.2cm}\\
$\mu_\text{k}^*=-\frac{1}{\nu_\kappa^*}\phi \chi \left(1+\frac{1}{2}\phi\partial_\phi\ln \chi\right)\Big\{\frac{3}{8}
\al (1-\al^2)-\frac{\Delta_\text{M}^*}{\sqrt{2\pi}}\phi \chi \left[2\Delta_\text{M}^{*2}-3\left(
\frac{1}{2}-\al^2\right)+\frac{3}{2}\sqrt{2\pi}\al\Delta_\text{M}^*\right]\Big\}$,\\
\vspace{0.2cm}\\
$\nu_\eta^*=\frac{3}{8}\chi \left[\left(\frac{7}{3}-\alpha\right)(1+\alpha)
+\frac{2\sqrt{2\pi}}{3}(1-\al)\Delta_\text{M}^*-\frac{2}{3}\Delta_\text{M}^{*2}\right]$,\\
\vspace{0.2cm}\\
$\nu_\kappa^*=\nu_\mu^*=\frac{1+\alpha}{2}\chi\left[
\frac{1}{2}+\frac{15}{8}(1-\alpha)\right]-\frac{\Delta_\text{M}^*}{16}\chi
\left[\sqrt{2\pi}(5\al-1)+10\Delta_\text{M}^{*}\right]$,\\
\vspace{0.2cm}\\
$\Delta_\text{M}^*(\al)=\frac{1}{2}\sqrt{\frac{\pi}{2}}\al\left[\sqrt{1+
\frac{4(1-\al^2)}{\pi \al^2}}-1\right]$,\\
\vspace{0.2cm}\\
$p^*=1+\phi \chi (1+\al)+2\sqrt{\frac{2}{\pi}}\phi \chi\Delta_\text{M}^*$,\\
\vspace{0.2cm}\\
$\chi=\frac{1-\frac{7}{16}\phi}{(1-\phi)^2}$.\\
\vspace{0.5cm}\\
\hline
\hline
\end{tabularx}
\end{table*}

It is quite apparent that to obtain analytical expressions for $\eta$ and $\eta_\text{b}$ one has to (i) consider the steady state ($\zeta^{(0)}=0$) and (ii) replace $\varphi$ by its Maxwellian form. Under these approximations, for a two-dimensional system, one gets the following expressions for the (dimensionless) shear $\eta^*=\eta/\eta_0$ and bulk $\eta_\text{b}^*=\eta_\text{b}/\eta_0$ viscosities: 
\beq
\label{4.35}
\eta^*=\left[1+\frac{1}{2}\phi \chi \left(1+\alpha+\sqrt{\frac{2}{\pi}}\Delta_\text{M}^*\right)\right]\eta_\text{k}^*
+\frac{1}{2}\eta_\text{b}^*,
\eeq
\beq
\label{4.36}
\eta_\text{b}^*=
\frac{8}{\pi}\phi^2 \chi \left(1+\alpha+\sqrt{\frac{\pi}{2}}\Delta_\text{M}^*\right),
\eeq
where
\beq
\label{4.37}
\eta_\text{k}^*=\nu_\eta^{^*-1}\left\{1-\frac{1}{4}\phi \chi\left[(1+\al)
(1-3 \alpha) -{4\sqrt{\frac{2}{\pi}}(1+2\al)}\Delta_\text{M}^*-4\Delta_\text{M}^{*2}\right]\right\},
\eeq
and
\beq
\label{4.38}
\nu_\eta^*=\frac{3}{8}\chi \left[\left(\frac{7}{3}-\alpha\right)(1+\alpha)
+\frac{2\sqrt{2\pi}}{3}(1-\al)\Delta_\text{M}^*-\frac{2}{3}\Delta_\text{M}^{*2}\right].
\eeq
\vicente{Here, we recall that $\Delta_\text{M}^*$ is given by Equation \eqref{3.18}}. When $\Delta_\text{M}^*=0$ in Equations \eqref{4.35}--\eqref{4.38}, one recovers the previous results derived for hard disks in the IHS model for vanishing energy rate ($\zeta^{(0)}=0$) \cite{L05,ACSGP13}.

\subsection{Thermal conductivity and diffusive heat conductivity coefficient} \label{sec.kappaandmu}

The determination of the thermal conductivity coefficient $\kappa$ and the diffusive heat conductivity coefficient $\mu$ follows analogous steps as those exposed before for the shear and bulk viscosities. Given that the calculations are very long, they will be omitted here. We refer to the interested reader to Refs.\ \cite{GBS18,GBS20,GBS26} for more technical details of these calculations. In any case, for the sake of completeness, the explicit expressions of the relevant (scaled) transport coefficients are displayed in Table \ref{table1} for a two-dimensional system as functions of the density and the coefficient of restitution. In Table \ref{table1}, $\kappa^*(\al)=\kappa(\al)/\kappa_0$ where $\kappa_0=(d(d+2)/2(d-1))(\eta_0/m)$ is the low-density value of the thermal conductivity of an elastic gas. Moreover, $\zeta_0^*=\zeta_\text{M}/n\sigma v_\text{th}$ and $\mu^*=n\mu/T\kappa_0$. Note that the coefficient $\mu^*$ vanishes for elastic collisions; for this reason $\mu^*(\al)$ has been scaled with respect to $\kappa^*(1)$ (the value of $\kappa^*$ for elastic collisions). \vicente{Furthermore, the (scaled) heat diffusive coefficient $\mu^*$ also vanishes in the low-density regime ($\phi=0$) when one neglects the contribution of the kurtosis $a_2$ since $\mu^*\propto a_2$ when $\phi=0$.}

Figure \ref{fig2} illustrates the dependence of the (scaled) shear viscosity coefficient $\eta^*(\al)/\eta^*(1)$ for $d=2$ and two values of the solid volume fraction $\phi$: $\phi=0.1$ (dilute granular gas) and $\phi=0.314$ (moderately dense granular gas). Here, $\eta^*(1)$ refers to the value of the (dimensionless) shear viscosity $\eta^*$ for elastic collisions. For the case of $\phi=0.314$, the theoretical results from Equations \eqref{4.25}--\eqref{4.38} are compared with the results from MD simulations \cite{SRB14}.
\vicente{It is worthwhile remarking that the molecular dynamics method is based on Newton's law equations and hence, it avoids the usual assumptions involved in the kinetic theory description (molecular chaos and Equation \eqref{7.15} for accounting for spatial correlations at moderate densities). In this context, a comparison between kinetic theory and MD simulations is a quite stringent test for the former theory.}

According to figure \ref{fig2}, we observe that, for a given value of $\al$, the shear viscosity (scaled with respect to its elastic value) decreases with increasing density. Moreover, for a given density, the shear viscosity decreases with increasing inelasticity. Regarding the comparison with MD simulations, we see that the (approximate) kinetic theory results qualitatively reproduce the simulation trends well, despite the relatively high gas density. At a more quantitative level, as inelasticity increases, the differences between theory and simulations become more significant, as expected.

Figure \ref{fig3} complements Figure \ref{fig2} by showing the $\al$-dependence of the (scaled) thermal conductivity coefficient $\kappa^*(\al)/\kappa^*(1)$ and the (scaled) diffusive heat conductivity coefficient $\mu^*(\al)/\kappa^*(1)$. Unlike the shear viscosity coefficient, we observe that the ratio $\kappa^*(\al)/\kappa^*(1)$ exhibits a non-monotonic dependence on $\al$.  Furthermore, dissipation and density have a greater impact on thermal conductivity than on shear viscosity. Regarding the coefficient $\mu^*$, we observe that 
this coefficient is negative and it is significantly affected by density. In any case, the magnitude of $\mu^*$ is quite small for any density and/or inelasticity; therefore, one can neglect the contribution to the heat flux coming from the density gradient. This means that, for practical purposes, one can assume that the heat flux obeys Fourier's law in the $\Delta$-model: $\mathbf{q}^{(1)}=-\kappa \nabla T$. This conclusion contrasts with the results obtained in the conventional IHS model \cite{BDKS98,GD99a,SMR99,G19}, since the coefficient $\mu$ is always positive, and its magnitude can exceed that of the thermal conductivity $\kappa$ for strong collisional dissipation.

\subsection{Stability of the HSS} \label{sec.stab.HSS}

Knowing the Navier--Stokes transport coefficients and the rate of energy makes it possible to analyze the stability of the HSS. This is a relevant state for confined, quasi-two-dimensional granular fluids. The HSS is a trivial solution of the Navier--Stokes hydrodynamic equations \eqref{2.17}--\eqref{2.19} characterized by a uniform state with $\mathbf{U}=\mathbf{0}$ (without loss of generality) and a steady temperature $T_\text{H}$ determined from the equation $\zeta^{(0)}(n_\text{H},T_\text{H})=0$. Here, the subscripts H denote the homogeneous steady state. This state has been widely studied in several previous papers \cite{BRS13,SRB14,BGMB13} and the theoretical results compare quite well with computer simulations. Since the HSS has been proven stable for \textit{dilute} granular gases \cite{BBGM16}, 
this subsection aims to investigate the stability of the HSS with respect to long enough wavelength perturbations at sufficiently high densities. To answer the above question, we will perform a linear stability analysis of the nonlinear Navier--Stokes hydrodynamic equations  \eqref{2.17}--\eqref{2.19} with respect to the HSS for \emph{small} initial perturbations. The Navier--Stokes equations are obtained by substituting the constitutive equations \eqref{4.19} and \eqref{4.20} into the balance equations \eqref{2.17}--\eqref{2.20}.

Near the HSS, we assume that the deviations $\delta y_{\alpha}({\bf r},t)=y_{\alpha}({\bf r},t)-y_{\text{H} \alpha}$ are small, where
$\delta y_{\alpha}({\bf r},t)$ denotes the deviation of $\{n, {\bf U}, T,\}$ from their values in the HSS. \vicente{To compare with the results obtained in the IHS model \cite{G05}}, we consider here the same time and space variables: $\tau=\frac{1}{2}\nu_{\text{H}}t$ and ${\boldsymbol{\ell}}=\frac{1}{2}(\nu_{\text{H}}/v_{0\text{H}})\mathbf{r}$, where $\nu_{\text{H}}=n_\text{H}T_{\text{H}}/\eta_{0\text{H}}$ and $v_{0\text{H}}=\sqrt{T_{\text{H}}/m}$. Here, $\eta_{0\text{H}}$ is given by Equation \eqref{4.30} with the replacement $T\to T_{\text{H}}$.  
The dimensionless time scale $\tau$ is a measure of the average number of collisions per particle in the time interval between $0$ and $t$. The unit length $v_{0,\text{H}}/\nu_{\text{H}}$ is proportional to the time-independent mean free path of gas particles.

As usual, the linearized hydrodynamic equations for the perturbations 
\beq
\label{4.39}
\left\{\delta n(\mathbf{r}; t), \delta \mathbf{U}(\mathbf{r}; t), \delta T(\mathbf{r}; t)\right\}
\eeq
are written in the Fourier space. A set of Fourier transformed dimensionless variables are then introduced as
$\rho_{{\bf k}}(\tau)=\delta n_{{\bf k}}(\tau)/n_{\text{H}}$, $ {\bf w}_{{\bf k}}(\tau)=\delta {\bf U}_{{\bf k}}(\tau)/v_{0\text{H}}$,
$\theta_{{\bf k}}(\tau)=\delta T_{{\bf k}}(\tau)/T_{\text{H}}$, where $\delta y_{{\bf k}\alpha}\equiv \{\delta \rho_{{\bf k}},{\bf w}_{{\bf k}}(\tau), \theta_{{\bf k}}(\tau)\}$ is defined as
\begin{equation}
\label{4.40}
\delta y_{{\bf k}\alpha}(\tau)=\int d{\boldsymbol {\ell}}\;
e^{-i {\bf k}\cdot {\boldsymbol {\ell}}}\delta y_{\alpha}
({\boldsymbol {\ell}},\tau).
\end{equation}
Note that in Equation (\ref{4.40}) the wave vector ${\bf k}$ is dimensionless.

As occurs in the previous studies on molecular \cite{RL77} and granular \cite{BDKS98,G05} fluids, linearization of the Navier--Stokes equations in $\rho_{{\bf k}}$, $\mathbf{w}_{{\bf k}}$, and $\theta_{{\bf k}}$ shows that the $d-1$ transverse velocity components ${\bf w}_{{\bf k}\perp}={\bf w}_{{\bf k}}-({\bf w}_{{\bf k}}\cdot
\widehat{{\bf k}})\widehat{{\bf k}}$ (orthogonal to the wave vector ${\bf k}$)
decouple from the other three modes. They obey the autonomous differential equation
\beq
\label{4.41}
{\bf w}_{{\bf k}\perp}(\tau)={\bf w}_{{\bf k}\perp}(0) \text{e}^{-\frac{1}{2}\eta^* k^2 \tau},
\eeq
where we have taken into account that $\eta^*$ does not depend on time in the HSS. Thus, since $\eta^*>0$ (see Equation \eqref{4.25}), then the $d-1$ transversal shear modes ${\bf w}_{{\bf k}\perp}( \tau)$ are \textit{linearly} stable.

The remaining (longitudinal) modes correspond to $\rho_{{\bf k}}$, $\theta_{{\bf k}}$, and
the longitudinal velocity component of the velocity field, $w_{{\bf k}||}={\bf w}_{{\bf
k}}\cdot \widehat{{\bf k}}$ (parallel to ${\bf k}$). These modes are coupled and obey the equation
\begin{equation}
\frac{\partial \delta y_{{\bf k}\lambda }(\tau )}{\partial \tau }=M_{\lambda \beta}
 \delta y_{{\bf k}\beta }(\tau ),
\label{4.42}
\end{equation}
where $\delta y_{{\bf k}\alpha }(\tau )$ denotes now the set  $\left\{ \rho _{{\bf k}},\theta _{{\bf k}},
 w_{{\bf k}||}\right\}$ and $\mathsf{M}$ is the square matrix \cite{GBS21a}
\begin{equation}
\mathsf{M}=
\left(
\begin{array}{ccc}
0 & 0 & -i k \\
-\frac{d+2}{2(d-1)}\mu^*k^2&-2\bar{\zeta}_0-\frac{d+2}{2(d-1)}\kappa^*k^2 & -i k\left(\frac{2}{d}p_\text{M}^*+\zeta_U\right)\\
-i k p_\text{M}^* C_\rho & -i k \left(p_\text{M}^*+\Psi_p\right) &-\frac{d-1}{d}\eta^*k^2-\frac{1}{2}\eta_\text{b}^*k^2
\end{array}
\right).
\label{4.43}
\end{equation}
Here, $C_\rho(\phi)=1+(1+\phi \partial_\phi \ln \chi)(1-p_\text{M}^{*-1})$, and it is understood that $p_\text{M}^*$, $\eta^*$, $\eta_\text{b}^*$, $\kappa^*$, $\mu^*$, and $\zeta_U$ are evaluated in the HSS. While $\kappa^*$ and $\mu^*$ were determined in Refs.\ \cite{GBS18,GBS20,GBS26}, the first-order contribution to the rate of energy $\zeta_U$ was not evaluated. A good approximation to it for $d=2$ is \cite{GBS21a}
\beq
\label{4.44}
\zeta_U=\phi \chi \Bigg[2\Delta_\text{M}^{*2}+\frac{2^{5/2}}{\sqrt{\pi}}\al \Delta_\text{M}^*-\frac{3}{2}(1-\al^2)\Bigg].
\eeq
 Additionally, for a two-dimensional system, in Equation \eqref{4.43} we have introduced the dimensionless quantities
\beq
\label{4.45}
\bar{\zeta}_0\equiv T_\text{H}\Bigg(\frac{\partial \zeta_0^*}{\partial T}\Bigg)=\chi \Delta^*\Bigg(\frac{1}{2}\sqrt{\frac{\pi}{2}}\al+\Delta_\text{M}^{*}\Bigg),
\eeq
\beq
\label{4.46}
\Psi_p\equiv T_\text{H}\Bigg(\frac{\partial p_\text{M}^*}{\partial T}\Bigg)=-\vicente{\sqrt{\frac{2}{\pi}}}\phi \chi \Delta_\text{M}^*.
\eeq
Here, \vicente{we recall that $\zeta_0^*=\zeta_\text{M}/n_\text{H}\sigma \sqrt{2T_\text{H}/m}$ where $\zeta_\text{M}$ is given by Equation \eqref{3.16}. }
For dilute granular gases ($\phi=0$), Equations \eqref{4.43}--\eqref{4.46} are consistent with the results derived in Ref.\ \cite{BBGM16} in the low-density limit.

\begin{figure}[h!]
\centering
\includegraphics[width=0.45\textwidth]{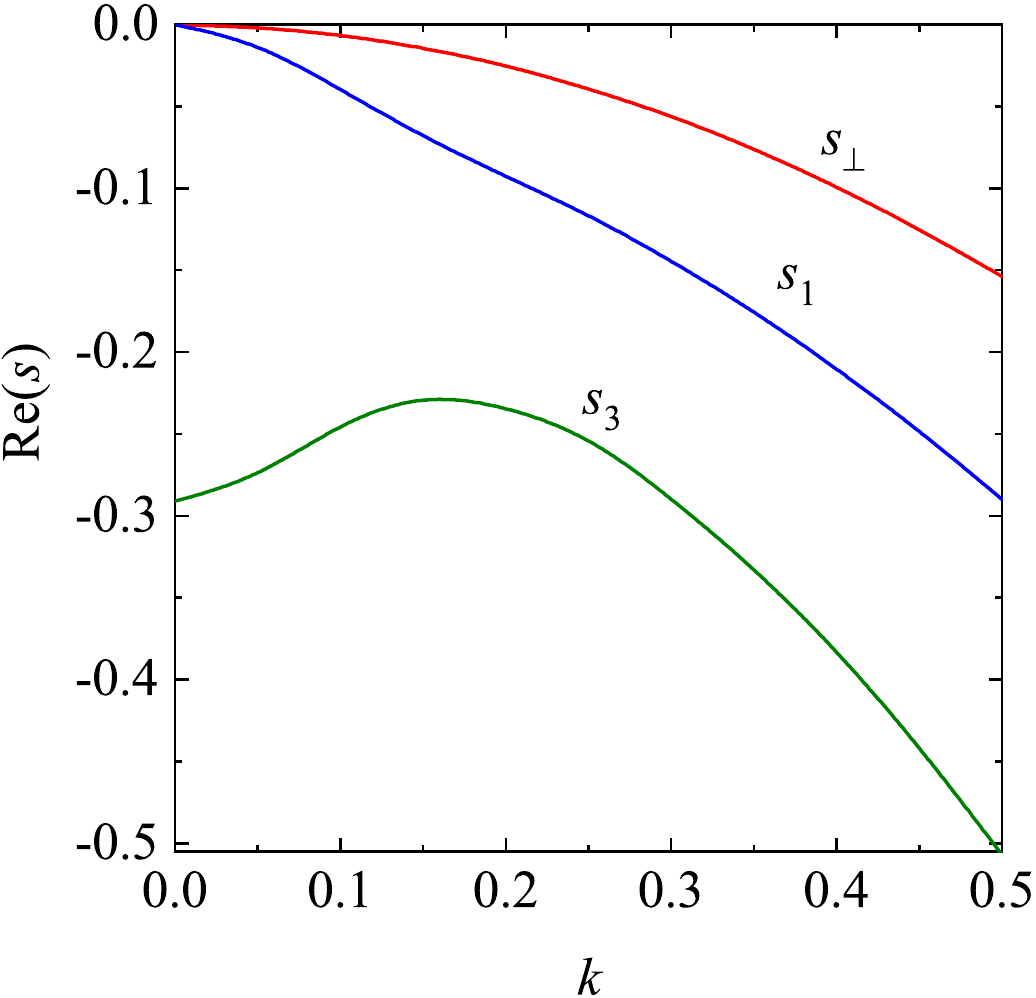}
\caption{Dispersion relations for a granular two-dimensional fluid ($d=2$) with $\alpha=0.8$ and
$\phi=0.2$. From top to bottom the curves correspond to the real parts of the
shear (transversal) mode $s_\perp$ and
the remaining three  longitudinal modes ($s_1=s_2$ and $s_3$).  
}
\label{fig4}
\end{figure}

The longitudinal three modes have the form $\exp[s_n(k) \tau]$ for $n=1,2,3$, where $s_n(k)$ are the eigenvalues of the matrix $\mathsf{M}$. For given values of the coefficient of restitution $\al$ and the density $\phi$, for $k\neq 0$ we find that one of the modes is real while the other two are a complex conjugate pair of propagating modes. Furthermore, an analysis of the eigenvalues of the matrix $\mathsf{M}$ for finite $k$ and moderate densities shows that in general $\text{Re}(s_n)\leq 0$ and hence the HSS is linearly \emph{stable} in the complete range of values of the wave number $k$ studied. As an illustration, the dispersion relations $s_{n}(k)$ for a two-dimensional granular fluid with $\alpha=0.8$ and $\phi=0.2$ are plotted in Figure \ref{fig4}. Only the real parts of the eigenvalues are represented. \vicente{We observe that the real part of the "heat" mode $s_3$ exhibits a a non-monotonic dependence on the (dimensionless) wave number $k$ while the other modes ($s_1=s_2$ and $s_\perp$) decrease with increasing $k$}.

\section{Granular mixtures}\label{sec6}

\subsection{Enskog kinetic equation}

Granular materials are usually present in nature or industry as polydisperse systems. The extension of the Enskog equation to granular mixtures within the context of the $\Delta$-model is straightforward. We consider an $s$-multicomponent granular mixture of inelastic, smooth hard disks ($d=2$) or spheres ($d=3$) of masses $m_i$ and diameters $\sigma_i$. Collisions among all pairs are inelastic and characterized by independent coefficients of normal restitution $\al_{ij}=\al_{ji}$, where $\al_{ij}$ is the coefficient of restitution for collisions between particles of species $i$ and $j$. For moderately dense systems, 
in the presence of the gravity field $m_i \mathbf{g}$, the set of Enskog kinetic equations are
\beq
\label{6.1}
\frac{\partial}{\partial t}f_i+\mathbf{v}\cdot \nabla f_i+\mathbf{g}\cdot \frac{\partial f_i}{\partial \mathbf{v}}
=\sum_{j=1}^s\; J_{\text{E},ij}[\mathbf{r},\mathbf{v}|f_i,f_j],
\eeq
where the Enskog collision operators $J_{\text{E},ij}$ for collisions $i$-$j$ in the $\Delta$-model read \cite{BSG20}
\beqa
\label{6.2}
& & J_\text{E,ij}[\mathbf{r},\mathbf{v}_1|f_i,f_j]\equiv \sigma_{ij}^{d-1}\int d{\bf v}_{2}\int d \widehat{\boldsymbol{\sigma}}
\Theta (-\widehat{{\boldsymbol {\sigma }}}\cdot {\bf g}_{12}-2\Delta_{ij})
(-\widehat{\boldsymbol {\sigma }}\cdot {\bf g}_{12}-2\Delta_{ij})
 \nonumber\\
& & \times \al_{ij}^{-2} f_{2,ij}(\mathbf{r}, \mathbf{r}+{\boldsymbol {\sigma }}_{ij},\mathbf{v}_1'', \mathbf{v}_2'';t)-\sigma_{ij}^{d-1}\int\ d{\bf v}_{2}\int d\widehat{\boldsymbol{\sigma}}
\Theta (\widehat{{\boldsymbol {\sigma}}}\cdot {\bf g}_{12})
(\widehat{\boldsymbol {\sigma}}\cdot {\bf g}_{12})\nonumber\\
& & \times f_{2,ij}(\mathbf{r}, \mathbf{r}+{\boldsymbol {\sigma}}_{ij},\mathbf{v}_1, \mathbf{v}_2;t).
\eeqa
Here, $\boldsymbol {\sigma}_{ij}=\sigma_{ij}\widehat{\boldsymbol {\sigma}}$, $\sigma_{ij}=(\sigma_i+\sigma_j)/2$, and  
\beq
\label{6.3}
f_{2,ij}(\mathbf{r}_1,\mathbf{r}_2,\mathbf{v}_1,\mathbf{v}_2; t)\equiv 
\chi_{ij}(\mathbf{r}_1,\mathbf{r}_2) f_i(\mathbf{r}_1,\mathbf{v}_1;t)f_j(\mathbf{r}_2,\mathbf{v}_2;t),
\eeq
$\chi_{ij}(\mathbf{r}_1,\mathbf{r}_2)$ being the pair correlation function for collisions $i-j$.
In Equation \eqref{6.2}, the collision rules for the restituting collisions $\left(\mathbf{v}_1'',\mathbf{v}_2''\right)\to \left(\mathbf{v}_1,\mathbf{v}_2\right)$ with the same collision vector $\widehat{{\boldsymbol {\sigma }}}$ are defined as
\beq
\label{6.4}
\mathbf{v}_1''=\mathbf{v}_1-\mu_{ji}\left(1+\alpha_{ij}^{-1}\right)(\widehat{{\boldsymbol {\sigma }}}\cdot \mathbf{g}_{12})\widehat{{\boldsymbol {\sigma }}}-2\mu_{ji}\Delta_{ij}\al_{ij}^{-1} \widehat{{\boldsymbol {\sigma }}},
\eeq
\beq
\label{6.5}
\mathbf{v}_2''=\mathbf{v}_2+\mu_{ij}\left(1+\alpha_{ij}^{-1}\right)(\widehat{{\boldsymbol {\sigma }}}\cdot \mathbf{g}_{12})\widehat{{\boldsymbol {\sigma}}}+2\mu_{ij}\Delta_{ij}\al_{ij}^{-1} \widehat{{\boldsymbol {\sigma }}},
\eeq
where $\mu_{ij}=m_i/(m_i+m_j)$. Analogously, the direct collisions $\left(\mathbf{v}_1,\mathbf{v}_2\right)\to \left(\mathbf{v}_1',\mathbf{v}_2'\right)$ are defined as 
\beq
\label{6.6}
\mathbf{v}_1'=\mathbf{v}_1-\mu_{ji}\left(1+\alpha_{ij}\right)(\widehat{{\boldsymbol {\sigma }}}\cdot \mathbf{g}_{12})\widehat{{\boldsymbol {\sigma }}}-2\mu_{ji}\Delta_{ij} \widehat{{\boldsymbol {\sigma }}},
\eeq
\beq
\label{6.6.1}
\mathbf{v}_{2}'=\mathbf{v}_2+\mu_{ij}\left(1+\alpha_{ij}\right)(\widehat{{\boldsymbol {\sigma }}}\cdot \mathbf{g}_{12})\widehat{{\boldsymbol {\sigma }}}+2\mu_{ij}\Delta_{ij} \widehat{{\boldsymbol {\sigma }}}.
\eeq
As in the case of monocomponent granular fluids, the property \eqref{2.13} for the Enskog collision operators $J_{\text{E},ij}[\mathbf{r},\mathbf{v}|f_i,f_j]$ still applies except that $f_{2}(\mathbf{r}, \mathbf{r}+{\boldsymbol {\sigma}},\mathbf{v}_1, \mathbf{v}_2;t)$ must be replaced by $f_{2,ij}(\mathbf{r}, \mathbf{r}+{\boldsymbol {\sigma}}_{ij},\mathbf{v}_1, \mathbf{v}_2;t)$ and $\mathbf{v}_1'$ is given by Equation \eqref{6.6}.  

The use of the property \eqref{2.13} for granular mixtures allows to derive the corresponding balance equations for the densities of mass, momentum, and energy. As expected, their forms are similar to those obtained in the IHS model \cite{GDH07} and are given by 
\begin{equation}
D_{t}n_{i}+n_{i}\nabla \cdot {\bf U}+\frac{\nabla \cdot {\bf j}_{i}}{m_{i}}
=0,  \label{6.7}
\end{equation}
\begin{equation}
D_{t}{\bf U}+\rho ^{-1}\nabla \cdot \mathsf{P}=\mathbf{g},  \label{6.8}
\end{equation}
\begin{equation}
D_{t}T-\frac{T}{n}\sum_{i=1}^s\frac{\nabla \cdot {\bf j}_{i}}{m_{i}}+\frac{2}{dn}
\left( \nabla \cdot {\bf q}+\mathsf{P}:\nabla {\bf U}\right) =-\zeta \,T.
\label{6.9}
\end{equation}
In Equations \eqref{6.7}--\eqref{6.9},
\begin{equation}
n_{i}=\int d{\bf v}f_{i}({\bf v})
\label{6.10}
\end{equation}
is the number density of species $i$,
\beq
\label{6.11}
\mathbf{U}=\rho^{-1}\sum_{i=1}^s m_{i}\int d {\bf v}{\bf v}f_{i}({\bf v})
\eeq
is the mean flow velocity, and
\begin{equation}
T=\frac{1}{d n}\sum_{i=1}^sm_{i}\int d{\bf
v}V^{2}f_{i}({\bf v}) \label{6.12}
\end{equation}
is the (global) granular temperature. In addition, $\rho=\sum_i \rho_i=\sum_i m_i n_i$ is the total mass density, and we recall that ${\bf V}={\bf v}-{\bf U}$ is the peculiar velocity. Apart from the granular temperature $T$, at a kinetic level it is convenient to introduce the partial temperatures $T_i$ for each species; they measure their mean kinetic energies. They are defined as
\beq
\label{6.13}
n_i T_i=\frac{m_{i}}{d}\int d{\bf v}V^{2}f_{i}({\bf v}),
\eeq
and hence, $nT=\sum_i n_i T_i$.

In the balance equations \eqref{6.7}--\eqref{6.9},
\begin{equation}
{\bf j}_{i}=m_{i}\int d{\bf v}_{1}\,{\bf V}_{1}\,f_{i}({\bf v}_{1})
\label{6.14}
\end{equation}
is the mass flux for the species $i$ relative to the local flow. The mass flux $\mathbf{j}_i$ has only \emph{kinetic} contributions.  The kinetic contributions to the pressure tensor $\mathsf{P}$ and the heat flux $\mathbf{q}$ are given as usual by 
\begin{equation}
\mathsf{P}_\text
{k}=\sum_{i=1}^s\,\int d{\bf v}\,m_{i}{\bf V}{\bf V}\,f_{i}({\bf v}),
\label{6.15}
\end{equation}
\begin{equation}
{\bf q}_\text{k}=\sum_{i=1}^s\,\int d{\bf v}\,\frac{1}{2}m_{i}V^{2}{\bf V}\,f_{i}({\bf v}).
\label{6.16}
\end{equation}

The collisional transfer contributions for the pressure tensor and the heat flux can be derived by following similar steps as those made in the $\Delta$-model for monocomponent granular gases \cite{GBS18}. Their expressions  are
\beqa
\label{6.17}
\mathsf{P}_{\text{c}}&=&\sum_{i=1}^s\sum_{j=1}^s\frac{1+\alpha_{ij}}{2}m_{ij}\sigma_{ij}^{d}
\int d\mathbf{v}_1\int d\mathbf{v}_2
\int d\widehat{\boldsymbol {\sigma}}\,\Theta (\widehat{{\boldsymbol {\sigma}}}
\cdot \mathbf{g}_{12})(\widehat{\boldsymbol {\sigma }}\cdot {\bf g}_{12})
\widehat{\boldsymbol {\sigma}}
\widehat{\boldsymbol {\sigma }}
\nonumber \\
& & \times\left[(\widehat{\boldsymbol {\sigma}}\cdot {\bf g}_{12})+\frac{2\Delta_{ij}}{1+\al_{ij}}\right] \int_{0}^{1}\; d\lambda f_{2,ij}\Big(\mathbf{r}-\lambda \boldsymbol{\sigma}_{ij},
\mathbf{r}+(1-\lambda)\boldsymbol{\sigma}_{ij},\mathbf{v}_1,\mathbf{v}_2,t\Big),
\eeqa
\beqa
\label{6.18}
{\bf q}_\text{c}&=&\sum_{i=1}^s\sum_{j=1}^s\frac{1+\alpha_{ij}}{8}m_{ij} \sigma_{ij}^{d}
\int d\mathbf{v}_{1}\int d\mathbf{v}_{2}\int
d\widehat{\boldsymbol {\sigma}}\Theta (\widehat{\boldsymbol{\sigma}}\cdot
\mathbf{g}_{12})(\widehat{\boldsymbol {\sigma}}\cdot \mathbf{g}_{12})^{2}\widehat{\boldsymbol {\sigma}}\Big[4
(\widehat{\boldsymbol {\sigma}}\cdot {\bf G}_{ij})\nonumber\\
& &+(\mu_{ji}-\mu_{ij})(1-\al_{ij})
(\widehat{\boldsymbol{\sigma}}\cdot
\mathbf{g}_{12})\Big]\int_{0}^{1}d\lambda f_{2,ij}\Big(\mathbf{r}-
\lambda{\boldsymbol{\sigma}}_{ij},\mathbf{r}+(1-\lambda)
{\boldsymbol {\sigma}}_{ij},\mathbf{v}_{1},\mathbf{v}_{2},t\Big)
\nonumber\\
& &-\sum_{i=1}^s\sum_{j=1}^s \frac{m_i}{4} \sigma_{ij}^d \Delta_{ij}
\int d\mathbf{v}_{1}\int d\mathbf{v}_{2}\int
d\widehat{\boldsymbol {\sigma}}\Theta (\widehat{\boldsymbol{\sigma}}\cdot
\mathbf{g}_{12})(\widehat{\boldsymbol {\sigma}}\cdot \mathbf{g}_{12})
\widehat{\boldsymbol {\sigma}}
\Big[4\mu_{ji}^2\Delta_{ij} \nonumber\\
& & +4\mu_{ji}^2\al_{ij} (\widehat{\boldsymbol {\sigma}}\cdot \mathbf{g}_{12})-
4\mu_{ji} (\widehat{\boldsymbol {\sigma}}\cdot \mathbf{G}_{ij})\Big]
\int_{0}^{1}d\lambda 
f_{2,ij}\Big(\mathbf{r}-\lambda {\boldsymbol{\sigma}}_{ij},\mathbf{r}+(1-\lambda)
{\boldsymbol {\sigma}}_{ij},\mathbf{v}_{1},\mathbf{v}_{2},t\Big). \nonumber\\
\eeqa
The energy rate $\zeta$ is
\beqa
\label{6.19}
\zeta&=&-\frac{2}{d n T}\sum_{i=1}^s\sum_{j=1}^s\sigma_{ij}^{d-1}m_{ij}\int \dd \mathbf{v}_1\int \dd \mathbf{v}_2
\int d\widehat{\boldsymbol {\sigma }}\,\Theta (\widehat{{\boldsymbol {\sigma}}}
\cdot \mathbf{g}_{12})(\widehat{\boldsymbol {\sigma}}\cdot {\bf g}_{12})\nonumber\\
& & \times \Big[\Delta_{ij}^2+\al_{ij}\Delta_{ij}(\widehat{{\boldsymbol {\sigma}}}\cdot \mathbf{g}_{12})-\frac{1-\al_{ij}^2}{4}(\widehat{{\boldsymbol {\sigma}}}\cdot \mathbf{g}_{12})^2\Big]
f_{2,ij}(\mathbf{r},\mathbf{r}+\boldsymbol{\sigma}_{ij},\mathbf{v}_1,\mathbf{v}_2,t).
\eeqa
In Equations \eqref{6.18}--\eqref{6.19}, $m_{ij}=m_im_j/(m_i+m_j)$ is the reduced mass and  $\mathbf{G}_{ij}=\mu_{ij}\mathbf{V}_1+\mu_{ji}\mathbf{V}_2$ is the center-of-mass velocity.

\subsection{Homogeneous time-dependent state}
\label{sec7}

As in the case of monocomponent gases, we consider first spatially homogeneous isotropic states for which the Enskog equation \eqref{6.1} in the absence of gravity becomes  
\beq
\label{7.1}
\frac{\partial}{\partial t}f_i(\mathbf{v};t)=\sum_{j=1}^s\; J_{\text{E},ij}[\mathbf{v}|f_i,f_j],
\eeq
where $J_{\text{E},ij}[f_i,f_j]$ is defined by Equation \eqref{6.2} with the replacements $\chi_{ij}(\mathbf{r},\mathbf{r}\pm\boldsymbol{\sigma}_{ij})\to \chi_{ij}$ and $f_{2,ij}(\mathbf{r},\mathbf{r}+\boldsymbol{\sigma}_{ij},\mathbf{v}_1, \mathbf{v}_2; t)\to \chi_{ij} f_i(\mathbf{v}_1;t) f_j(\mathbf{v}_2;t)$. Here, $\chi_{ij}$ is the (homogeneous) pair correlation function at contact for collisions $i$-$j$.

In the homogeneous time-dependent state, the only nontrivial balance equation is that for the temperature $T$. Since the mass and heat fluxes vanish and $\mathbf{U}=\mathbf{0}$, the equation for $T(t)$ of a multicomponent granular mixture is still given by Equation \eqref{3.3} with 
\beqa
\label{7.2}
\zeta&=&-\frac{2}{d n T}\sum_{i=1}^s \sum_{j=1}^s\sigma_{ij}^{d-1}m_{ij}\chi_{ij} 
\int d\mathbf{v}_{1}\int d\mathbf{v}_{2}
\Big(B_1 g_{12}\Delta_{ij}^2+B_2 g_{12}^2 \al_{ij} \Delta_{ij}-B_3 g_{12}^3 \frac{1-\al_{ij}^2}{4}\Big)
\nonumber\\
& & \times f_i(\mathbf{v}_1,t)f_j(\mathbf{v}_2,t),
\eeqa
where the coefficients $B_k$ are defined in Equation \eqref{3.5}. The time evolution of the partial temperatures $T_i$ can be directly obtained from the Enskog equation \eqref{7.1} and the definition \eqref{6.13}:
\beq
\label{7.3}
\frac{\partial T_i}{\partial t}=-\zeta_i T_i,
\eeq
where 
\beq
\label{7.4}
\zeta_i=\sum_{j=1}^s \zeta_{ij}=
-\frac{1}{d n_i T_i}\sum_{j=1}^s \int d\mathbf{v} m_i v^2 J_{\text{E},ij}[f_i,f_j].
\eeq
According to Equations \eqref{7.3} and \eqref{7.4}, 
\beq
\label{7.5}
\zeta=\sum_{i=1}^s\; x_i \gamma_i \zeta_i,
\eeq
where $x_i=n_i/n$ is the concentration or mole fraction of species $i$ and $\gamma_i=T_i/T$ is the temperature ratio of species $i$. The deviation of $\gamma_i$ from 1 provides a measure of the departure from energy equipartition (i.e., when $T_i=T$ for any component $i$). The time evolution of the temperature ratios $\gamma_i(t)$ can be easily obtained from Equations \eqref{3.3} and \eqref{7.3} as
\beq
\label{7.5.1}
\frac{\partial}{\partial t}\ln \gamma_i=\zeta-\zeta_i.
\eeq

As in the monocomponent case, after a transient period, one expects that the velocity distribution functions $f_i(\mathbf{v},t)$ adopt a \textit{normal} form where the time-dependence of them is only through the global granular temperature $T(t)$. This means that $f_i(\mathbf{v},t)$ is given by the scaling distribution
\begin{equation}
\label{7.6}
f_i(\mathbf{v},t)=n_iv_{\text{th}}^{-d}(t)\varphi_i\left(\mathbf{c}, \Delta_{\ell j}^*\right),\quad \ell, j=1,\ldots, s,
\end{equation}
where we recall that $\mathbf{c}\equiv \mathbf{v}/v_{\text{th}}$ and $v_{\text{th}}(t)=\sqrt{2T(t)/\overline{m}}$ is a thermal velocity
defined in terms of the temperature of the mixture $T(t)$. In addition, $\overline{m}=\sum_i m_i/s$ and $\Delta_{ij}^*(t)\equiv \Delta_{ij}/v_{\text{th}}(t)$. According to Equation \eqref{7.6}, $\partial_t f_i=-\zeta T \partial_T f_i$ and so, 
\beq
\label{7.7}
\frac{\partial f_i}{\partial t}=\frac{1}{2}\zeta\frac{\partial }{\partial \mathbf{v}}\cdot \left(\mathbf{v}f_i\right)
+\frac{1}{2}\zeta\sum_{j=1}^s\sum_{\ell=1}^s\; \Delta_{\ell j}^*\frac{\partial f_i}{\partial \Delta_{\ell j}^*}.
\eeq
Thus, in dimensionless form, the set of $s$ coupled Enskog equations \eqref{7.1} for the homogeneous time-dependent problem can be written as
\beq
\label{7.8}
\frac{1}{2}\zeta^*\left(\frac{\partial}{\partial \mathbf{c}}\cdot \left(\mathbf{c}\varphi_i\right)
+\sum_{j=1}^s\sum_{\ell=1}^s\; \Delta_{\ell j}^*\frac{\partial \varphi_i}{\partial \Delta_{\ell j}^*}\right)=\sum_{j=1}^s\; J_{\text{E},ij}^*[\mathbf{c}|\varphi_i,\varphi_j],
\eeq
where $\zeta^*=\zeta/\nu$ and $J_{\text{E},ij}^*[\mathbf{c}|\varphi_i,\varphi_j]=(v_\text{th}^d/n_i\nu)J_{\text{E},ij}[\mathbf{v}|f_i,f_j]$. Here, $\nu=n\overline{\sigma}^{d-1}v_\text{th}$ is an effective collision frequency and $\overline{\sigma}=\sum_i \sigma_i/s$.

Since $f_i(\mathbf{v})$ depends on $\mathbf{v}$ through its modulus, then the mass and heat fluxes vanish and $P_{\lambda \beta}=p \delta_{\lambda \beta}$. The hydrostatic pressure $p=n T p^*$, where the coefficient $p^*$ for a multicomponent granular mixture is  
\beqa
\label{7.9}
p^*&=&1+\frac{\pi^{d/2}}{d\Gamma\left(\frac{d}{2}\right)}\sum_{i=1}^s\sum_{j=1}^s\mu_{ji}n\sigma_{ij}^d\chi_{ij}x_ix_j\Bigg[(1+\al_{ij})\gamma_i+
\frac{2}{\sqrt{\pi}}\frac{\Gamma\left(\frac{d}{2}\right)}{\Gamma\left(\frac{d+1}{2}\right)}
\frac{m_i}{\overline{m}}\Delta_{ij}^*\nonumber\\
& & \times \int d\mathbf{c}_1 \int d\mathbf{c}_2 g_{12}^*\varphi_i (\mathbf{c}_1)\varphi_j (\mathbf{c}_2)\Bigg].
\eeqa
For mechanically equivalent particles ($m_i=m$, $\sigma_i=\sigma$, $\al_{ij}=\al$, and $\Delta_{ij}^*=\Delta^*$), $\gamma_i=\theta_i=1$ and Equation \eqref{7.9} agrees with Equation \eqref{3.12}. 

\subsection{Homogeneous steady states. Maxwellian approximation}

We consider here the steady state solution to Equation \eqref{7.8}. In this case, for given values of $\Delta_{ij}^*$, $\partial_t T_i(t)=0$ and according to Equation \eqref{7.3}
\beq
\label{7.10}
\zeta=\zeta_1=\zeta_2=\cdots=\zeta_s=0.
\eeq
As expected, the determination of $\zeta_i$ requires the knowledge of the scaling distributions $\varphi_i$, whose exact form is not known to date. As in the conventional IHS model \cite{GD99b}, the distributions $\varphi_i$ can be expanded in a series of Sonine polynomials, the coefficients (cumulants) of the series being the corresponding velocity moments of $\varphi_i$. Here, as in Section \ref{sec3}, to estimate the partial energy rates $\zeta_i$, we take the simplest Maxwellian approximation $\varphi_{i,\text{M}}(\mathbf{c})$ to $\varphi_i(\mathbf{c})$, namely,
\beq
\label{7.11}
\varphi_i(\mathbf{c})\to \varphi_{i,\text{M}}(\mathbf{c})=\pi^{-d/2} \theta_i^{d/2}\; e^{-\theta_i c^2},
\eeq
where $\theta_i=m_i/(\overline{m}\gamma_i)$. As in previous works on granular mixtures \cite{GD99b}, for the sake of convenience, $\varphi_{i,\text{M}}$ is defined in terms of the partial temperature $T_i$ instead of the (global) granular temperature $T$. 

With the Maxwellian approximation \eqref{7.11}, the partial energy rate $\zeta_i\to \zeta_{i,\text{M}}$ can be computed. In dimensionless form, it can be written as $\zeta_{i,\text{M}}=\zeta_{i,\text{M}}^* \nu$ where \cite{BSG20} 
\beqa
\label{7.12}
\zeta_{i,\text{M}}^*&=&\frac{4\pi^{(d-1)/2}}{d\Gamma\left(\frac{d}{2}\right)}\sum_{j=1}^s
x_j\chi_{ij}
\left(\frac{\sigma_{ij}}{\overline{\sigma}}\right)^{d-1}\mu_{ji}(1+\al_{ij})\theta_i^{-1/2}
\left(1+\theta_{ij}\right)^{1/2}\nonumber\\
& & \times 
\left[1-\frac{1}{2}\mu_{ji}(1+\alpha_{ij})(1+\theta_{ij}) \right]-\frac{4\pi^{d/2}}{d\Gamma\left(\frac{d}{2}\right)}\sum_{j=1}^s x_j\chi_{ij}
\left(\frac{\sigma_{ij}}{\overline{\sigma}}\right)^{d-1}\mu_{ji}\Delta_{ij}^*
\nonumber\\
& & \times
\left[
\frac{2\mu_{ji}\Delta_{ij}^*}{\sqrt{\pi}}\theta_i^{1/2}\left(1+\theta_{ij}\right)^{1/2}
-1+\mu_{ji}(1+\al_{ij})\left(1+\theta_{ij}\right)\right].
\eeqa
Here,  $\theta_{ij}=\theta_i/\theta_j=m_i\gamma_j/m_j\gamma_i$ gives the ratio between the mean-square
velocity of the particles of the species $j$ relative to that of the particles of the species $i$. Moreover, taking the Maxwellian approximation \eqref{7.11}, the expression \eqref{7.9} for $p^*$ reduces to 
\beq
\label{7.12.1}
p_\text{M}^*=1+\frac{\pi^{d/2}}{d\Gamma\left(\frac{d}{2}\right)}\sum_{i=1}^s\sum_{j=1}^s\mu_{ji}n\sigma_{ij}^d\chi_{ij}
x_ix_j\Bigg[(1+\al_{ij})\gamma_i
+
\frac{2}{\sqrt{\pi}}\frac{m_i}{\overline{m}}\Delta_{ij}^*\left(\frac{\theta_i+\theta_j}{\theta_i\theta_j}\right)^{1/2}
\Bigg].
\eeq
 In the limit of mechanically equivalent particles, Equations \eqref{7.12} and \eqref{7.12.1} agree with Equations \eqref{3.16} and \eqref{3.17}, respectively, as expected.

\subsection{Binary mixtures. Comparison between kinetic theory and computer simulations}

To illustrate the dependence of the partial temperatures on the parameters of the mixture, we consider a binary mixture ($s=2$) for the sake of simplicity. In this case, the relevant dimensionless quantities in the steady state are the scaled temperature $T^*$ (defined below) and the temperature ratio $T_1/T_2$. Both quantities are determined from the constraints \eqref{7.10}:
\beq
\label{7.13}
\zeta_1^*=0, \quad \zeta_2^*=0.
\eeq
The solution to Equations \eqref{7.13} with the expression \eqref{7.12} for the energy rates provides $T^*$ and $T_1/T_2$ in terms of the parameter space of the problem. This is constituted by the ratio of masses $m_1/m_2$, the ratio of diameters $\sigma_1/\sigma_2$, the concentration $x_1$, the volume fraction or density $\phi$, the coefficients of restitution $\al_{11}$, $\al_{22}$, and $\al_{12}$, and the dimensionless velocities $\Delta_{11}^*$, $\Delta_{22}^*$, and $\Delta_{12}^*$. In the case of a two-dimensional ($d=2$) system, the volume fraction $\phi$ is defined as
\beq
\label{7.14}
\phi=\sum_{i=1}^2\; \frac{\pi}{4}n_i \sigma_i^2,
\eeq
while a good approximation for the pair distribution function is \cite{JM87}
\beq
\label{7.15}
\chi_{ij}=\frac{1}{1-\phi}+\frac{9}{16}\frac{\phi}{(1-\phi)^2}\frac{\sigma_i\sigma_j M_1}
{\sigma_{ij}M_2},
\eeq
where $M_\ell=\sum_i x_i \sigma_i^\ell$. Finally, the reduced (steady) temperature $T^*$ is defined as
\beq
\label{7.16}
T^*=\frac{T}{\overline{m}\overline{\Delta}^2/2},
\eeq
where $\overline{\Delta}=\sqrt{\Delta_{11}^2+\Delta_{22}^2+
\Delta_{12}^2}.$

\begin{figure}[h!]
\centering
\includegraphics[width=0.45\textwidth]{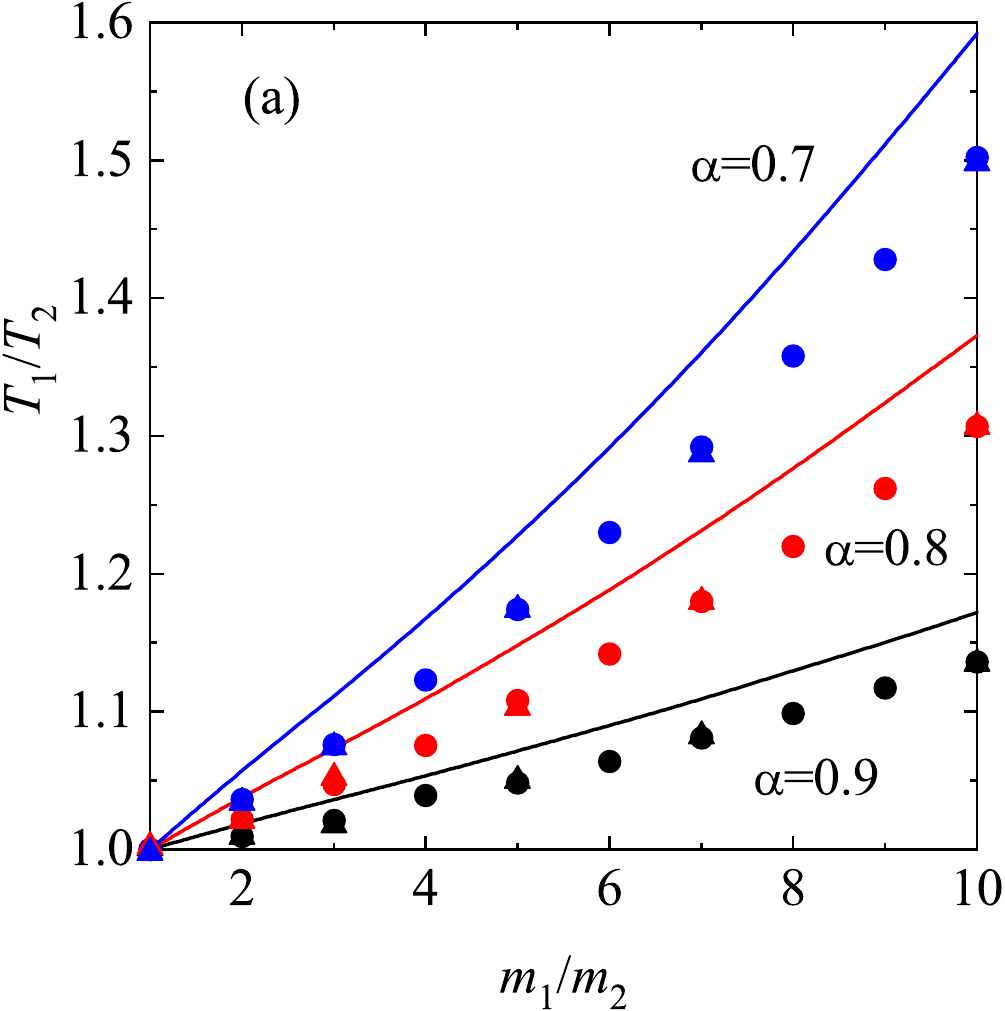}
\includegraphics[width=0.46\textwidth]{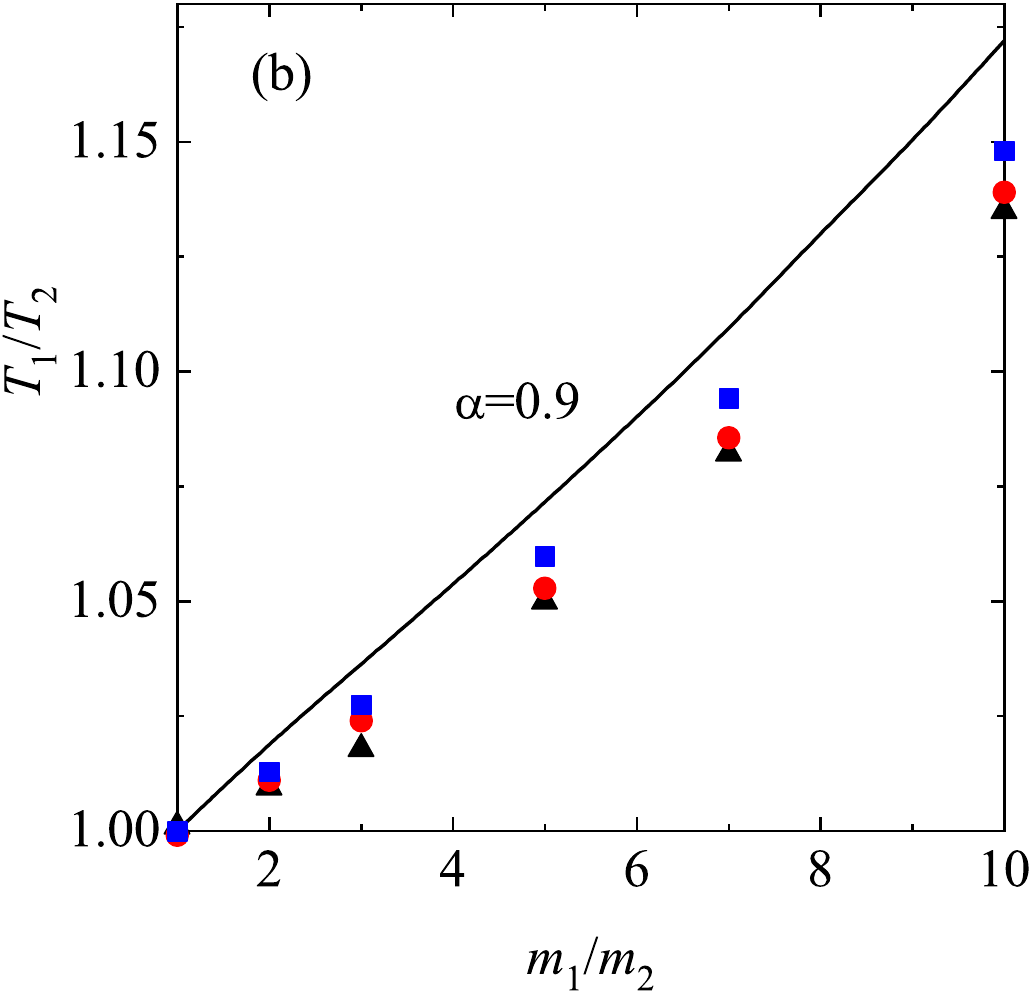}
\caption{Panel (a): Plot of the temperature ratio $T_1/T_2$ versus the mass ratio $m_1/m_2$ for $\sigma_1=\sigma_2$, and three different values of the (common) coefficient of restitution $\alpha$: $\al=0.9$, 0.8 and 0.7. The lines refer to the Enskog theoretical results while the symbols correspond to the results obtained by numerically solving the Enskog equation by means of the DSMC method (circles) and by performing MD simulations for $\phi=0.0016$ (triangles). Panel (b): Plot of the temperature ratio $T_1/T_2$ versus the mass ratio $m_1/m_2$ for $\sigma_1=\sigma_2$, $\al=0.7$, and three different values of the volume fraction $\phi$: $\phi=0.0016$ (triangles), 0.1 (circles) and 0.2 (squares). Symbols refer to MD simulations and the line to the Enskog theoretical result. We assume in both panels that $\Delta_{11}=\Delta_{22}=\Delta_{12}$. 
Reprinted figure with permission from  R.\ Brito, R.\ Soto, and V.\ Garz\'{o},  Phys.~Rev.~E \textbf{2020}, \textit{102}, 052904 \cite{BSG20}. Copyright (2020) by the American Physical Society.
}
\label{fig6}
\end{figure}

Since there are relatively many parameters involved in the problem, we take a common coefficient of restitution $\al_{11}=\al_{22}=\al_{12}\equiv\al$, as usual. Moreover, we consider two-dimensional granular mixtures with the concentration $x_1=\frac{1}{2}$. The (approximate) theoretical results for $T^*$ and $T_1/T_2$ will be compared with two different standard simulation methods. The first one is the direct simulation Monte Carlo (DSMC) method \cite{B94} introduced years ago by Bird for dilute molecular gases. Here, we have adapted this method to a gas of hard disks with inelastic collisions. The DSMC method numerically solves the inelastic Boltzmann equation and assumes the molecular chaos hypothesis, namely the absence of velocity correlations between particles about to collide. However, the method does not assume the existence of the normal solution \eqref{7.6}, and it goes beyond the Maxwellian approximation to $f_i$ (it determines the ``exact'' velocity distribution functions). 
Thus, comparing analytical and DSMC results for very dilute systems ($\phi\to 0$) can be used to evaluate the reliability of the scaling solution \eqref{7.6} and the accuracy of the expression \eqref{7.12} for estimating the partial energy rates (this expression is obtained by replacing the true $\varphi_i$ by its Maxwellian form \eqref{7.11}). \vicente{As previously noted in section \ref{sec4}, since MD simulations avoids the  assumptions of the Enskog kinetic theory}, comparing the theory to MD is more stringent than comparing it to DSMC results. In both simulation methods (DSMC and MD), particles move in two dimensions with collisions given by the rules \eqref{6.6} and \eqref{6.6.1} of the $\Delta$-model.


\begin{figure}[h!]
\centering
\includegraphics[width=0.45\textwidth]{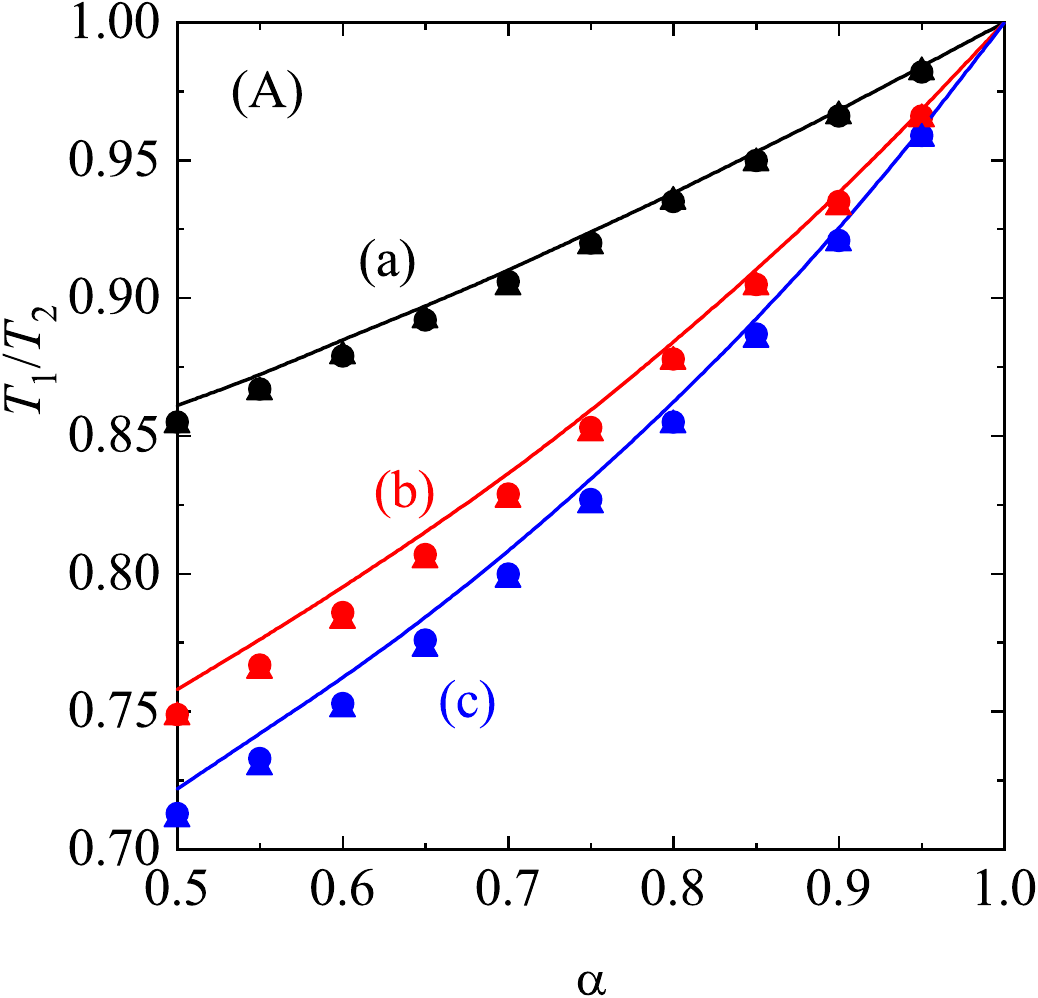}
\includegraphics[width=0.46\textwidth]{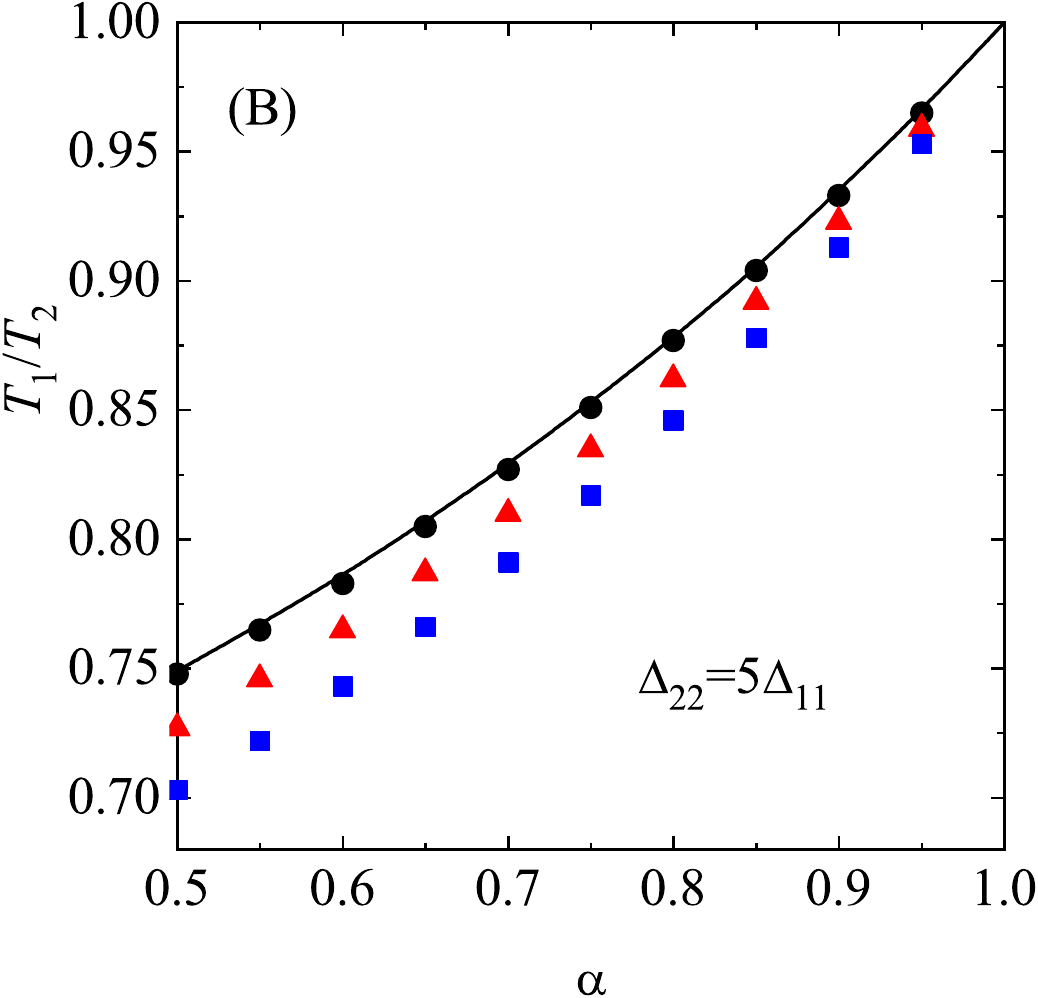}
\caption{Panel (A): Plot of the temperature ratio $T_1/T_2$ versus the (common) coefficient of restitution $\al$ for $\sigma_1=\sigma_2$ and $m_1=m_2$.  We assume here that $\Delta_{22}=\lambda \Delta_{11}$ and $\Delta_{12}=(\Delta_{11}+\Delta_{22})/2$. Three different values of $\lambda$ have been considered: $\lambda=2$ (a), $\lambda=5$ (b), and $\lambda=10$ (c). Symbols refer to DSMC results (circles) and MD simulations (triangles) for $\phi=0.01$ while the lines correspond to the Enskog theoretical results. Panel (B): Plot of the temperature ratio $T_1/T_2$ versus the (common) coefficient of restitution $\al$ for for $\sigma_1=\sigma_2$ and $m_1=m_2$. We assume here that $\lambda=5$ and so, $\Delta_{22}=5\Delta_{11}$ and $\Delta_{12}=3\Delta_{11}$. Three different values of the solid volume fraction are considered: $\phi=0.01$ (solid line and circles), $\phi=0.1$ (triangles), and $\phi=0.2$ (squares). Symbols refer to MD simulations and the line to the Enskog theoretical result. Reprinted figure with permission from  R.\ Brito, R.\ Soto, and V.\ Garz\'{o}, Phys.~Rev.~E \textbf{2020}, \textit{102}, 052904  \cite{BSG20}. Copyright (2020) by the American Physical Society.
}
\label{fig7}
\end{figure}

We consider first the usual case of binary mixtures where their constituents differ only by their diameters and masses but the energy injection parameters are the same for all types of collisions (i.e., $\Delta_{11}=\Delta_{22}=\Delta_{12}$).
To assess the departure from energy equipartition, panels (a) and (b) of Figure \ref{fig6} show the temperature ratio $T_1/T_2$ as a function of the mass ratio. In panel (a), we consider different values of the (common) coefficient of restitution $\al$. As occurs in the conventional IHS model \cite{GD99b,MG02,BT02,DHGD02,Barrat2002}, $T_1/T_2$ increases with increasing mass ratio $m_1/m_2$ and hence, the temperature of the heavier particles is larger than that of the lighter ones. In any case, the departure from energy equipartition is less significant in the Delta-collisional model than in the IHS model. While both simulation methods agree again with great accuracy, they deviate from the theoretical results as the mass ratio grows. \vicente{As is well known, in the case of elastic collisions \cite{FK72}, the use of the simplest leading-order truncation to evaluate the transport coefficients is accurate to approximately 5\%. However, there are exceptions, such as extreme mass ratios (e.g., electron-proton systems). For inelastic collisions, the discrepancy between kinetic theory and simulations for disparate-mass binary mixtures may also originate from the use of the Maxwellian approximation \eqref{7.11} to $\varphi_i$ (leading-order truncation) to determine the temperature ratio. Apart from this source of discrepancy, one could argue that molecular chaos is more likely broken in highly asymmetric mixtures.}

To gauge the impact of density on $T_1/T_2$, we plot it versus $m_1/m_2$ for three different values of the solid volume fraction $\phi$ in the panel (b) of Figure \ref{fig6}. Since $\sigma_1=\sigma_2$ and $x_1=\frac{1}{2}$, Equation \eqref{7.15} yields $\chi_{11}=\chi_{22}=\chi_{12}$ and hence, they factor in Equation \eqref{7.13}. Consequently, the Enskog kinetic theory does not predict any dependence of $T^*$ and $T_1/T_2$ on the density $\phi$. However, beyond the Enskog equation, density corrections to $T_1/T_2$ can exist if there are position correlations not accounted for in the approximation \eqref{7.15} for $\chi_{ij}$. The comparison with MD simulations carried in the panel (b) tests this prediction. We observe from the panel (b) of Figure \ref{fig6} that the dependence of $T_1/T_2$ on $\phi$ is very weak (mostly appears at high mass ratio), validating the results derived from the Enskog equation.

Now, we consider the case in which the only difference between the two species is the energy injection at collisions. Namely, the two species are mechanically equivalent ($\sigma_1 = \sigma_2$ and $m_1 = m_2$), but $\Delta_{11}\neq \Delta_{22}\neq \Delta_{12}$. Specifically, we assume that $\Delta_{11}<\Delta_{22}$ and $\Delta_{12}=(\Delta_{11}+\Delta_{22})/2$. Since  $\Delta_{11}<\Delta_{22}$  then $\Delta_{12}>\Delta_{11}$. This means that the particles of species 1 (2) have a higher (smaller) temperature than if they were alone because the energy injected in the 1-2 collisions is higher (smaller) than in the 1-1 (2-2) collisions. Figure \ref{fig7} illustrates the $\al$-dependence of the temperature ratio for different systems and densities. In general, we observe that the Enskog theoretical predictions agree with the DSMC and MD simulations at low density. Panel (A) highlights a significant departure from energy equipartition, as $T_1/T_2$ differs greatly from 1. Clearly, the energy injection for species 1 is smaller than for species 2, as evidenced by the fact that the temperature ratio $T_1/T_2<1$.
The effect of density on $T_1/T_2$ is illustrated in the panel (B). We observe that the influence of $\phi$ on $T_1/T_2$ is more pronounced in this case  than in the scenario depicted in Panel (b)
of Figure \ref{fig6}.

\vicente{In summary, the comparison carried out in this section between kinetic theory and computer simulations shows that the failure of the Enskog kinetic theory at high densities can be expected based on previous results obtained for ordinary (elastic) mixtures \cite{FK72}. At high densities, effects such as multiparticle collisions are  not accounted for in the Enskog collision operator. These effects are expected to be more pronounced in granular fluids than in the conventional fluids since the colliding pairs tend to be more focused. Consequently, the range of densities for which the Enskog theory is expected to provide accurate results diminishes with increasing collisional dissipation.}

\section{Navier--Stokes transport coefficients for binary granular mixtures. Low-density regime}
\label{sec8}

\subsection{Kinetic and balance equations}

As in the monocomponent gas case, once the homogeneous-time dependent state for multicomponent systems is characterized, the next step is to use this state as the reference state in the Chapman--Enskog perturbation solution of the Enskog kinetic equation \eqref{6.1}. However, studying the transport properties of multicomponent granular mixtures is much more complicated than studying those of a single granular gas. This is because the number of transport coefficients in a mixture is larger than in a monocomponent gas, and these coefficients depend on parameters such as diameters, masses, concentration, and coefficients of restitution. Due to these difficulties, our analysis in this section is restricted to the \textit{low-density} regime of a granular binary mixture ($s=2$). In this regime, $\chi_{ij}=1$ and the Enskog collision operators $J_{\text{E},ij}[f_i,f_j]$ reduce to the Boltzmann operators
\beqa
\label{8.1}
& & J_{ij}[\mathbf{v}_1|f_i,f_j]\equiv \sigma_{ij}^{d-1}\int d{\bf v}_{2}\int d \widehat{\boldsymbol{\sigma}}
\Theta (-\widehat{{\boldsymbol {\sigma }}}\cdot {\bf g}_{12}-2\Delta_{ij})
(-\widehat{\boldsymbol {\sigma }}\cdot {\bf g}_{12}-2\Delta_{ij})
 \nonumber\\
& & \times \al_{ij}^{-2} f_{i}(\mathbf{r},{v}_1'',t)f_{j}(\mathbf{r}, \mathbf{v}_2'',t) 
-\sigma_{ij}^{d-1}\int\ d{\bf v}_{2}\int d\widehat{\boldsymbol{\sigma}}
\Theta (\widehat{{\boldsymbol {\sigma}}}\cdot {\bf g}_{12})
(\widehat{\boldsymbol {\sigma}}\cdot {\bf g}_{12})f_{i}(\mathbf{r}, \mathbf{v}_1,t)f_{j}(\mathbf{r},\mathbf{v}_2,t).
\nonumber\\
\eeqa
Moreover, in the low-density regime, the collisional contributions to the transport coefficients are much more smaller than their kinetic forms and hence, they can be neglected.

The determination of the Chapman--Enskog solution to first order in spatial gradients in the $\Delta$-model follows similar mathematical steps as those made in the conventional IHS model \cite{GD02,GM07,GDH07,G19}. One subtle point in implementing the Chapman--Enskog method to the $\Delta$-model for mixtures is that there are nonzero first-order contributions to the partial temperatures and the energy rate. Most of the technical details involved in this derivation can be found in Ref.\ \cite{GBS21}. 

As discussed in Section \ref{sec4}, we assume that after a transient regime the granular mixture achieves a hydrodynamic state characterized by the fact that the distributions $f_i(\mathbf{r},\mathbf{v},t)$ depend on space and time through a \textit{functional} dependence on the hydrodynamic fields (\textit{normal} solution). Here, as in the case of the IHS model \cite{GD02}, we take the concentration $x_1$, the pressure $p=n T$, the temperature $T$, and the mean flow velocity $\mathbf{U}$ as the hydrodynamic fields of the binary mixture. For small spatial gradients, $f_i$ can be written as a series expansion in powers of the nonuniformity parameter $\epsilon$,       
\begin{equation}
f_{i}=f_{i}^{(0)}+\epsilon \,f_{i}^{(1)}+\epsilon^2 \,f_{i}^{(2)}+\cdots \;,
\label{8.2}
\end{equation}
where the formal parameter $\epsilon$ is taken to be equal to 1 at the end of the calculations. As expected, the zeroth-order distribution $f_{i}^{(0)}$ is nothing more than the local version of the homogeneous-time dependent distribution \eqref{7.6} studied in Section \ref{sec7}.  

The balance hydrodynamic equations to first order are 
\beq
\label{8.3}
D_t^{(1)}x_1=0, \quad D_t^{(1)}U_\lambda=-\rho^{-1}\nabla_\lambda p+g_\lambda,
\eeq
\beq
\label{8.4}
D_t^{(1)}p=-\frac{d+2}{d}p \nabla\cdot \mathbf{U}-p\zeta^{(1)}, \quad
D_t^{(1)}T=-\frac{2}{d}T \nabla\cdot \mathbf{U}-T\zeta^{(1)},
\eeq
where $D_t^{(1)}=\partial_t^{(1)}+\mathbf{U}\cdot \nabla$ and $\zeta^{(1)}=\zeta_U \nabla \cdot \mathbf{U}$ is the first-order contribution to the energy rate. The kinetic equation verifying the first-order distribution $f_i^{(1)}(\mathbf{r},\mathbf{v},t)$ can be obtained by employing the balance equations \eqref{8.3}--\eqref{8.4}. The solution to this kinetic equation is given by \cite{GBS21}
\beqa
\label{8.5}
f_{i}^{(1)}(\mathbf{V})&=&{\boldsymbol {\mathcal {A}}}_{i}(\mathbf{V})\cdot \nabla x_{1}+{\boldsymbol {\mathcal {B}}}_{i}(\mathbf{V})\cdot
\nabla p+{\boldsymbol {\mathcal {C}}}_i(\mathbf{V})\cdot \nabla T\nonumber\\
& &+
\mathcal{D}_{i,\lambda\beta}(\mathbf{V})\frac{1}{2}\left(\nabla_{\lambda}U_{\beta}+\nabla_{\beta}U_{\lambda}-\frac{2}{d}\delta _{\lambda\beta}\nabla \cdot
\mathbf{U} \right)
+\mathcal{E}_i(\mathbf{V}) \nabla\cdot \mathbf{U}\;.
\eeqa

In Equation \eqref{8.5}, the quantities ${\boldsymbol {\mathcal {A}}}_{i}(\mathbf{V})$, ${\boldsymbol {\mathcal {B}}}_{i}(\mathbf{V})$, ${\boldsymbol {\mathcal {C}}}_{i}(\mathbf{V})$, $\mathcal{D}_{i,\beta\lambda}(\mathbf{V})$, and $\mathcal{E}_i(\mathbf{V})$ obey the following linear set of coupled integral equations:
\begin{equation}
\left[-\zeta ^{(0)}\left( T\partial _{T}+p\partial _{p}\right) +{\cal L}_{i}
\right] {\boldsymbol {\mathcal {A}}}_{i}+{\cal M}_{i}{\boldsymbol {\mathcal {A}}}_{j}={\bf A}_{i}+\left(
\frac{\partial \zeta ^{(0)}}{\partial x_{1}}\right) _{p,T}\left( p{\boldsymbol {\mathcal {B}}}_{i}+T{\boldsymbol {\mathcal {C}}}_{i}\right),
\label{8.6}
\end{equation}
\begin{equation}
\left[-\zeta ^{(0)}\left( T\partial _{T}+p\partial _{p}\right)
+{\cal L}
_{i}-2\zeta ^{(0)}\right] {\boldsymbol {\mathcal {B}}}_{i}+{\cal M}_{i}{\boldsymbol {\mathcal {B}}}_{j}={\bf B}_{i}+
\frac{T\zeta ^{(0)}}{p}{\boldsymbol {\mathcal {C}}}_{i},  \label{8.7}
\end{equation}
\beqa
& &\left[-\zeta ^{(0)}\left( T\partial _{T}+p\partial _{p}\right) +{\cal L}
_{i}-\frac{1}{2}\zeta ^{(0)}\left(1-\Delta^*\frac{\partial \ln \zeta_0^*}{\partial \Delta^*}\right)\right] {\boldsymbol {\mathcal {C}}}_{i}+{\cal M}_{i}{\boldsymbol {\mathcal {C}}}_{j}={\bf C}_{i}
\nonumber\\
& & 
-\frac{p\zeta ^{(0)}}{2T}\left(1+\Delta^*\frac{\partial \ln \zeta_0^*}{\partial \Delta^*}\right)
{\boldsymbol {\mathcal {B}}}_{i},  \label{8.8}
\eeqa
\begin{equation}
\left[-\zeta ^{(0)}\left( T\partial_{T}+p\partial_{p}\right) +{\cal
L}_{i}\right] \mathcal{D}_{i,\beta \lambda}+{\cal M}_{i}\mathcal{D}_{j,\beta\lambda}=D_{i,\beta\lambda},  \label{8.9}
\end{equation}
\begin{equation}
\left[-\zeta ^{(0)}\left( T\partial_{T}+p\partial_{p}\right) +{\cal
L}_{i}\right] \mathcal {E}_{i}+{\cal M}_{i}\mathcal {E}_{j}=E_{i},  \label{8.10}
\end{equation}
where $\zeta_0^*=\zeta^{(0)}/\nu$ and we have introduced the linearized Boltzmann operators
\begin{equation}
{\cal L}_{i}f_{i}^{(1)}=-\left(
J_{ii}[f_{i}^{(0)},f_{i}^{(1)}]+J_{ii}[f_{i}^{(1)},f_{i}^{(0)}]+
J_{ij}[f_{i}^{(1)},f_{j}^{(0)}]\right),
\label{8.11}
\end{equation}
\begin{equation}
{\cal M}_{i}f_{j}^{(1)}=-J_{ij}[f_{i}^{(0)},f_{j}^{(1)}].  \label{8.12}
\end{equation}
\vicente{In Equations \eqref{8.6}--\eqref{8.12}, the index $j$ refers to other species in the binary mixture, that is, $j\neq i$.}
Note that in Equation \eqref{8.8}  we have introduced the shorthand notation
\beq
\label{8.12correct}
\Delta^*\frac{\partial}{\partial \Delta^*}\equiv \left(\Delta_{11}^*\frac{\partial}{\partial \Delta_{11}^*}+\Delta_{22}^*\frac{\partial}{\partial \Delta_{22}^*}+\Delta_{12}^*\frac{\partial}{\partial \Delta_{12}^*}\right).
\eeq
In the particular case $\Delta_{11}^*=\Delta_{22}^*=\Delta_{12}^*=\Delta^*$, only one of the three terms of the identity \eqref{8.12correct} must be considered.

The coefficients of the field gradients on the right side of Equations \eqref{8.6}--\eqref{8.10} are functions of the peculiar velocity and the hydrodynamic fields. They are given by
\begin{equation}
{\bf A}_{i}({\bf V})=-\left(\frac{\partial}{\partial x_{1}}
f_{i}^{(0)}\right)_{p,T}{\bf V}, \quad  {\bf B}_{i}({\bf V})=-\frac{\partial f_{i}^{(0)}}{\partial p}{\bf V}-\rho^{-1}
\frac{\partial f_{i}^{(0)}}{\partial {\bf V}}, \label{8.13}
\end{equation}
\begin{equation}
{\bf C}_{i}({\bf V})=-\frac{\partial f_{i}^{(0)}}{\partial T}{\bf V}, \quad D_{i,\lambda\beta}({\bf V})=V_\lambda \frac{\partial f_i^{(0)}}{\partial V_\beta},
\label{8.14}
\end{equation}
\beq
\label{8.15}
E_i(\mathbf{V})=-\frac{1}{d}\Delta^*\frac{\partial f_i^{(0)}}{\partial \Delta^*}-\frac{1}{2}\zeta_U \left[\frac{\partial}
{\partial \mathbf{V}}\cdot \left(\mathbf{V}f_i^{(0)}\right)+\Delta^*\frac{\partial f_i^{(0)}}{\partial \Delta^*}\right].
\eeq

As in the case of monocomponent gases, the Navier--Stokes transport coefficients of the granular mixture can be expressed in terms of the solutions to the set of coupled linear integral equations \eqref{8.6}--\eqref{8.10}. However, as usual, to obtain explicit forms for these transport coefficients one has to resort to the leading terms in a Sonine polynomial expansion of the unknowns ${\boldsymbol {\mathcal {A}}}_{i}$, ${\boldsymbol {\mathcal {B}}}_{i}$, ${\boldsymbol {\mathcal {C}}}_{i}$, $\mathcal{D}_{i,\lambda\beta}$, and $\mathcal{E}_{i}$. Given that this task is relatively long and tedious, for the sake of illustration, we offer here the determination of the mass flux transport coefficients with some detail.  

\subsection{Diffusion transport coefficients}

To first-order, the mass flux $\mathbf{j}_1^{(1)}$ is 
\beq
\label{8.16}
{\bf j}_{1}^{(1)}=-\frac{m_{1}m_{2}n}{\rho} D\nabla x_{1}-\frac{\rho}{p}D_{p}\nabla p-\frac{\rho}{T}D_T\nabla T,
\eeq
where $D$ is the diffusion coefficient, $D_p$ is the pressure diffusion coefficient, and $D_T$ is the thermal diffusion coefficient. According to the definition \eqref{6.14} of the mass flux, the diffusion transport coefficients are identified as  
\begin{equation}
D=-\frac{1}{d}\frac{\rho}{m_{2}n}\int d{\bf v}\,{\bf V}\cdot {\boldsymbol {\mathcal {A}}}_{1}, \quad 
D_{p}=-\frac{1}{d}\frac{m_{1}p}{\rho}\int d{\bf v}\,{\bf V}\cdot {\boldsymbol {\mathcal {B}}}_{1},
\quad D_T=-\frac{1}{d}\frac{m_{1}T}{\rho}\int d{\bf v}\,{\bf V}\cdot {\boldsymbol {\mathcal {C}}}_{1}.
\label{8.17}
\end{equation}

\vicente{From the comparison of Equations \eqref{8.5} and \eqref{8.6}--\eqref{8.8}, it is expected that the vectorial quantities ${\boldsymbol {\mathcal {A}}}_{i}$ , ${\boldsymbol {\mathcal {B}}}_{i}$, and ${\boldsymbol {\mathcal {C}}}_{i}$ are proportional to $\mathbf{A}_i$, $\mathbf{B}_i$, and $\mathbf{C}_i$, respectively. Thus, they are directed along $\mathbf{V}$ [see Equations \eqref{8.13} and \eqref{8.14}]. As a consequence, 
} to get the diffusion transport coefficients we consider the following lowest order Sonine polynomial approximations for ${\boldsymbol {\mathcal {A}}}_{i}$ , ${\boldsymbol {\mathcal {B}}}_{i}$, and ${\boldsymbol {\mathcal {C}}}_{i}$:
\beq
\label{8.20}
\left(
\begin{array}{c}
{\boldsymbol {\mathcal {A}}}_{i}\\
{\boldsymbol {\mathcal {B}}}_{i}\\
{\boldsymbol {\mathcal {C}}}_{i}
\end{array}
\right)\longrightarrow f_{i,\text{M}}\mathbf{V}\left(
\begin{array}{c}
a_{i}\\
b_{i}\\
c_{i}
\end{array}
\right),
\eeq
where
\beq
\label{8.21}
f_{i,\text{M}}({\bf V})=n_i\left(\frac{m_i}{2\pi T_i^{(0)}}\right)^{d/2}\exp\left(-
\frac{m_i V^2}{2T_i^{(0)}}\right)
\eeq
is the Maxwellian distribution characterized by the zeroth-order partial temperature $T_i^{(0)}$. The coefficients $a_i$, $b_i$, and $c_i$ are related in this approximation to the transport coefficients $D$, $D_p$, and $D_T$ through Equations \eqref{8.17} as
\begin{equation}
\label{8.22}
a_1=-\frac{n_2T_2^{(0)}}{n_1T_1^{(0)}}a_2=-\frac{m_1m_2n}{\rho n_1T_1^{(0)}}D,
\end{equation}
\begin{equation}
\label{8.23}
b_1=-\frac{n_2T_2^{(0)}}{n_1T_1^{(0)}}b_2=-\frac{\rho}{p n_1T_1^{(0)}}D_p,
\end{equation}
\begin{equation}
\label{8.24}
c_1=-\frac{n_2T_2^{(0)}}{n_1T_1^{(0)}}c_2=-\frac{\rho}{T n_1T_1^{(0)}}D_T.
\end{equation}
In Equations \eqref{8.22}--\eqref{8.24}, we have taken into account the constraint $n_1T_1^{(0)}+n_2T_2^{(0)}=nT=p$.

The coefficients $D$, $D_p$, and $D_T$ can be determined by substitution of Equation \eqref{8.20} into the integral equations \eqref{8.6}--\eqref{8.8}. Next, one multiplies both sides of these equations by $m_i \mathbf{V}$ and integrates over $\mathbf{v}$. After some algebra, one gets
\beqa
\label{8.25}
\left[-\frac{1}{2}\zeta^{(0)}\left(1-\Delta^*\frac{\partial \ln D^*}{\partial \Delta^*}\right)+\nu_D\right]D&=&\frac{\rho}{m_1m_2 n}\Bigg[\left(\frac{\partial}{\partial x_1}n_1 T_1^{(0)}\right)_{p,T}\nonumber\\
& & 
+\rho
\left(\frac{\partial \zeta ^{(0)}}{\partial x_{1}}\right)_{p,T}\left(D_p+D_T\right)\Bigg],
\eeqa
\beq
\label{8.26}
\left[\frac{1}{2}\zeta^{(0)}\left(1+\Delta^*\frac{\partial \ln D_p^*}{\partial \Delta^*}\right)-2\zeta^{(0)}+\nu_D\right]D_p=\frac{n_1T_1^{(0)}}{\rho }\left(1-\frac{m_1 nT}{\rho T_1^{(0)}}\right)+\zeta^{(0)} D_T,
\eeq
\beq
\label{8.27}
\Bigg[\frac{1}{2}\zeta^{(0)}\Delta^*\left(\frac{\partial \ln D_T^*}{\partial \Delta^*}+\frac{\partial \ln \zeta_0^*}{\partial \Delta^*}\right)+\nu_D\Bigg]D_T=-\frac{n_1T}{2\rho}\Delta^*\frac{\partial \gamma_1}{\partial \Delta^*}-\frac{\zeta^{(0)}}{2}\left(1+\Delta^*\frac{\partial \ln \zeta_0^*}{\partial \Delta^*}\right)D_p.
\eeq
In Equations \eqref{8.25}--\eqref{8.27}, $\zeta_0^*=\zeta^{(0)}/\nu$, the collision frequency $\nu_D$ is defined as
\beq
\label{8.28}
\nu_D=-\frac{1}{dn_1T_1^{(0)}}\int d\mathbf{v}_1 m_1 \mathbf{V}_1 \cdot \left(J_{12}[f_{1,M}\mathbf{V}_1,f_{2}^{(0)}]-\frac{n_1 T_1^{(0)}}{n_2T_2^{(0)}}J_{12}[f_1^{(0)},f_{2,M}\mathbf{V}_2]\right),
\eeq
and the derivatives with respect to $x_1$ at constant pressure and temperature are given by
\beq
\label{8.29}
\left(\frac{\partial}{\partial x_1}n_1 T_1^{(0)}\right)_{p,T}=p\Big(\gamma_1+x_1\frac{\partial \gamma_1}{\partial x_1}\Big),
\eeq
\beq
\label{8.30}
\left(\frac{\partial \zeta^{(0)}}{\partial x_1}\right)_{p,T}=\nu\Bigg[\left(\frac{\partial \zeta_0^*}{\partial x_1}\right)_{\gamma_1}+
\frac{\partial \zeta_0^*}{\partial \gamma_1}\frac{\partial \gamma_1}{\partial x_1}\Bigg].
\eeq
Note that $\gamma_2=(1-x_1\gamma_1)/(1-x_1)$ and hence, $\partial_{x_1}\gamma_2$ can be easily expressed in terms of $\partial_{x_1}\gamma_1$. In addition, upon obtaining Equations \eqref{8.25}--\eqref{8.27}, we have introduced the dimensionless transport coefficients
\beq
\label{8.31}
D^*=\frac{m_{1}m_{2}\nu}{\rho T}D,\quad
D_{p}^*=\frac{\rho \nu}{nT}D_{p},\quad D_T^*=\frac{\rho \nu}{nT}D_T,
\eeq
and have used the relations
\beq
\label{8.32}
\left(T\frac{\partial}{\partial T}+p\frac{\partial}{\partial p}\right)D=\frac{D}{2}\left(1-\Delta^*\frac{\partial \ln D^*}{\partial \Delta^*}\right),
\eeq
\beq
\label{8.33}
\left(T\frac{\partial}{\partial T}+p\frac{\partial}{\partial p}\right)\frac{\rho}{p}D_p=-\frac{\rho}{2p}D_p\left(1+\Delta^*\frac{\partial \ln D_p^*}{\partial \Delta^*}\right),
\eeq
\beq
\label{8.34}
\left(T\frac{\partial}{\partial T}+p\frac{\partial}{\partial p}\right)\frac{\rho}{T}D_T=-\frac{\rho}{2T}D_T\left(1+\Delta^*\frac{\partial \ln D_T^*}{\partial \Delta^*}\right).
\eeq

\subsection{Steady state conditions}

As in the case of the monocomponent granular gas, to achieve analytical expressions of the transport coefficients in the Delta-collisional model one has to consider the steady state conditions. They are defined by the constraints $\zeta_1^{(0)}=\zeta_2^{(0)}=\zeta^{(0)}=0$.  Thus, in the steady state, Equations \eqref{8.25}--\eqref{8.27} become simply linear algebraic equations whose solutions for the dimensionless transport coefficients are
\beq
\label{8.35}
D_p^*=\frac{x_1}{\nu_D^*}\Bigg(\gamma_1-\frac{\mu}{x_2+\mu x_1}\Bigg),
\eeq
\beq
\label{8.36}
D_T^*=-\frac{x_1 \Delta^*\left(\frac{\partial \gamma_1}{\partial \Delta^*}\right)+\Delta^*\left(\frac{\partial \zeta_0^*}{\partial \Delta^*}\right) D_p^*}{2\nu_D^*+\Delta^*\left(\frac{\partial \zeta_0^*}{\partial \Delta^*}\right)},
\eeq
\beq
\label{8.37}
D^*=\frac{\gamma_1+x_1\left(\frac{\partial \gamma_1}{\partial x_1}\right)+\left(D_p^*+D_T^*\right)\left(\frac{\partial \zeta_0^*}{\partial x_1}\right)}{\nu_D^*},
\eeq
where $\nu_D^*=\nu_D/\nu$ and $\mu=m_1/m_2$ is the mass ratio. The derivatives $\partial_{x_1}\gamma_1$, $\partial_{x_1}\zeta_0^*$, $\partial_{\Delta^*}\gamma_1$, and $\partial_{\Delta^*}\zeta_0^*$ are determined in the Appendix \ref{appA} in the particular case $\Delta_{11}=\Delta_{22}=\Delta_{12}$. Additionally, the expression of $\nu_D^*$ when $f_i^{(0)}$ is replaced by its Maxwellian form $f_{i,\text{M}}$ is \cite{GBS21}
\beq
\label{8.38}
\nu_D^*=\frac{2\pi^{(d-1)/2}}{d\Gamma\left(\frac{d}{2}\right)}\left(x_1\mu_{12}+x_2\mu_{21}\right)
\left[\left(\frac{\theta_1+\theta_2}{\theta_1\theta_2}\right)^{1/2}(1+\al_{12})+\sqrt{\pi}\Delta_{12}^*\right].
\eeq
The constraints $\mathbf{j}_1^{(1)}=-\mathbf{j}_2^{(1)}$ and $\nabla x_1=-\nabla x_2$ necessarily imply that $D$ must be symmetric while $D_p$ and $D_T$ must be antisymmetric with respect to the change $1\leftrightarrow 2$. This can be easily verified from Equations \eqref{8.35}--\eqref{8.37} by noting that $x_1\gamma_1+x_2\gamma_2=1$ and $x_1 \partial \gamma_1/\partial \Delta^*=-x_2 \partial \gamma_2/\partial \Delta^*$. This shows the self-consistency of the expressions found for the diffusion transport coefficients.

\subsection{Pressure tensor and heat flux}

The first-order contribution to the pressure tensor is 
\beq
\label{8.39}
P_{\lambda\beta}^{(1)}=-\eta\left( \nabla_{\lambda}U_{\beta}+\nabla_{\beta}U_{\lambda}-\frac{2}{d}\delta _{\lambda\beta}\nabla \cdot
\mathbf{U} \right),
\eeq
while the heat flux $\mathbf{q}^{(1)}$ is  
\beq
\label{8.40}
\mathbf{q}^{(1)}=-T^2 D''\nabla x_1-L \nabla p-\kappa \nabla T.
\eeq
In Equations \eqref{8.39} and \eqref{8.40}, $\eta$ is the shear viscosity coefficient, $D''$ is the Dufour coefficient, $L$ is the pressure energy coefficient, and $\kappa$ is thermal conductivity coefficient.     

The shear viscosity $\eta$ is defined as
\beq
\label{8.40.1}
\eta=-\frac{1}{(d-1)(d+2)} \sum_{i=1}^2\; \int d{\bf v}\, V_\lambda  V_\beta \mathcal{D}_{i,\lambda\beta}(\mathbf{V}).
\eeq
The evaluation of $\eta$ follows similar mathematical steps as those made for the diffusion transport coefficients, although the calculations are a bit more complex. The shear viscosity of the mixture is $\eta=\eta_1+\eta_2$, where the partial contributions $\eta_i$ are defined as  
\beq
\label{8.41}
\eta_i=-\frac{1}{(d-1)(d+2)} \int d{\bf v}\, V_\lambda  V_\beta \mathcal{D}_{i,\lambda\beta}(\mathbf{V}).
\eeq
The leading Sonine approximation for the unknown $\mathcal {D}_{i, \lambda \beta}$ is 
\beq
\label{8.42}
\mathcal{D}_{i,\lambda\beta}(\mathbf{V}) \rightarrow -\frac{m_i\eta_i}{n_i T_i^{(0)2}} \left(V_\lambda V_\beta-\frac{1}{d}\delta_{\lambda\beta}V^2\right)f_{i,\text{M}}(\mathbf{V}).
\eeq
The next step is to substitute Equation \eqref{8.42} into the integral equation \eqref{8.9} and then, multiply it by $m_i(V_\lambda V_\beta-\frac{1}{d}\delta_{\lambda\beta}V^2)$, \vicente{sum over the repeated indices $\lambda$ and $\beta$,}  and integrate over velocity.  After some algebra, in the steady state, the dimensionless shear viscosity coefficient $\eta^*=(\nu/p)\eta$ is \cite{GBS21} 
\beq
\label{8.43}
\eta^*=\frac{\left(\tau_{22}^*-\tau_{21}^*\right)x_1\gamma_1+\left(\tau_{11}^*-\tau_{12}^*\right)x_2\gamma_2}
{\tau_{11}^*\tau_{22}^*-\tau_{12}^*\tau_{21}^*},
\eeq
where the expressions of $\tau_{ij}^*$ are given by Equations (C6)--(C10) of Ref.\ \cite{GBS21}.

\begin{figure}[h!]
\centering
\includegraphics[width=0.45\textwidth]{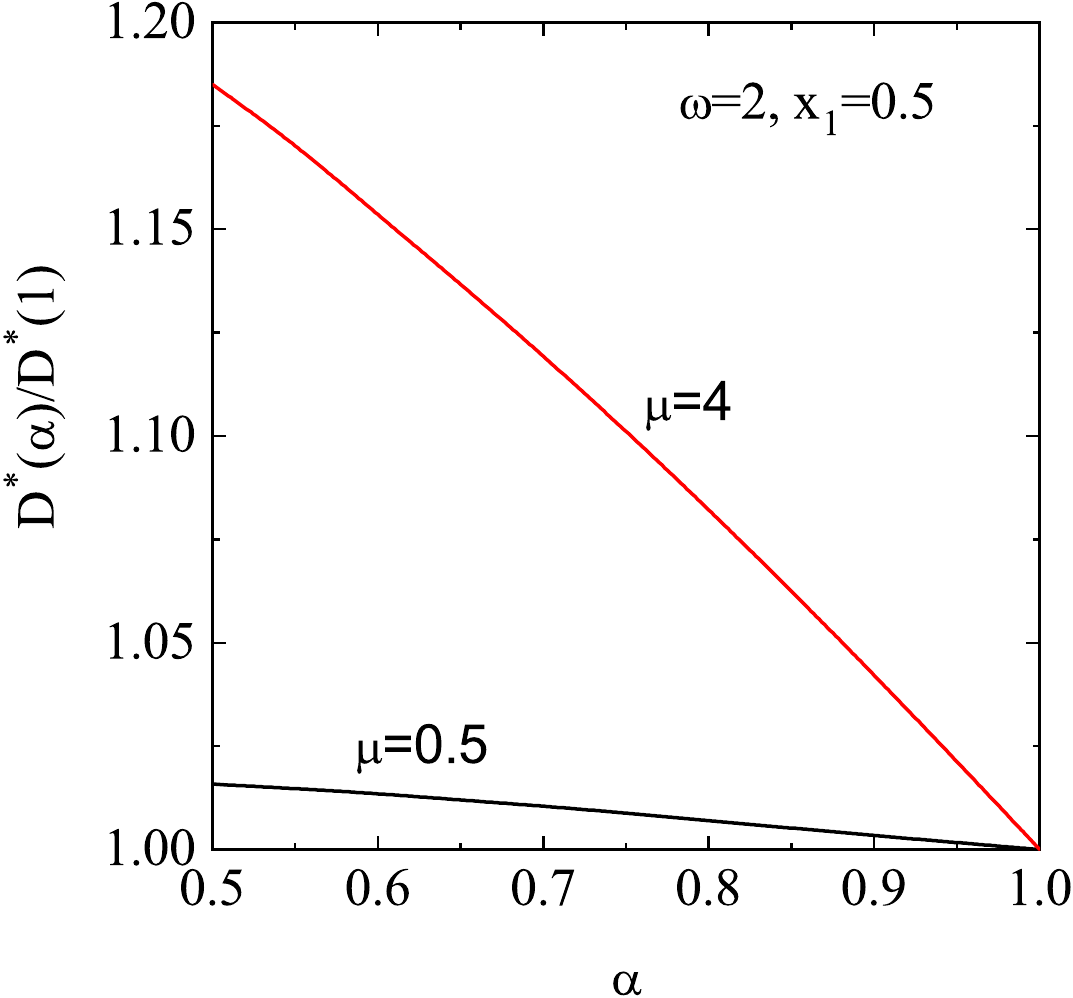}
\includegraphics[width=0.46\textwidth]{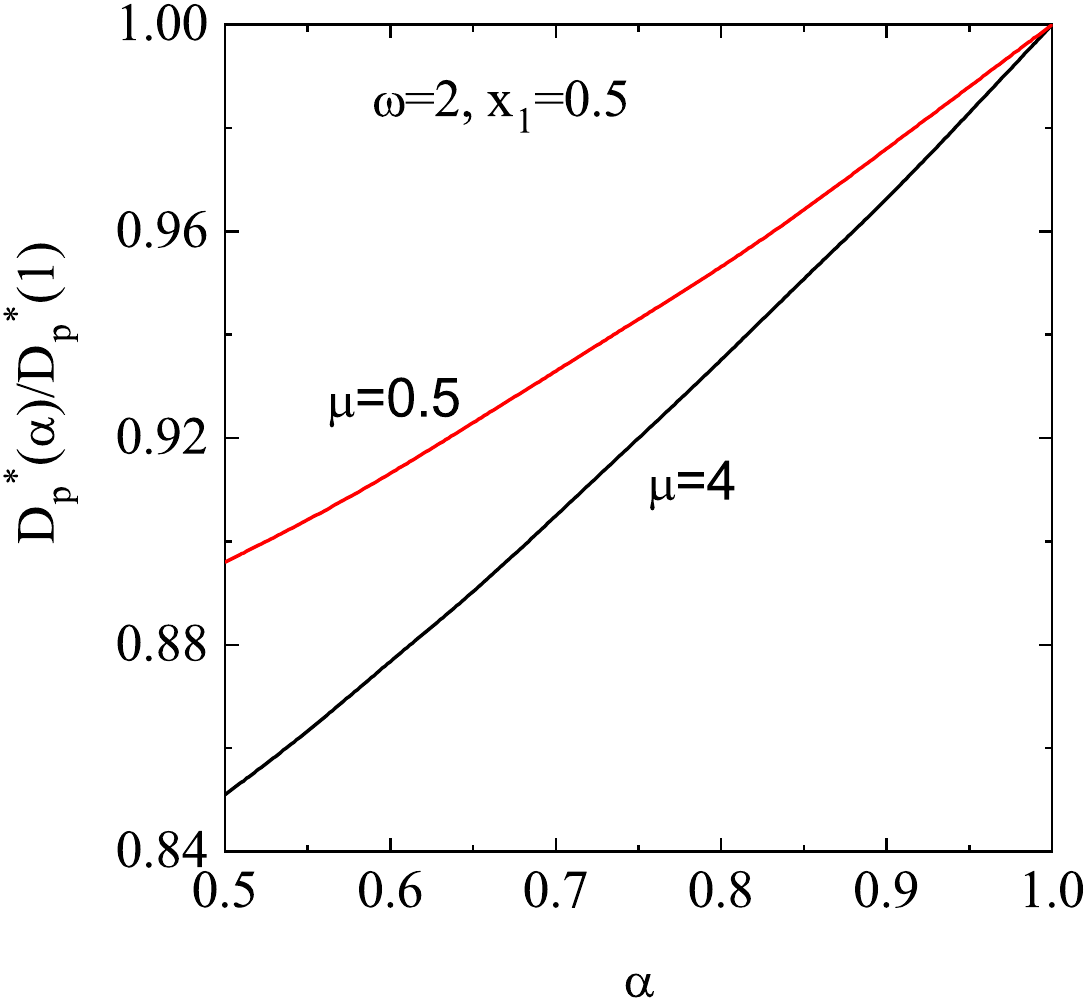}
\includegraphics[width=0.45\textwidth]{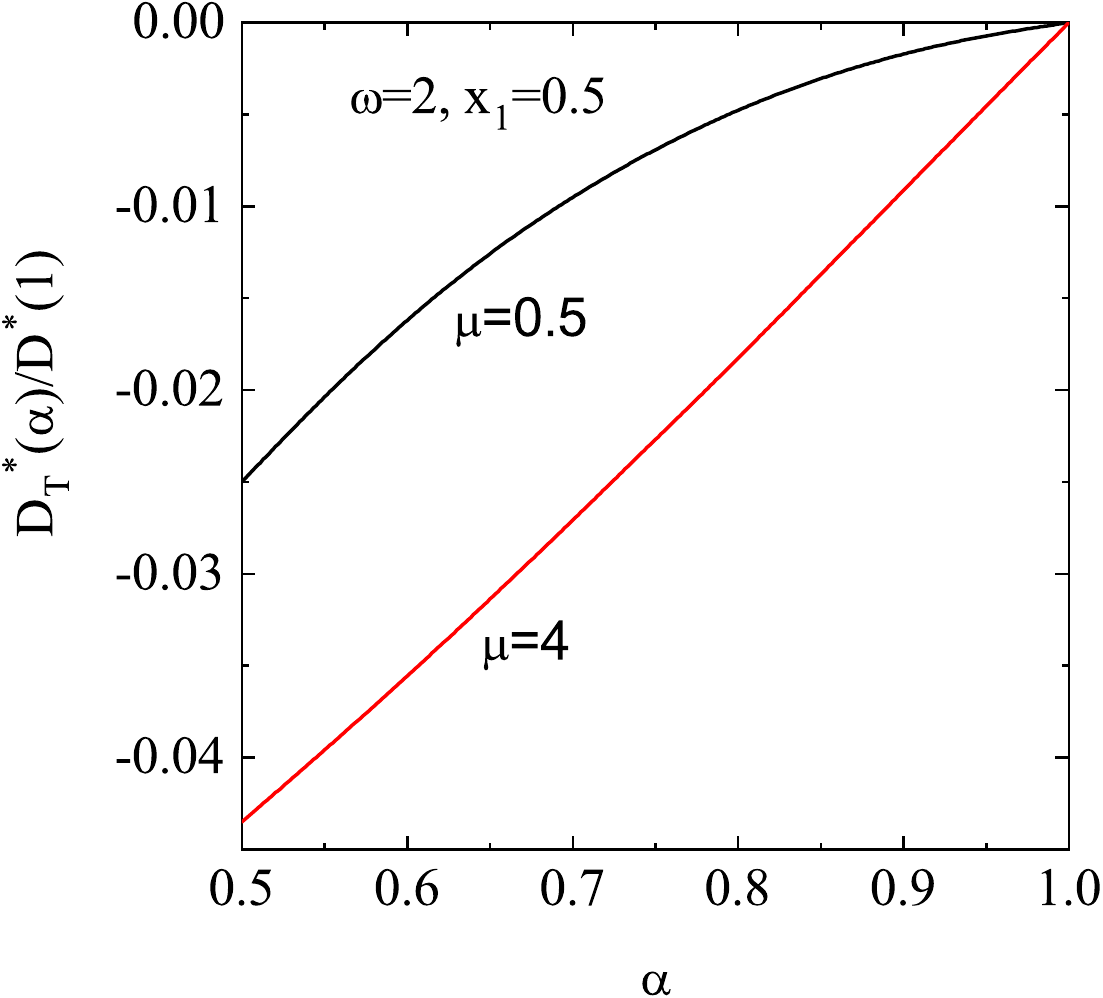}
\caption{Plots of the (scaled) diffusion transport coefficients $D^*(\al)/D^*(1)$, $D_p^*(\al)/D_p^*(1)$, and $D_T^*(\al)/D^*(1)$ versus the (common) coefficient of restitution $\al$ for $d=2$, $\omega=2$, $x_1=\frac{1}{2}$, and two different values of the mass ratio $\mu$: $\mu=0.5$ and $\mu=4$.  
}
\label{fig8}
\end{figure}

The evaluation of the heat flux transport coefficients $D''$, $L$, and $\kappa$ is more involved since it requires going up to the second Sonine approximation. However, it is still  possible to obtain simple expressions for these coefficients when the first Sonine approximations \eqref{8.20} are considered for ${\boldsymbol {\mathcal {A}}}_{i}$, ${\boldsymbol {\mathcal {B}}}_{i}$, and ${\boldsymbol {\mathcal {C}}}_{i}$, respectively. In this approximation, the heat flux transport coefficients are proportional to the diffusion transport coefficients and their forms are
\beq
\label{8.44}
\left\{D'', L, \kappa \right\}=\frac{d+2}{2}\left(\frac{\gamma_1}{m_1}-\frac{\gamma_2}{m_2}\right)
\left\{\frac{n m_1m_2}{\rho T}D, \frac{\rho}{n} D_p, \rho D_T \right\}.
\eeq
According to Equations \eqref{8.44}, for mechanically equivalent components, energy equipartition holds ($\gamma_1=\gamma_2$) \cite{BSG20} and so the first Sonine approximation to the heat transport coefficients vanishes ($D''=L=\kappa=0$). Hence, the forms \eqref{8.44} are not able to reproduce the expression of the heat flux for a single granular gas. Nevertheless, these expressions are consistent in the order of approximation used to obtain the mass flux transport coefficients and hence, they can be employed in several applications for granular confined mixtures.

\subsection{Some illustrative systems}

To illustrate the dependence of the transport coefficients on inelasticity, it is more convenient to plot the transport coefficients in their dimensionless forms. The expressions of the diffusion transport coefficients are given by Equations \eqref{8.35}--\eqref{8.37} while the shear viscosity is given by Equation \eqref{8.43}.\footnote{As noted in Ref.\ \cite{GBS24a}, there is a misprint in the last term of Equation (A8) of Ref.\ \cite{GBS21} since the term $\partial \zeta_1^*/\partial \gamma_1$ must be replaced by the term $\partial \zeta_1^*/\partial \Delta^*$. This modification causes changes in some of the Figures plotted in Ref.\ \cite{GBS21}. The plots presented here correct the Figures 1, 3, and 7 of Ref.\ \cite{GBS21}.} Moreover, for the sake of simplicity, we will assume the case $\Delta_{11}^*=\Delta_{22}^*=\Delta_{12}^*\equiv \Delta^*$ and will take a common coefficient of restitution $\al_{11}=\al_{22}=\al_{12}\equiv \al$ in a two-dimensional mixture ($d=2$). Since in the steady state, $\Delta^*$ is a function of $\al$, $x_1$, and the mechanical parameters of the mixture, then the parameter space is reduced to three quantities: $\left\{\mu\equiv m_1/m_2, \omega\equiv \sigma_1/\sigma_2, x_1\right\}$.    

\begin{figure}[h!]
\centering
\includegraphics[width=0.45\textwidth]{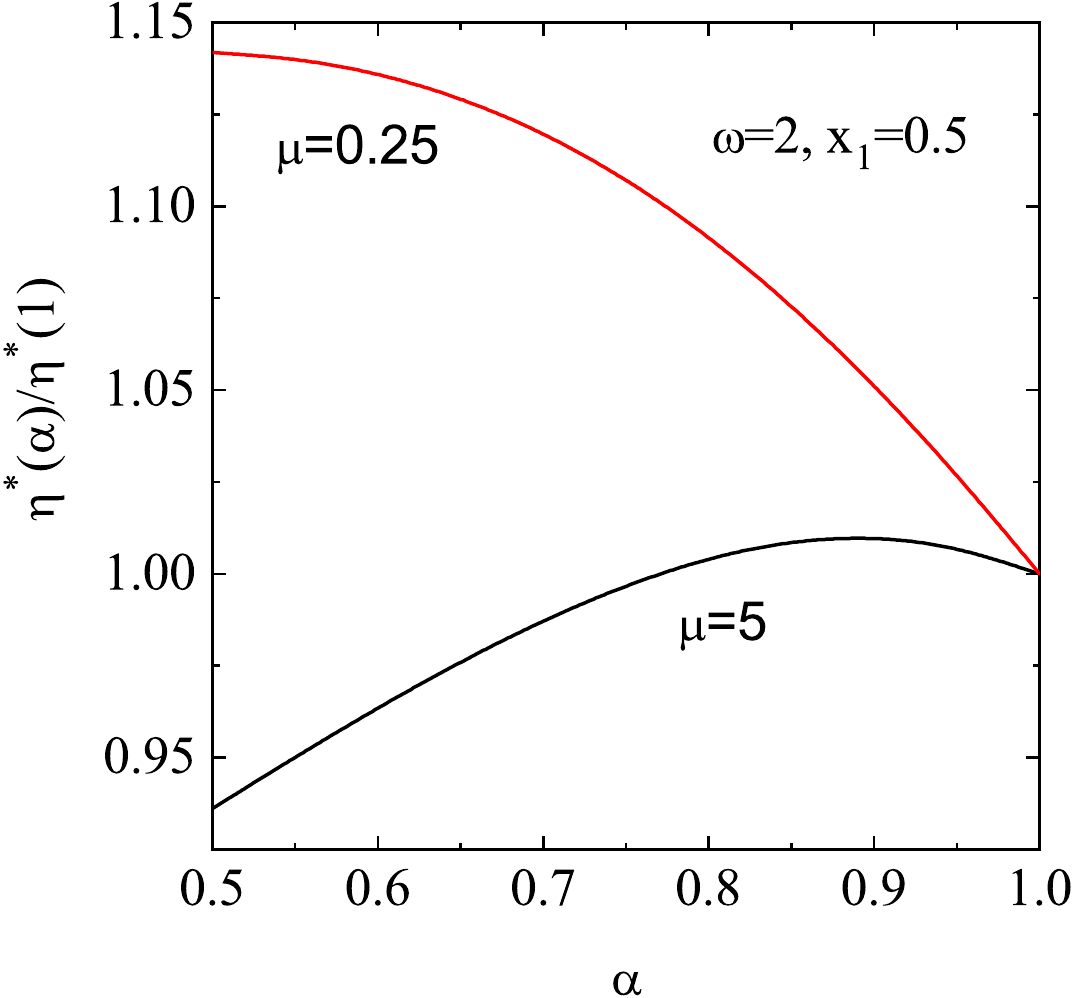}
\caption{Plot of the (scaled) shear viscosity coefficient $\eta^*(\alpha)/\eta^*(1)$ as a function of the (common) coefficient of restitution $\alpha$ for $d=2$, $\omega=2$, $x_1=0.5$, and two different values of the mass ratio $\mu$: $\mu=0.25$ and $\mu=5$.  
}
\label{fig9}
\end{figure}

For elastic collisions, $\Delta^*(1)=0$, $T_1/T_2=1$, $D_T^*=0$, and 
\beq
\label{8.47}
D_p^*(1)=\frac{x_1x_2}{\nu_D^*(1)}\frac{1-\mu}{1+(\mu-1)x_1}, \quad D^*(1)=\frac{1}{\nu_D^*(1)}, \quad \nu_D^*(1)=\sqrt{2\pi}\frac{x_1\mu_{12}+x_2\mu_{21}}{\sqrt{\mu_{12}\mu_{21}}}.
\eeq
Since we want to assess the effect of inelasticity on transport properties, as usual we normalize here the transport coefficients with respect to their values for elastic collisions. Since the thermal diffusion coefficient $D_T^*$ vanishes for elastic collisions in the first Sonine approximations, we have normalized it with respect to $D^*(1)$.     

Figure \ref{fig8} shows the $\al$-dependence of the scaled coefficients 
$D^*(\al)/D^*(1)$, $D_p^*(\al)/D_p^*(1)$, and $D_T^*(\al)/D^*(1)$ 
for several systems. 
In general, the effect of inelasticity on mass transport is significant, as the reduced coefficients clearly deviate from their forms for elastic collisions. However, as in the monocomponent limiting case, these deviations are generally less significant than those in the conventional IHS model (see for instance, Figs.\ 1-3 of Ref.\ \cite{GMD06}). Regarding the dependence on the mass ratio, we observe that, at a given value of $\al$, while the value of $D^*(\al)/D^*(1)$ increases with increasing $\mu$, the opposite occurs for the ratio $D_p^*(\al)/D_p^*(1)$. Figure \ref{fig8} also shows that the thermal diffusion coefficient $D_T^*$ is always negative (at least in the case $\Delta_{11}^*=\Delta_{22}^*=\Delta_{12}^*$, illustrated here), although its 
magnitude is quite small. The fact that $D_T^*$ can be negative is consistent with the results derived in the conventional IHS model. The sign of the coefficient $D_T^*$ is important in problems such as granular segregation by thermal diffusion \cite{JY02,BRM05,BRM06,SGNT06,G06,G08a,BEGS08,G09,G11}.

The ratio $\eta^*(\al)/\eta^*(1)$ is plotted in Fig.\ \ref{fig9} as a function of the coefficient of restitution $\al$ for two different mixtures. As before, $\eta^*(1)$ refers to the shear viscosity coefficient for elastic collisions. As with the diffusion coefficients, we observe that the effect of inelasticity on the shear viscosity is smaller than in the conventional IHS model (see for instance, Fig.\ 5 of Ref. \cite{GM07}). Additionally, depending on the mass ratio, the normalized shear viscosity decreases (increases) when decreasing $\al$ when the mass ratio is larger (smaller) than 1.

In summary, in the context of the Delta-collisional model, the mass and momentum transport coefficients of a (confined) granular mixture  differ from those of a molecular mixture, although these deviations are generally less significant than in the IHS model. In most cases, differences with the molecular results increase with increasing dissipation, and depending on the transport coefficient, mass ratio has a significant influence.

\section{Some applications of the kinetic theory for granular 
mixtures in the $\Delta$-model}
\label{sec9}
    
As in the monocomponent case, different problems can be analyzed by using the explicit forms of the Navier-Stokes transport coefficients. In this Section, we will study two interesting problems among them. First, we will quantify the violation of Onsager's well-known reciprocity relations \cite{GM84} for confined granular mixtures. Since time reversal invariance does not hold for granular gases, we expect that Onsager's relations will not be verified for finite inelasticity. However, it remains challenging to quantify the deviations from these relations as dissipation increases. Second, we will analyze the stability of the HSS in a granular mixture. As with monocomponent granular gases, our results show that the HSS is linearly stable with respect to perturbations with wavelengths long enough. As expected, however, the forms of the $d-1$ transversal shear modes and the four longitudinal modes (i.e., those associated with concentration, hydrostatic pressure, temperature, and the longitudinal component of flow velocity) differ from those obtained in the HCS for a granular mixture \cite{GMD06}.

\subsection{Violation of  Onsager's reciprocity relations}

In the usual language of the linear irreversible thermodynamics
for ordinary fluids, to first order in spatial gradients, the constitutive equations for
the mass and heat fluxes of a binary mixture are written as \cite{GM84}
\begin{equation}
\mathbf{j}_{i}=-\sum_{j}L_{ij}\left( \frac{\nabla \mu _{j}}{T}\right)
_{T}-L_{iq}\frac{\nabla T}{T^{2}}-C_{p}\nabla p,  \label{9.1}
\end{equation}
\beq
\mathbf{J}_{q}=\mathbf{q}-\frac{d+2}{2}T\frac{m_2-m_1}{m_1m_2}\mathbf{j}_1=-L_{qq}\nabla T-\sum_{i}L_{qi}\left( \frac{\nabla \mu _{i}}{T}
\right)_{T}-C_{p}^{\prime}\nabla p,  \label{9.2}
\eeq
where $\mu_{i}$ is the chemical potential of the species $i$ \vicente{per unit mass}. In the low-density regime,
\begin{equation}
\left( \frac{\nabla \mu_{i}}{T}\right)_{T}=\frac{1}{m_{i}}\nabla \ln
(x_{i}p).  \label{9.3}
\end{equation}
The reason for introducing the heat flow $\mathbf{J}_{q}$ is because for elastic collisions this flow is conjugate to the temperature gradient in the form of entropy production \cite{GM84}. The difference between $\mathbf{q}$ and $\mathbf{J}_{q}$ is a term associated with diffusion.

\begin{figure}[h!]
\centering
\includegraphics[width=0.45\textwidth]{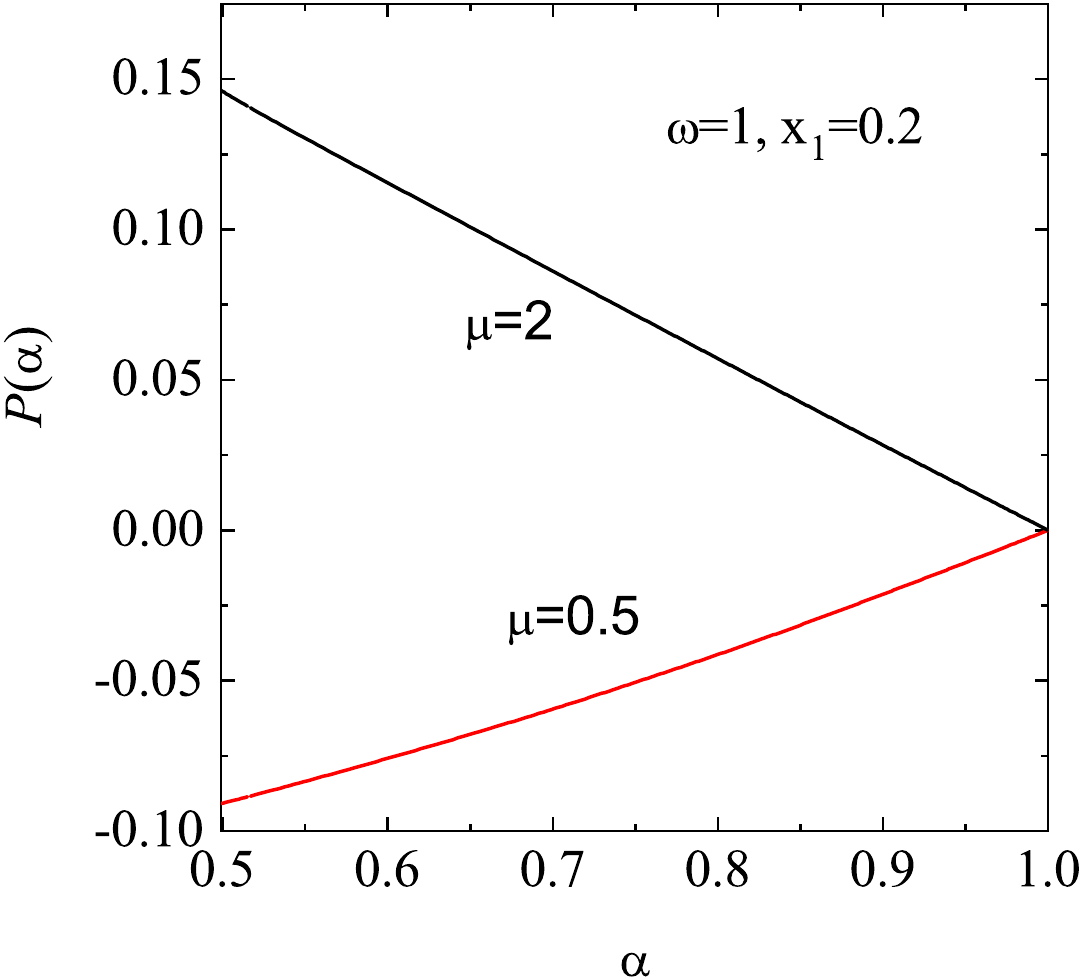}
\includegraphics[width=0.46\textwidth]{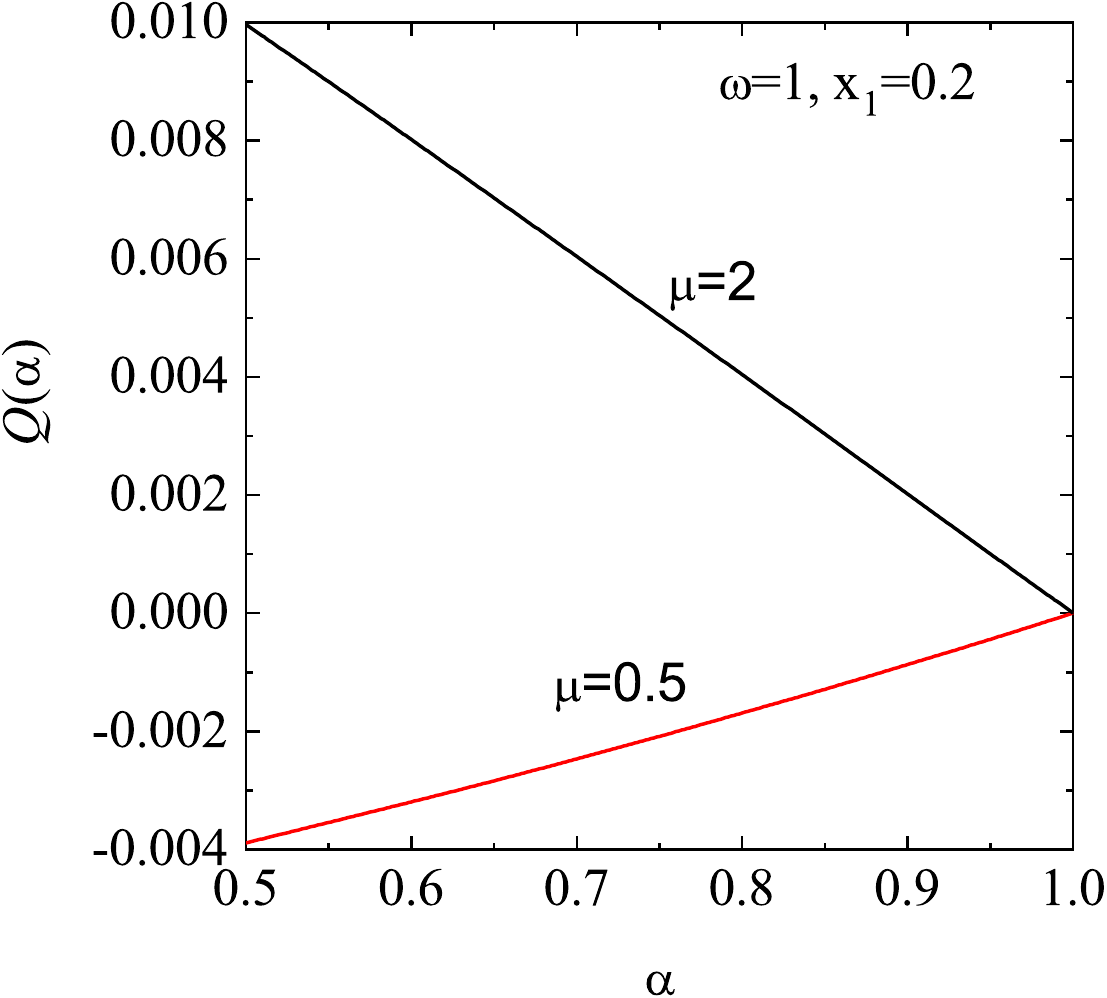}
\includegraphics[width=0.45\textwidth]{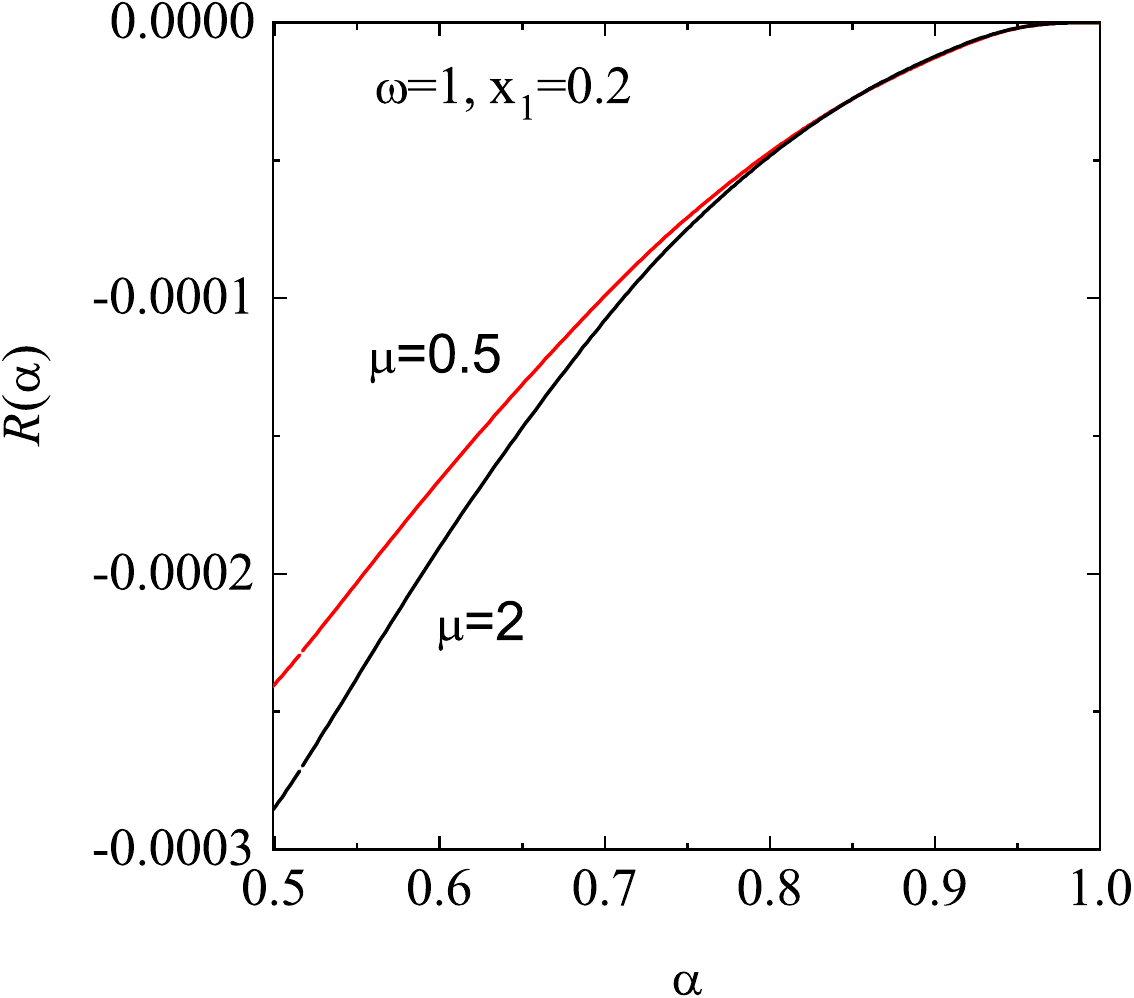}
\caption{Plot of the dimensionless coefficients $P(\al)$, $Q(\al)$, and $R(\al)$ versus the (common) coefficient of restitution $\alpha_{ij}\equiv \alpha$ for $d=2$, $x_1=0.2$, $\omega=1$, and two different values of the mass ratio $\mu$: $\mu=0.5$ and $\mu=0.2$.  
}
\label{fig10}
\end{figure}

The coefficients $L_{ij}$, $L_{iq}$, $C_p$, $L_{qq}$, $L_{qi}$, and $C_p'$ are the so-called Onsager phenomenological
coefficients. For ordinary or molecular fluids ($\al_{ij}=1$), Onsager showed that time
reversal invariance of the underlying microscopic equations of motion leads to the relations:
\begin{equation}
L_{ij}=L_{ji},\quad L_{iq}=L_{qi},\quad C_{p}=C_{p}^{\prime}=0.  \label{9.4}
\end{equation}
The first two symmetries are called reciprocal relations as they relate
transport coefficients for different processes. The coefficients $L_{qi}$ link the mass flux to the thermal gradient while the coefficients $L_{iq}$ link the heat flux to the gradient of the chemical potentials. The last two identities ($C_p=0$ and $C_p'=0$) are statements
that the pressure gradient does not appear in any of the fluxes even though it is admitted by symmetry. In particular, the condition $C_p'=0$ is important for monocomponent elastic gases since it yields Fourier's law for heat flux ($\mathbf{q}^{(1)}\propto \nabla T$) and hence, there is no  contribution to the heat flux proportional to the density gradient $\nabla n$. On the contrary, for the IHS model, $C_p'\neq0$ and there is an additional contribution to the heat flux proportional to $\nabla n$, as discussed in Sections \ref{sec5} and \ref{sec.kappaandmu}~\cite{SMR99,BDKS98,G19,CW07}.

To quantify the possible violation of Onsager's relations, we have to express first the Onsager coefficients ($L_{ij}$, $L_{1q}$, $C_p$, $L_{qq}$, $L_{q1}$, and $C_p'$) in terms of both the diffusion ($D$, $D_p$, $D_T$) and heat flux ($D''$, $L$, $\kappa$) transport coefficients. To do it, since $\nabla x_{1}=-\nabla x_{2}$, then Equation~\eqref{9.3} yields the relation
\begin{equation}
\frac{(\nabla \mu_{1})_{T}-(\nabla \mu_{2})_{T}}{T}=\frac{n\rho }{\rho
_{1}\rho _{2}}\left[ \nabla x_{1}+\frac{n_{1}n_{2}}{n\rho }
(m_{2}-m_{1})\nabla \ln p\right] .  \label{9.5}
\end{equation}
The relationships between the Onsager coefficients $L_{ij}$ and the diffusion and heat flux transport coefficients are
\begin{equation}
L_{11}=-L_{12}=-L_{21}=\frac{m_{1}m_{2}\rho _{1}\rho _{2}}{\rho ^{2}}D,\quad
L_{1q}=\rho T D_T,  \label{9.6}
\end{equation}
\begin{equation}
\label{9.7}
L_{q1}=-L_{q2}=\frac{T^{2}\rho _{1}\rho_{2}}{n\rho}D^{\prime \prime}-
\frac{d+2}{2}\frac{T\rho_{1}\rho_{2}}{\rho^{2}}(m_{2}-m_{1})D,
\eeq
\beq
\label{9.8}
L_{qq}=\kappa -\frac{d+2}{2}\rho \frac{m_{2}-m_{1}}{m_{1}m_{2}}D_T, \quad C_{p}\equiv \frac{\rho}{p} D_{p}-\frac{\rho_{1}\rho_{2}}{p\rho^{2}}
(m_{2}-m_{1})D,
\end{equation}
\begin{equation}
C_{p}^{\prime}\equiv L-\frac{d+2}{2}T\frac{m_{2}-m_{1}}{m_{1}m_{2}}
C_{p}- \frac{n_{1}n_{2}}{np\rho }T^{2}(m_2-m_1)D^{\prime \prime}.
\label{9.9}
\end{equation}
As said before, since $D$ is symmetric under the change $1\leftrightarrow 2$, then Onsager's relation $L_{12}=L_{21}$ trivially holds. To analyze the other relations, we 
define the dimensionless functions
\beq
\label{9.10}
P(\al_{ij})\equiv \left(\frac{\gamma_1}{\mu_{12}}-\frac{\gamma_2}{\mu_{21}}-\frac{m_2^2-m_1^2}{m_1m_2}\right)D^*
 -\frac{2}{d+2}\frac{(m_1+m_2)n\rho}{\rho_1\rho_2}D_T^*,
\eeq
and 
\beq
\label{9.11}
Q(\al_{ij})\equiv D_p^*-\frac{\rho_1\rho_2}{n \rho}\frac{m_2-m_1}{m_1m_2}D^*.
\eeq
Note that for obtaining Equations \eqref{9.10} and \eqref{9.11} use of Equation \eqref{8.44}
has been made. The function $P$ vanishes when $L_{1q}=L_{q1}$ while the function $Q$ vanishes when $C_p=0$. Finally, when $C_p=0$ and $C_p'=0$, the function
\beq
\label{9.13}
R(\al_{ij})\equiv \Big[\mu_{21}(1-\gamma_1)-\mu_{12}(1-\gamma_2)\Big]Q(\al_{ij})
\eeq
equals zero.

For elastic collisions, $D_T^*=0$ and $D_p^*$ and $D^*$ are given by Equations \eqref{8.47}. Using these expressions yields the expected results: $P(1)=Q(1)=R(1)=0$. Also, for mechanically equivalent particles with arbitrary $\alpha$, $D_p^*=D_T^*=0$ and therefore, $P$, $Q$, and $R$ vanish as well. However, beyond these two limiting cases, Onsager's relations do not apply as
expected. The origin of this failure is essentially due to (i) the absence of energy equipartition in granular mixtures ($T_1^{(0)}\neq T_2^{(0)}$) and (ii) the   
homogeneous time-dependent reference state which gives contributions to diffusion coefficients coming from the derivatives $(\partial_{x_1}\gamma_1)_s$, $(\partial_{\Delta^*}\gamma_1)_s$, $(\partial_{x_1}\zeta_i^*)_s$, and $(\partial_{\Delta^*}\zeta_i^*)_s$. \vicente{Because energy non-equipartition is involved in determining the above derivatives, it is difficult to disentangle the impact of each effect on violating Onsager's relations.}

To illustrate the deviations from Onsager's relations, Figure \ref{fig10} shows   
the dependence of the quantities $P$, $Q$, and $R$ on the (common) coefficient of restitution $\alpha_{ij}\equiv \alpha$ for $d=2$, $x_1=0.2$, $\omega=1$, and two different values of the mass ratio. 
Violation of Onsager's relations is especially evident in the case of the function $P$. The departure from zero is very small for $Q$ and $R$, even in cases of strong dissipation. This implies that $C_p$ and $C_p'$ are small. The main conclusion of this subsection is that deviations from Onsager's relations in the Delta-collisional model are much smaller than in the IHS model for the same systems (see Figs.\ 7, 8, and 9 of Ref.~\cite{GMD06}).

\subsection{Stability analysis of the HSS in granular mixtures}
\label{sec.stab.HSS.mixtures}

As a second application, we perform a linear stability analysis of the Navier-Stokes hydrodynamic equations in the case of a binary granular mixture. Thus, the question is whether and to what extent the conclusions about the stability of the HSS in the monocomponent case can be changed for mixtures.

As in Section \ref{sec5}, we assume that the deviations $\delta y_{\beta}({\bf r},t)=y_{\beta}({\bf r},t)-y_{\text{H} \beta}$ are small, where
$\delta y_{\beta}({\bf r},t)$ denotes the deviation of $\{x_1, \mathbf{U},p, T,\}$ from their values in the HSS. For the sake of convenience \vicente{and to compare with the stability analysis performed from the IHS model \cite{GMD06}}, we introduce the same time and space dimensionless variables as in Ref.\ \cite{GMD06}: $\tau=\nu_\text{H}t$ and ${\boldsymbol{\ell}}=n_\text{H}\sigma_{12}^{d-1}\mathbf{r}$. Here,  $\nu_\text{H}=n_\text{H} \sigma_{12}^{d-1}v_{\text{th,H}}(T_\text{H})$, where  
$v_{\text{th,H}}=\sqrt{2T_\text{H}/\overline{m}}$
is the thermal velocity of the binary mixture. \vicente{Note that, for mechanically equivalent particles, these dimensionless variables differ from those used in the stability analysis of subsection 4.5.}

As usual, the linearized hydrodynamic equations for the perturbations 
$$
\left\{\delta x_1(\mathbf{r}; t), \delta \mathbf{U}(\mathbf{r}; t), \delta p(\mathbf{r}; t), \delta T(\mathbf{r}; t)\right\}
$$ 
are written in the Fourier space. A set of Fourier transformed dimensionless variables are introduced as
\beq
\label{9.14}
\rho_{\mathbf{k}}(\tau)=\frac{\delta x_{1\mathbf{k}}(\tau)}{x_{1\text{H}}}, \quad \mathbf{w}_{\mathbf{k}}(\tau)=\frac{\delta \mathbf{U}_{\mathbf{k}}(\tau)}{v_{\text{th,H}}}, \quad \Pi_{\mathbf{k}}(\tau)=\frac{\delta p_{{\bf k}}(\tau)}{p_\text{H}}, \quad
\theta_{{\bf k}}(\tau)=\frac{\delta T_{\mathbf{k}}(\tau)}{T_{\text{H}}},
\eeq
where $p_\text{H}=n_\text{H}T_\text{H}$. Here, $\delta y_{\mathbf{k}\beta}\equiv \{\delta x_{1\mathbf{k}}(\tau),{\bf w}_{{\bf k}}(\tau), \Pi_\mathbf{k}(\tau), \theta_{{\bf k}}(\tau)\}$ is defined as
\begin{equation}
\label{9.16}
\delta y_{\mathbf{k}\beta}(\tau)=\int d \mathbf{r}'\;
e^{-\imath \mathbf{k}\cdot \mathbf{r}'}\delta y_{\beta}
(\mathbf{r}',\tau).
\end{equation}
\vicente{As in the stability analysis carried out in subsection 4.5 for monocomponent gases}, the subscripts H denotes the HSS.

After writing the corresponding linearized version of the Navier-Stokes hydrodynamic equations in the Fourier space, it is quite apparent that the $d-1$ transverse velocity components ${\bf w}_{{\bf k}\perp}={\bf w}_{{\bf k}}-({\bf w}_{{\bf k}}\cdot
\widehat{{\bf k}})\widehat{{\bf k}}$ (orthogonal to the wave vector ${\bf k}$)
decouple from the other four modes and they verify $d-1$ differential equations given by
\beq
\label{9.16correct}
\frac{\partial \mathbf{w}_{{\bf k}\perp}}{\partial \tau}+\frac{1+\mu}{4(x_1 \mu+x_2)}\eta^* k^2 \mathbf{w}_{{\bf k}\perp}=0.
\eeq
Since $\eta^*$ does not depend on time in the HSS, the solution to Equation \eqref{9.16correct} is
\beq
\label{9.16solution}
\mathbf{w}_{{\bf k}\perp}(\tau)=\mathbf{w}_{{\bf k}\perp}(0)\exp\left(-\frac{1+\mu}{4(x_1 \mu+x_2)}\eta^* k^2\tau\right).
\eeq
Thus, the $d-1$ transversal shear modes ${\bf w}_{{\bf k}\perp}(\tau)$ are linearly stable because the shear viscosity $\eta^*$ is always positive [see Equation \eqref{8.43}].

The set of differential equations for the four longitudinal modes $\rho_\mathbf{k}$, $\theta_\mathbf{k}$, $\Pi_\mathbf{k}$, and $\mathbf{w}_{{\bf k}||}$ (parallel to $\mathbf{k}$) is more intricate. In matrix form, this set can be written as \cite{GBS24}
\begin{equation}
\frac{\partial \delta z_{\mathbf{k}\alpha }(\tau )}{\partial \tau }=\left(
M_{\alpha \beta }^{(0)}+\imath kM_{\alpha \beta }^{(1)}+k^{2}M_{\alpha \beta
}^{(2)}\right) \delta z_{\mathbf{k}\beta }(\tau ),  \label{9.17}
\end{equation}
where now $\delta z_{\mathbf{k}\alpha }(\tau )$ denotes the four variables $
\left( \rho _{\mathbf{k}},\theta _{\mathbf{k}},\Pi _{\mathbf{k}},w_{\mathbf{
k }||}\right) $. The matrices in Equation \eqref{9.17} are
\begin{equation}
M^{(0)}=\left(
\begin{array}{cccc}
0 & 0 & 0 & 0 \\
-A & B& 0&0\\
-A & B& 0&0\\
0 & 0 & 0 &0
\end{array}
\right) ,  \label{9.18}
\end{equation}
\begin{equation}
M^{(1)}=\left(
\begin{array}{cccc}
0 & 0 & 0 & 0 \\
0 & 0 & 0 & -\left(\frac{2}{d}+\zeta_U\right) \\
0 & 0 & 0 & -\left(\frac{d+2}{d}+\zeta_U\right) \\
0 & 0 & -\frac{1}{4}\frac{1+\mu}{x_{1}\mu +x_{2}} & 0
\end{array}
\right) ,  \label{9.19}
\end{equation}
\begin{equation}
M^{(2)}=\left(
\begin{array}{cccc}
-\frac{1}{4}\frac{\mu x_{1}+x_{2}}{\mu_{12}}D^{\ast} & -\frac{1}{4 x_1}\frac{\mu x_{1}+x_{2}}{\mu_{12}}D_T^* & -\frac{1}{4x_1}\frac{\mu x_{1}+x_{2}}{\mu_{12}}D_{p}^{\ast} & 0 \\
x_{1}\left(\frac{1-\mu}{4\mu_{12}}D^*- \frac{1}{d}D^{\prime \prime }{}^{\ast} \right)
& \frac{1-\mu}{4\mu_{12}}D_T^{\ast}-\frac{1}{d}
\kappa^{\ast } & \frac{1-\mu}{4 \mu_{12}}D_p^*-
\frac{1}{d}L^{\ast}& 0 \\
-\frac{1}{d}x_{1}D^{\prime \prime }{}^{\ast} & -\frac{1}{d} \kappa^{\ast}
& -\frac{1}{d}L^{\ast } & 0 \\
0 & 0 & 0 & -\frac{d-1}{2d}\frac{1+\mu}{\mu x_1+x_2}\eta^{\ast}
\end{array}
\right) .  \label{9.20}
\end{equation}
In Equation \eqref{9.18}, we have introduced the (dimensionless) quantities
\beq
\label{9.21}
A=x_1\Bigg\{\left(\frac{\partial \zeta_2^*}{\partial x_1}\right)+x_1\gamma_1\left[\left(\frac{\partial \zeta_1^*}{\partial x_1}\right)-\left(\frac{\partial \zeta_2^*}{\partial x_1}\right)\right]\Bigg\},
\eeq
\beq
\label{9.22}
B=\frac{1}{2}\Delta_s^*\Bigg\{\left(\frac{\partial \zeta_2^*}{\partial \Delta^*}\right)+x_1\gamma_1\left[\left(\frac{\partial \zeta_1^*}{\partial \Delta^*}\right)-\left(\frac{\partial \zeta_2^*}{\partial \Delta^*}\right)\right]\Bigg\}.
\eeq
Additionally, the dimensionless heat flux transport coefficients are defined as
\beq
\label{9.23}
D^{\prime \prime }{}^{\ast}= \frac{\overline{m}\nu}{n}D'', \quad L^*=\frac{\overline{m}\nu}{T}L,
\quad \kappa^*=\frac{\overline{m}\nu}{p}\kappa.
\eeq
As in the case of the transverse modes, the subscript H has been
suppressed in Equations \eqref{9.18}--\eqref{9.23} for the sake of brevity. For mechanically equivalent particles, $D_p^*=D_T^*=0$, which implies $L^*=\kappa^*=0$ in the first Sonine approximation. Moreover, in this limiting case,  $A=0$, $B=(\Delta^*/2)(\partial \zeta_0^*/\partial \Delta^*)$, and the results are consistent with those obtained in subsection 4.5 for monocomponent granular gases \cite{GBS21a}.

\begin{figure}[h!]
\centering
\includegraphics[width=0.45\textwidth]{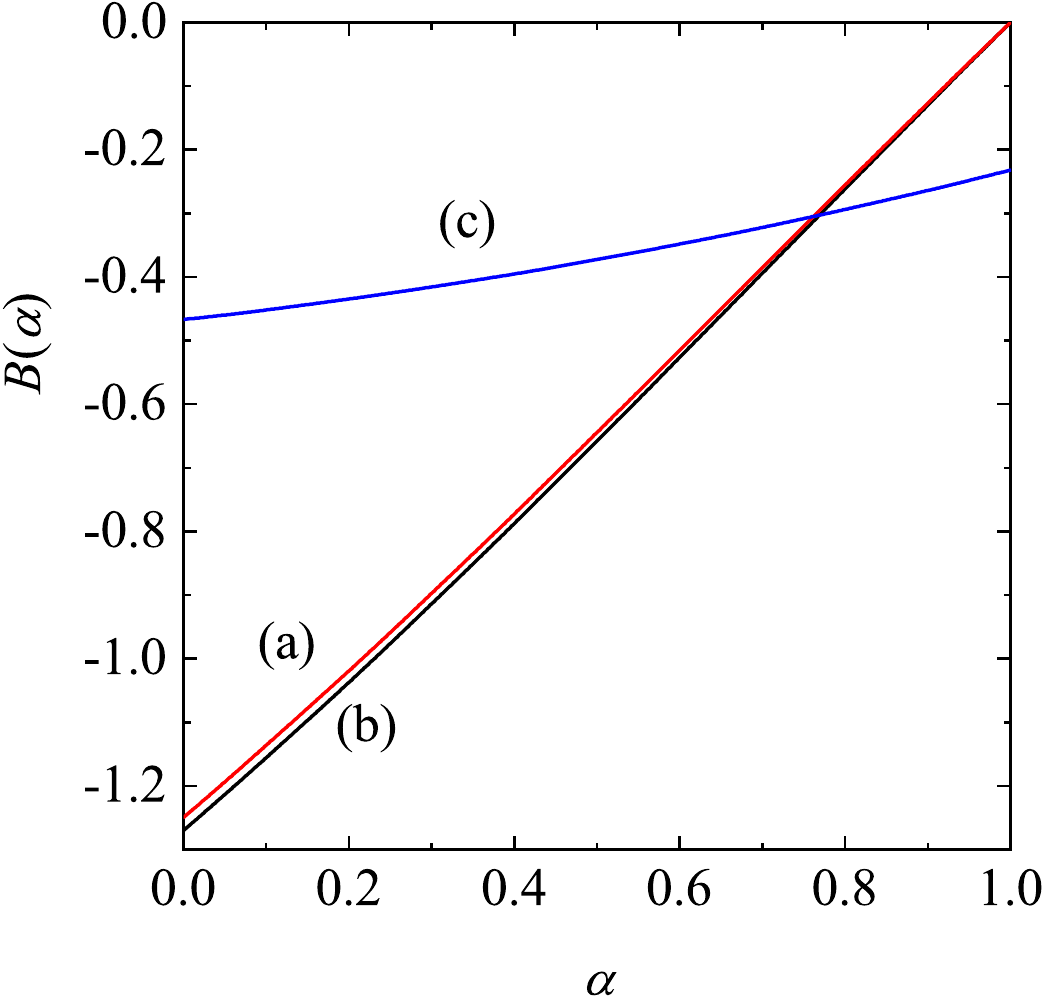} 
\caption{Dependence of the eigenvalue $B$ on  coefficient of restitution $\al$ for a two-dimensional system and three different granular binary mixtures: $x_1=0.5$, $\omega=0.5$, $\mu=0.75$, and $\al_{ij}\equiv \al$ (a); $x_1=0.5$, $\omega=2$, $\mu=2$, and $\al_{ij}\equiv \al$ (b); and $x_1=0.2$, $\omega=1$, $\mu=0.2$, $\al_{22}$ =0.8, $\al_{11}\equiv \al$, and $\al_{12}=(\al_{22}+\al)/2$ (c).
}
\label{fig11}
\end{figure}

The time evolution of the four longitudinal modes has the form $e^{s_n(k) \tau}$ ($n=$1, 2, 3, and 4), where the quantities $s_n(k)$ are the eigenvalues of the matrix $M_{\al \beta}\equiv M_{\al \beta}^{(0)}+\imath k M_{\al \beta}^{(1)}+k^2 M_{\al \beta}^{(2)}$. In other words, the eigenvalues $s_n(k)$ are the solutions of the quartic equation
\beq
\label{9.24}
\det \left(\mathsf{M}-s \boldsymbol{1} \right)=0,
\eeq
where $\boldsymbol{1}$ is the matrix identity. The determination of the dependence of the eigenvalues $s_n(k)$ on the (dimensionless) wave vector $k$ and the parameters of the mixture is a quite intricate problem. Therefore, to gain some insight into the general problem, it is convenient to study first the solution to the quartic equation \eqref{9.24} in the extreme long wavelength limit, $k=0$.

When $k=0$, the square matrix $\mathsf{M}$ reduces to $\mathsf{M}^{(0)}$ whose eigenvalues are   
$s_{||}=\left\{0,0,0,B\right\}$. According to Equation \eqref{9.22}, the dependence of $B$ on the parameter space of the system is in general complex. A simple situation corresponds to the case of mechanically equivalent particles where $\zeta_1^*=\zeta_2^*=\zeta_0^*$ and so, in the steady state 
\beq
\label{9.25}
\frac{\partial \zeta_0^*}{\partial \Delta^*}=-\frac{1}{2}\left(\sqrt{2\pi}\al+4\Delta^*\right).
\eeq
Thus, in this limiting case
\beq
\label{9.26}
B=-\frac{1}{4}\Delta^*\left(\sqrt{2\pi}\al+4\Delta^*\right)<0,
\eeq
and the longitudinal modes are linearly stable in agreement with previous results \cite{GBS21a}. In the case of mechanically different particles, a detailed study of the dependence of the quantity $B$ on the parameters of the mixture for the choice $\Delta_{ij}=\Delta$ shows that $B$ is always negative. As a consequence, all the longitudinal modes in a granular mixture are stable when $k=0$ in the $\Delta$-model. As an illustration, in Figure \ref{fig11} we plot the dependence of $B$ on the coefficient of restitution $\al$ for three different mixtures. We clearly observe that the eigenvalue $B$ is always negative; its magnitude increases with decreasing $\al$.

\begin{figure}[h!]
\centering
\includegraphics[width=0.45\textwidth]{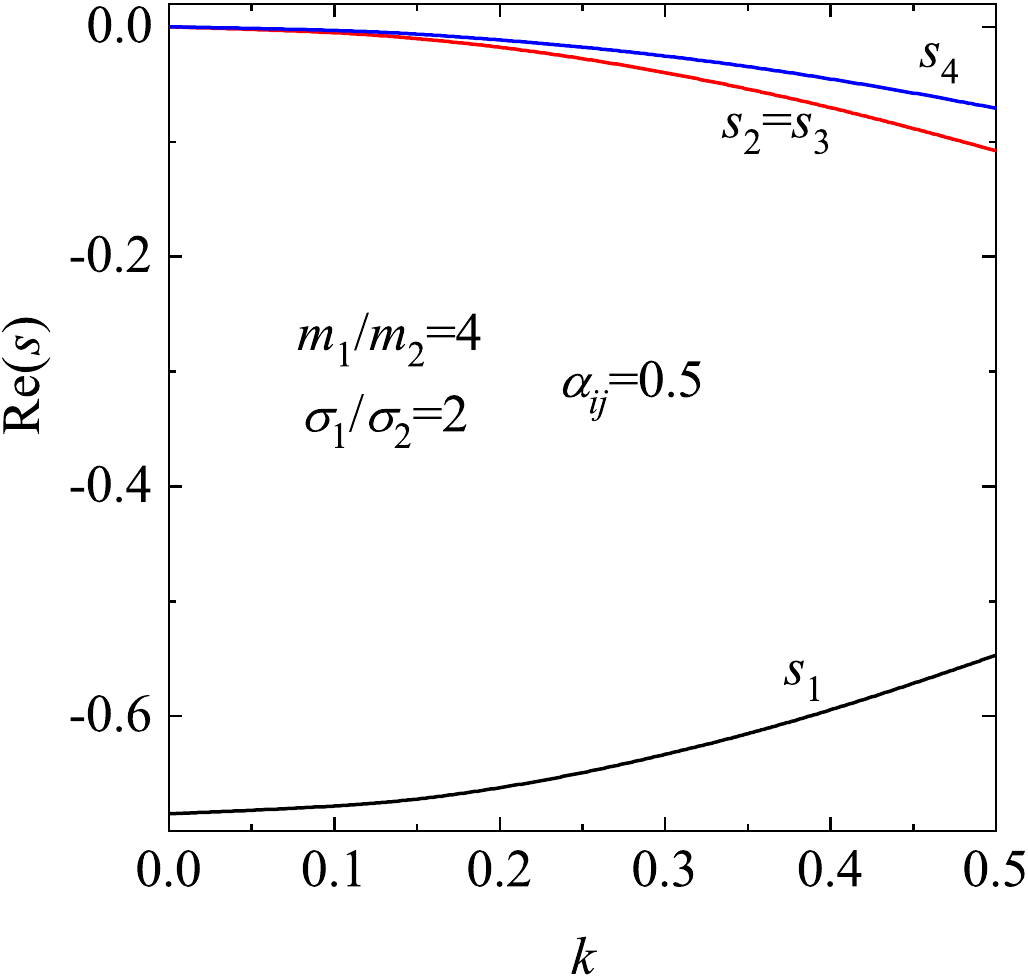} 
\caption{Real parts of the longitudinal eigenvalues $s_i$ as functions of the wave number $k$ for a two-dimensional granular binary mixture with $x_1=0.5$, $\omega=2$, $\mu=4$ and the (common) coefficient of restitution $\al_{ij}\equiv 0.5$.
}
\label{fig12}
\end{figure}

Beyond the limit $k\to 0$, the eigenvalues of $\mathsf{M}$ must be numerically determined.
In the case $\Delta_{ij}\equiv \Delta$, a careful study of the dependence of the eigenvalues of the matrix $\mathsf{M}$ on the parameters of the mixture shows that the real part of \emph{all} the eigenvalues is \emph{negative} and hence, the HSS is linearly stable in the complete range of values of the wave number $k$ studied \cite{GBS24}. This result contrasts with that obtained in the conventional IHS model \cite{GMD06}.  As an illustration, Fig.\ \ref{fig12} shows the real parts of the eigenvalues $s_i$ ($i=1, 2, 3, 4$) as functions of the (dimensionless) wave number $k$ for a two-dimensional granular binary mixture with a concentration $x_1=0.5$, a diameter ratio $\sigma_1/\sigma_2=2$, a mass ratio $m_1/m_2=4$, and a (common) coefficient of restitution $\al_{ij}=0.5$.  As in the case of monocomponent gases, we observe that two of the modes (denoted as $s_2$ and $s_3$) are a complex conjugate pair of propagating modes [$\text{Re}(s_2)=\text{Re}(s_3)$] while the other two modes ($s_1$ and $s_4$) are real for all values of the wave number. We see that the magnitude of $s_4$ is very small while the mode $s_1$ ($s_2$) increases (decreases) with increasing the wave number. Although not shown in the Figure, we also observe that the influence of the disparity in masses and/or diameters does not play a relevant role on the dependence of the eigenvalues $s_i$ on the wave number $k$ since the results for monocomponent granular gases are quite close to the ones found for bidisperse systems.

\section{Concluding remarks}
\label{sec10}

In this review, we have presented a comprehensive account of the kinetic theory of the $\Delta$-collisional model for driven granular gases. The model was originally introduced as a minimal way to incorporate collisional energy injection into the dynamics of granular particles, particularly in situations that mimic vertically vibrated and confined systems. The main feature of the $\Delta$-model  is that energy injection is implemented directly at the level of binary collisions, in a way that exactly conserves momentum, in contrast with other thermostats, like Gaussian or random thermostats. This mechanism leads to a nonequilibrium steady state that is found to be homogeneous, where collisional dissipation (determined by the normal restitution coefficient, $\alpha$) and energy injection (modeled by the parameter $\Delta$) balance each other. This homogeneous steady state provides a reference state for hydrodynamic expansions and stability analyses. This is in contrast with freely cooling granular gases, whose intrinsic time dependence, due to dissipation of energy, complicates the derivation of transport properties. The structure of the injection term via binary collisions, retains the usual form of the collision operator for hard spheres in the kinetic equations, supplemented with terms dependent on the $\Delta$ parameter. This formal similarity has enabled the application of standard tools of kinetic theory, including Sonine polynomial expansions and Chapman–Enskog methods, to derive Navier–Stokes transport coefficients and characterize the hydrodynamic fields.  As usual in granular gases, there is a new contribution to the heat current. It is proportional to the gradient of density and the transport coefficient is the {\em diffusive heat conductivity} coefficient, $\mu$. This term is in addition to the usual term, 
proportional to the gradient of temperature with  thermal conductivity coefficient $\kappa$. Analytical expressions for all transport coefficients are obtained by considering the leading terms of the Sonine polynomial expansions. 
Of course, the resulting transport coefficients depend in a nontrivial way on both the coefficient of restitution and the driving parameter. There may be alternative routes to calculate transport coefficients, using Green-Kubo formulas adapted to dissipative dynamics~\cite{GreenKubo05,Ernst_2006} or Helfand formulas~\cite{DBB08}, and it would be interesting to compare those results with the ones derived here. 

Linear stability analysis of the hydrodynamic modes carried out in Sections~\ref{sec.stab.HSS} and \ref{sec.stab.HSS.mixtures} 
demonstrates that the collisional driving modifies the long-wavelength behavior of the system, leading to the conclusion that the homogeneous state is stable. The spectrum of hydrodynamic modes reflects the stabilizing role of collisional injection. This is in contrast with freely cooling systems, where clustering and shear instabilities emerge in the system. The $\Delta$ parameter stabilizes the system and prevents clusters or other instabilities to appear. 
Physically, the origin of the stability at large wavelengths is that in the $\Delta$-model the stationary temperature turns out to be density independent. This results in that the pressure, which is the product of the temperature and a function of density, is a monotonically increasing function of density, hence displaying a positive compressibility. 
%
The linear stability of the HSS 
means that the $\Delta$-model in its present form cannot reproduce clustering instabilities observed experimentally in driven granular monolayers. Understanding whether clustering can emerge through finite-size effects, boundary-induced inhomogeneities, or regimes beyond the Navier–Stokes approximation is an intriguing open question. Extensions incorporating a density-dependent or velocity-dependent $\Delta$  parameter could potentially make the stationary temperature  decrease with increasing density, generating a van der Waals loop, thereby admitting clustering while retaining much of the analytical structure. Some steps in this direction are already given in Ref.\ \cite{RSG18}.

The model allows a generalization to include different types of particles to study the behavior of mixtures, with a richer phenomenology. When two or more species are present, distinguished by any of their dynamical properties (mass, diameter, restitution coefficient or energy injection one), the system exhibits a breakdown of energy equipartition: each species reaches a different granular temperature in the steady state, a typical signature (or consequence) of the nonequilibrium nature of the granular fluids. The Enskog equation is generalized to mixtures and allows to compute partial temperatures, which, in the present review are compared with computer simulations of both Event Driven type (MD simulations) and DSMC method for a two component system.  

In the case of mixtures, there is a new balance equation for the density of each species in addition to momentum and energy (or temperature) balance equations. The new equations (or their transforms into a concentration and pressure equations) introduce three diffusivities: a diffusion coefficient, a pressure diffusion coefficient and a thermal diffusion coefficient, whose explicit definitions are given in this review. The determination of these transport coefficients opens the way to extending the analysis to segregation phenomena under gravity~\cite{GGBS24}. In particular, there we study the Brazil nut effect of motion of large particles in the direction of the external gravity field or against it.  An specific problem addressed in this review is the violation of Onsager's reciprocity relations. In the case of elastic collisions (molecular mixtures), such reciprocity relations are derived under the assumption  of the time reversibility of microscopic dynamics. However, this is not the case in granular systems.
However, and probably due to the fact that there is a homogeneous steady state in the $\Delta$-model, the violation of Onsager's reciprocal relations is much weaker than the case of a pure IHS dynamics, where such stationary state does not exist. 

An important aspect emphasized in some parts of this review is the interplay between analytical theory and numerical simulations. Event Driven MD and DSMC simulations have been used to validate the predictions of kinetic theory. In particular, simulations allow us to identify the limits of common approximations (such as effects of truncation to low-order Sonine expansions or spatial correlations), and to explore regimes beyond strict hydrodynamic conditions or larger densities, where Enskog equation may fail. Generally speaking, numerical simulations agree well with the analytical results derived from the Enskog equation.

Although the results reported here for transport properties in granular mixtures have been restricted to the dilute regime, progress has recently been made regarding moderate densities. Thus, the tracer diffusion coefficients have been explicitly determined \cite{GGBS24} by considering the lowest Sonine polynomial approximation. The extension of these results to arbitrary concentrations has been recently worked out \cite{GMG26}, and the forms of the diffusion coefficients, as well as the shear and bulk viscosities have been obtained. An interesting future project is to determine the heat flux transport coefficients. Knowing the complete set of Navier--Stokes transport coefficients will allow us to analyze the stability of the homogeneous steady state and/or to assess the violation of Onsager's reciprocity relations for dense granular mixtures, among other applications.  

\vicente{For the sake of simplicity, the results provided here for transport in binary mixtures have been restricted to the case $\Delta_{11}=\Delta_{22}=\Delta_{12}\equiv \Delta$.
The extension to the case $\Delta_{11}\neq\Delta_{22}\neq\Delta_{12}$ (namely, when the energy injection depends on the species) is simple but requires additional calculations. We plan to perform this calculation in the near future and assess how the choice of this case affects for instance to the stability of the HSS of a granular binary mixture.}

Comparison with realistic quasi-two-dimensional experiments and full three-dimensional simulations of the confined geometry also requires more systematic attention. In particular, the precise relationship between effective collisional driving, characterized by the $\Delta$ parameter, with realistic boundary forcing in vertically vibrated systems, like amplitude or frequency, deserves further clarification. 
It is worth noting that, compared to the full quasi two-dimensional system, the $\Delta$-model makes some idealized  approximations that might need to be reconsidered if the predictions of the model are to be compared with experiments. First, the motion of the grains is strictly restricted to the horizontal plane and the vertical motion is eliminated in favor of the added velocity at collisions. This implies that particles collide when their horizontal distance equals the diameter and there is no partial overlap due to three dimensionality (see Fig.~\ref{fig.setup}). As a result, the maximum planar density is slightly smaller than for the quasi two-dimensional system. Another effect of eliminating the vertical direction is that the particles cannot lock with the wall, which has been observed experimentally and can lead to absorbing and hyperuniform states~\cite{RPGRSCM11,neel2014dynamics,MPSTSF24,Maire_2025,le2025control}. Similarly, as there is no gradual increase of the energy in the vertical degrees of freedom and the value of $\Delta$ is fixed, some microscopic correlations are lost; for example in the Q2D  geometry, after a grain-grain collision they will have smaller vertical velocities, modifying for a short time their collision frequencies and the available energy to be exchanged in subsequent collisions.

Another promising direction concerns strongly inhomogeneous states, such as shear flows, temperature gradients, or confined geometries where boundary layers cannot be neglected. Although the homogeneous steady state provides a convenient reference, many experimentally relevant situations involve spatial gradients that challenge standard hydrodynamic expansions. Exploring non-Newtonian transport, rheological properties, and nonlinear instabilities within the $\Delta$-framework constitutes a natural continuation of the work reviewed here.

In summary, the $\Delta$-model has developed into a coherent and versatile theoretical framework for driven granular fluids. It is an example of how relatively simple modifications of microscopic rules can generate qualitatively new nonequilibrium behavior while remaining amenable to rigorous kinetic analysis. It provides a consistent kinetic and hydrodynamic description, accommodates mixtures and segregation phenomena, and connects naturally with experimental realizations of confined vibrated systems. The continued interest in the model, including recent theoretical and numerical studies, and extensions to the study of other phenomena, underscores its relevance as a reference system for exploring nonequilibrium statistical mechanics in dissipative  matter.

\authorcontributions{ All authors have read and agreed to the published version of the manuscript.}

\funding{R.B. acknowledges financial support from Grant No.~PID2023-147067NB-I00 funded by  
MCIU/AEI/10.13039/501100011033 and by ERDF/EU. 
R.S. acknowledges financial support from Fondecyt Grant No.~1220536 of ANID, Chile.
V.G. acknowledges financial support from Grant No. PID2024-156352NB-I00 funded by MCIU/AEI/10.13039/501100011033 and by ERDF/EU, and from Grant No.~GR24022 funded by Junta de Extremadura (Spain) and by European Regional Development Fund (ERDF) ``A way of making Europe''.
 
}

\institutionalreview{Not applicable}

\informedconsent{Not applicable}

\dataavailability{The data that support the findings of this study are available from the corresponding author upon reasonable request.} 


\conflictsofinterest{The authors report no conflict of interest.}

\appendixtitles{yes} 
\appendixstart
\appendix
\section[\appendixname~\thesection: Some expressions for transport in confined granular mixtures]{Some expressions for transport in confined granular mixtures}
\label{appA}

In this Appendix we give some expressions needed to evaluate the diffusion transport coefficients as well as to perform the linear stability analysis of the HSS in a granular binary mixture. In particular,  
the first-order contribution $\zeta_U$ to the energy rate can be written as
\beq
\label{a1}
\zeta_U=\sum_{i=1}^2\xi_i^*\varpi_i^*,
\eeq
where
\beqa
\label{a2}
\xi_i^{*}&=&\frac{3\pi^{(d-1)/2}}{d\Gamma\left(\frac{d}{2}\right)}\frac{x_i m_i}{\overline{m}\gamma_i}\sum_{j=1}^2  x_j \left(\frac{\sigma_{ij}}{\sigma_{12}}\right)^{d-1}\mu_{ji}(1-\al_{ij}^2)\left(\theta_i+\theta_j\right)^{1/2}\theta_i^{-3/2}\theta_j^{-1/2}\nonumber\\
& & -\frac{4\pi^{(d-1)/2}}{d\Gamma\left(\frac{d}{2}\right)}x_i \Delta^*\sum_{j=1}^2 x_j \left(\frac{\sigma_{ij}}{\sigma_{12}}\right)^{d-1}
\mu_{ji}\Bigg\{\sqrt{\pi}\al_{ij}+\left(\theta_i+\theta_j\right)^{-1/2}\theta_i^{3/2}\theta_j^{-1/2}\Delta^*
\nonumber\\
& & \times
\Big[d-d\left(\theta_i+\theta_j\right)\theta_i^{-1}+(d+1)\theta_i \theta_j^{-1}\Big]\Bigg\},
\eeqa
\beq
\label{a3}
\varpi_1^*=\frac{1}{d}\frac{\Delta^*\left(\frac{\partial \gamma_1}
{\partial \Delta^*}\right)}{\Lambda_1^*}, \quad \varpi_2^*=-\frac{x_1}{x_2}\varpi_1^*,
\eeq
\beq
\label{a4}
\Lambda_1^*=\omega_{11}^*-\frac{x_1}{x_2} \omega_{12}^*-\Bigg[\frac{1}{2}\Delta^*
\left(\frac{\partial \gamma_1}{\partial \Delta^*}\right)-\gamma_1\Bigg]\Bigg(\xi_1^*-\frac{x_1}{x_2} \xi_2^*\Bigg).
\eeq
In Equation \eqref{a10}, the expressions of $\omega_{11}^*=\omega_{11}/\nu$ and $\omega_{12}^*=\omega_{12}/\nu$ can be easily obtained from Equations (C12) and (C13), respectively, of the Appendix C of Ref.\ \cite{GBS21}.

According to Equations \eqref{8.35}--\eqref{8.37} and Equations \eqref{a3}--\eqref{a4}, it is quite apparent that the diffusion transport coefficients and the first-order contribution to the rate of energy are given in terms of the derivatives $(\partial_{x_1}\gamma_1)$, $(\partial_{\Delta^*}\gamma_1)$, $(\partial_{x_1}\zeta_i^*)$, and $(\partial_{\Delta^*}\zeta_i^*)$. \vicente{We recall that these derivatives are evaluated in the steady state.} The derivative $(\partial \gamma_1/\partial \Delta^*)$ is \cite{GBS21,GBS24a}
\beq
\label{a5}
\left(\frac{\partial \gamma_1}{\partial \Delta^*}\right)=\frac{\sqrt{Y^2-4N \Delta^* Z}-Y}{2N \Delta^*},
\eeq
where 
\beq
\label{a6}
Y=M \Delta^*-2 N \gamma_{1}+\gamma_{1}
\left(\frac{\partial \zeta_1^*}{\partial \gamma_1}\right), \quad Z=\gamma_{1}\left(\frac{\partial \zeta_1^*}{\partial \Delta^*}\right)-2 M \gamma_{1}.
\eeq
In Equations \eqref{a5} and \eqref{a6}, we have introduced the quantities
\beq
\label{a7}
M=\frac{1}{2}\Bigg[x_1 \gamma_1 \Big(\frac{\partial \zeta_1^*}{\partial \Delta^*}\Big)_{\gamma_1}
+x_2 \gamma_2 \Big(\frac{\partial \zeta_2^*}{\partial \Delta^*}\Big)_{\gamma_1}\Bigg], \quad N=\frac{1}{2}\Bigg(x_1 \gamma_1 \frac{\partial \zeta_1^*}{\partial \gamma_1}
+x_2 \gamma_2 \frac{\partial \zeta_2^*}{\partial \gamma_1}\Bigg).
\eeq
In addition, the derivatives $(\partial_{\Delta^*}\zeta_i^*)$ and $(\partial_{x_1}\zeta_i^*)$ are given by 
\beq
\label{a9}
\left(\frac{\partial \zeta_i^*}{\partial \Delta^*}\right)=\left(\frac{\partial \zeta_i^*}{\partial \Delta^*}\right)_{\gamma}+\left(\frac{\partial \zeta_i^*}{\partial \gamma_1}\right)\left(\frac{\partial \gamma_1}{\partial \Delta^*}\right), \quad \left(\frac{\partial \zeta_i^*}{\partial x_1}\right)=\left(\frac{\partial \zeta_i^*}{\partial x_1}\right)_{\gamma}+\left(\frac{\partial \zeta_i^*}{\partial \gamma_1}\right)\left(\frac{\partial \gamma_1}{\partial x_1}\right).
\eeq
Finally, the derivative $\partial \gamma_1/\partial x_1$ can be written as
\beq
\label{a10}
\left(\frac{\partial \gamma_1}{\partial x_1}\right)=-\frac{\gamma_{1}\frac{\partial \zeta_1^*}{\partial x_1}+Q\left[\Delta^* \left(\frac{\partial \gamma_1}{\partial \Delta^*}\right)-2\gamma_1\right]}
{\gamma_{1}\frac{\partial \zeta_1^*}{\partial \gamma_1}+N
\left[\Delta^* \left(\frac{\partial \gamma_1}{\partial \Delta^*}\right)-2\gamma_1\right]},
\eeq
where
\beq
\label{a11}
Q=\frac{1}{2}\left(x_1\gamma_{1}\frac{\partial \zeta_1^*}{\partial x_1}+x_2\gamma_{2}\frac{\partial \zeta_2^*}{\partial x_1}\right).
\eeq
In Equations \eqref{a10} and \eqref{a11}, the derivative $\partial_{x_1}\zeta_i^*$ is taken at $\gamma_i\equiv \text{const}$.

\reftitle{References}
\bibliography{diffusionDelta}

\end{document}